\documentclass[preprint,10pt,sort&compress]{elsarticle}
\usepackage[a4paper, total={6.5in, 8.6in}]{geometry}

\usepackage[utf8]{inputenc}

\usepackage{amsmath}
\usepackage{bbm}
\usepackage{mathtools}
\usepackage{amsfonts}
\usepackage{mathrsfs}
\usepackage{slashed}
\usepackage{tensor}
\usepackage{bbold}
\usepackage{MnSymbol}
\usepackage{textcomp}
\usepackage[export]{adjustbox}
\usepackage[]{mdframed}

\usepackage{graphicx}
\usepackage{float}
\usepackage{xcolor}
\usepackage{array}
\usepackage[abs]{overpic}

\usepackage{placeins}
\usepackage{makecell}
\usepackage{subcaption}

\usepackage{xspace}
\usepackage{siunitx}
\usepackage{xfrac}
\usepackage{hyperref}
\usepackage[nameinlink]{cleveref}
\usepackage{appendix}

\usepackage{listings}
\definecolor{backcolor}{rgb}{0.99,0.98,0.98}
\definecolor{string-color}{rgb}{0.3333, 0.5254, 0.345}
\definecolor{darkgrey}{rgb}{0.0627, 0.07, 0.082}
\definecolor{darkred}{rgb}{0.3, 0.05, 0.05}
\definecolor{codeblue}{rgb}{0.2,0.35,0.75}
\definecolor{codepurple}{rgb}{0.38,0.1,0.52}
\definecolor{codegray}{rgb}{0.5,0.5,0.5}
\definecolor{codegreen}{rgb}{0.05,0.3,0.05}
\definecolor{codered}{rgb}{0.6,0.2,0.1}
\definecolor{backgroundColour}{rgb}{0.99,0.99,0.98}
\lstdefinestyle{myStyle}{
    language = C++,
    basicstyle = {\ttfamily \small \color{darkgrey}},
    backgroundcolor = {\color{backcolor}},
	commentstyle=\color{codegreen},
    stringstyle = {\color{string-color}},
    keywordstyle = {\color{codeblue}},
    keywordstyle = [2]{\color{codepurple}},
    keywordstyle = [3]{\color{codered}},
    keywordstyle = [4]{\color{codegray}},
    keywordstyle = [5]{\color{codegreen}},
    otherkeywords = {<, >, :, ::, DiFfRG, constexpr, uint, size_t, &, get, vector, array, Tensor, Scalar, FunctionND},
    morekeywords = [2]{AbstractModel, FEFunctionDescriptor, VariableDescriptor, ExtractorDescriptor, ComponentDescriptor
	TimeStepperSUNDIALS_IDA, UMFPack, Point, real, AD, NoJacobians, FE_AD,LLFFlux,FlowBoundaries
	},
    morekeywords = [3]{DiFfRG, CG, DG, dDG, LDG, def, Variables, dealii, std, autodiff},
    morekeywords = [4]{<, >, :, ::, ;, &},
    morekeywords = [5]{},
	breakatwhitespace=false,
	breaklines=true,
	captionpos=b,
	keepspaces=true,
	numbers=left,
	numbersep=5pt,
	numberstyle=\scriptsize\color{darkred},
	showspaces=false,
	showstringspaces=false,
	showtabs=false,
	tabsize=2
}
\lstdefinestyle{genStyle}{
    basicstyle = {\ttfamily \small \color{darkgrey}},
    backgroundcolor = {\color{backcolor}},
	commentstyle=\color{codegreen},
    stringstyle = {\color{string-color}},
    keywordstyle = {\color{codeblue}},
	breakatwhitespace=false,
	breaklines=true,
	captionpos=b,
	keepspaces=true,
	numbers=left,
	numbersep=5pt,
	numberstyle=\scriptsize\color{darkred},
	showspaces=false,
	showstringspaces=false,
	showtabs=false,
	tabsize=2
}
\lstdefinelanguage{CMake}{%
	morekeywords={if, else, endif, project, cmake_minimum_required, set, find_package, add_executable},%
	sensitive=false,%
	morecomment=[l]{\#},%
	morecomment=[s]{/*}{*/},%
	morestring=[b]",%
	otherkeywords={add_flows, setup_application},%
	keywordstyle = [2]{\color{codepurple}},
	morekeywords = [2]{REQUIRED, HINTS, VERSION, SYSTEM},
}
\lstdefinelanguage{Bash}{%
	morekeywords={if, else, fi, mkdir, cd, cmake, bash, git},%
	sensitive=false,%
	morecomment=[l]{\#},%
	morecomment=[s]{/*}{*/},%
	morestring=[b]",%
    keywordstyle = {\color{codeblue}},
    keywordstyle = [2]{\color{codepurple}},
    morekeywords = [2]{$, /, ..},
}

\usepackage{mmacells}
\mmaDefineMathReplacement[≤]{<=}{\leq}
\mmaDefineMathReplacement[≥]{>=}{\geq}
\mmaDefineMathReplacement[≠]{!=}{\neq}
\mmaDefineMathReplacement[→]{->}{\to}[2]
\mmaDefineMathReplacement[⧴]{:>}{:\hspace{-.2em}\to}[2]
\mmaDefineMathReplacement{∉}{\notin}
\mmaDefineMathReplacement{∞}{\infty}
\mmaDefineMathReplacement{𝕕}{\mathbbm{d}}
\mmaSet{
	leftmargin=4.5em,
	morefv={gobble=1},
	linklocaluri=mma/symbol/definition:#1,
	morecellgraphics={yoffset=1.9ex},
	labelsep=0.5em
}

\usepackage{tikz}
\usetikzlibrary{backgrounds}
\usetikzlibrary{decorations.pathreplacing}
\usetikzlibrary{shapes, arrows,calc}
\tikzstyle{roundbox} = [rectangle, draw, text centered, rounded corners,
minimum height=2em, minimum width=2em, draw=black!10, fill=blue!4]
\tikzstyle{process} = [rectangle, draw, minimum height=1em,
minimum width=3em, text centered, draw=black!10, fill=green!4]
\tikzstyle{integration} = [ellipse, draw, text centered, minimum height=1em,
minimum width=3em, draw=black!10, fill=red!4]
\tikzset{font={\fontsize{9pt}{11}\selectfont}}

\usepackage{booktabs}
\usepackage{multirow}

\newcolumntype{C}{>{$}c<{$}}
\AtBeginDocument{
	\heavyrulewidth=.08em
	\lightrulewidth=.05em
	\cmidrulewidth=.03em
	\belowrulesep=.65ex
	\belowbottomsep=0pt
	\aboverulesep=.4ex
	\abovetopsep=0pt
	\cmidrulesep=\doublerulesep
	\cmidrulekern=.5em
	\defaultaddspace=.5em
}

\newcommand{\DiFfRG}{\texttt{DiFfRG}\xspace}
\newcommand{\cpp}[1]{\lstinline[language=C++,style=myStyle]|#1|}
\newcommand{\cmake}[1]{\lstinline[language=CMake]|#1|}
\newcommand{\mathem}[1]{\lstinline[language=Mathematica]|#1|}
\newcommand{\bash}[1]{\lstinline[language=Bash]|#1|}
\newcommand{\LEGO}{LEGO\textsuperscript{\textregistered}}


\graphicspath{{./figures/}}

\newcommand{\gettitle}{DiFfRG:\\A Discretisation Framework for functional Renormalisation Group flows}

\hypersetup{
	pdftitle={\gettitle},
	pdfauthor={Sattler},
	pdfkeywords={discontinuous galerkin} {finite element method} {functional renormalisation group} {effective potential} {phase transition} {numerical methods} {phase structure} {GPU computing} {parallel computing} {vertex expansion} {derivative expansion},
	bookmarksopen=true,
	bookmarksopenlevel=2,
	bookmarksnumbered=true,
	colorlinks,
	linkcolor={red!75!black},
	citecolor={blue!75!black},
	urlcolor={blue!75!black}
}

\journal{Computer Physics Communications}

\begin{document}

\renewcommand{\thefootnote}{\fnsymbol{footnote}}
\begin{frontmatter}

	\title{\gettitle}

	\author[a]{Franz R. Sattler 
		\footnote{\url{sattler@thphys.uni-heidelberg.de}}}

	\author[a,b]{Jan M. Pawlowski}

	\address[a]{Institut f\"ur Theoretische
		Physik, Universit\"at Heidelberg, Philosophenweg 16, 69120 Heidelberg,
		Germany}
	\address[b]{ExtreMe Matter Institute EMMI, GSI, Planckstr. 1, 64291 Darmstadt, Germany}

	\begin{abstract}

		We introduce \texttt{DiFfRG} (\texttt{Di}scretisation \texttt{F}ramework for \texttt{f}unctional \texttt{R}enormalisation \texttt{G}roup flows), a comprehensive computational \texttt{C++} framework for solving functional Renormalisation Group flows in very general truncation schemes.
		Its central features are threefold: Firstly, the use of Finite Element Methods (FEM) for efficient, easy to set up and quantitatively reliable computation of field dependences. Secondly, the (simultaneous) setup of large, fully momentum-dependent vertex expansions. Thirdly, efficient time-discretisation methods incorporating insights from studies of solving theories which exhibit spontaneous symmetry breaking, going hand in hand with shocks and an exponential increase of the information flow velocity in field space. 
		The framework provides a \texttt{Mathematica} package for automatic code generation of flow equations, finite and zero temperature integration routines with support for GPU hardware and extensive parallelisation capabilities. Detailed examples and tutorials are provided and discussed herein which showcase and introduce the framework to the user. We illustrate the capabilities of \DiFfRG with four examples, with the complete codes fully included: finite temperature O(N) theory, a Quark-Meson model, SU(3) Yang-Mills theory and four-Fermi flows in the QCD phase diagram.

	\end{abstract}

\end{frontmatter}
\renewcommand{\thefootnote}{\arabic{footnote}}

\vspace{0.25cm}
{\bf PROGRAM SUMMARY}\\\vspace{-0.2cm}

\begin{small}
	\noindent
	{\em Program Title:}  DiFfRG \\
	{\em Licensing provisions:} GPLv3                                   \\
	{\em Programming language:} C++, Mathematica, Python                                  \\
	{\em Computer:}  Any Linux/Mac machine                           \\
	{\em Nature of problem:}  Evaluate large systems of fRG flows with full effective potentials and/or vertex expansions\\
	{\em Solution method:} Combine finite element methods and efficient time-steppers with fast numerical integration of flow equations and momentum grids. \\
	{\em Unusual features:} The code-generation capabilities of the \texttt{Mathematica} subpackage allow for very fast development of extremly large systems of flow equations. Field-space discretisations can be easily defined by the user with a range of different finite element methods.\\
\end{small}

\clearpage
\tableofcontents

\section{Introduction}
\label{sec:introduction}

The functional Renormalisation Group (fRG) is a powerful non-perturbative approach for solving quantum field theories. It has found widespread successful applications in different areas ranging from condensed matter and statistical physics systems, nuclear physics, high-energy physics including beyond standard model applications, to quantum gravity, for a recent comprehensive review see \cite{Dupuis:2020fhh}. This versatility is based on the impressive flexibility of the fRG in terms of the mainly numerical implementation of different systematic expansion schemes. While in many cases the choice of an appropriate expansion scheme is key to even quantitative reliability in simple approximation, there are also many applications in complex systems, where high order approximations are required for even getting a qualitative understanding of the intricate physics at hand. One of the many examples for the latter are competing order phenomena present in many condensed matter systems or the change of the dynamical degrees of freedom as happens e.g.~in QCD at finite density. 

Consequently, the need for flexible and scalable codes for solving complex systems is paramount across the different application areas of the fRG. If using the fRG there, one either has to write dedicated codes or one relies on existing libraries. The former is both time-consuming and requires a thorough knowledge of programming, advanced numerical algorithms and techniques. Furthermore, making sure the code is bug-free is of crucial importance but automated testing presents another significant time-investment.
Thus, it is preferable to build on existing and tested code, as this can vastly shorten the time-cost of any research project which relies on numerical simulations using the fRG. To date, only very few codes which can be used to build general large-scale fRG simulations are openly available. Recently, the numerical libraries \texttt{DivERGe} \cite{10.21468/SciPostPhysCodeb.26}, developed for simulating fermionic models on arbitrary lattices, and \texttt{KeldyshQFT} \cite{Ritz:2024glj} for the real-frequency multiloop fRG, have been released. Also, in \cite{Ihssen:2022xkr} a framework has been open-sourced for the discretisation of effective potentials using finite element (FE) methods.

In the present work we provide a very general, extendable open-source computational toolkit for continuum fRG, not constrained to very specific use-cases. This library, the \textit{Discretisation Framework for the functional Renormalisation Group} (\DiFfRG) is designed with the necessity of combining multiple discretisation approaches and theoretical setups in mind. \texttt{DiFfRG} incorporates several established tools for the derivation of fRG flows, in particular the Mathematica packages \texttt{QMeS-derivation}\cite{Pawlowski:2021tkk}, \texttt{FormTracer}\cite{Cyrol:2016zqb} and \texttt{TensorBases}\cite{TensorBases}. 
As we aim at a broad application spectrum, \texttt{DiFfRG} is designed as a set of tools that can be combined in the way the user needs them rather than a straight-forward library which can be directly applied to a problem.

The code has already been utilised in \cite{Ihssen:2023xlp, Ihssen:2024miv}: in \cite{Ihssen:2023xlp} we have computed the full effective potential and the field-dependent Yukawa coupling of the Quark-Meson model, a low-energy effective model of QCD. In \cite{Ihssen:2024miv}, the \DiFfRG framework has been used to resolve 2+1-flavor 2QCD in the vacuum with a full mesonic potential. Importantly, with the publication of the \DiFfRG framework, these results are fully reproducible and hence testable. These example already indicate, that the full potential of \DiFfRG is used in situations, where the physics at hand requires flows of field-dependent quantities and/or momentum-dependent vertices. An example is the QCD-setup of \cite{Ihssen:2024miv}, where a fully field-dependent effective potential for the mesons has been calculated, together with the relevant gauge and quark correlation functions. A stand-alone feature of \DiFfRG is the easy and direct possibility of using hydrodynamic methods for the flows of full effective potentials or other field-dependent quantities. Applications of numerical fluid dynamics to fRG equations has so far been investigated in~\cite{Grossi:2019urj, Wink:2020tnu, Grossi:2021ksl, Koenigstein:2021syz, Koenigstein:2021rxj, Steil:2021cbu, Stoll:2021ori, Ihssen:2022xjv, Ihssen:2022xkr, Ihssen:2023qaq, Murgana:2023xrq, Ihssen:2023xlp, Ihssen:2024miv} and we have incorporated insights gained from these studies into \DiFfRG.

The \DiFfRG library is openly accessible on github \cite{DiFfRGgit}.
\DiFfRG is mainly object-oriented and uses templates to make any combinations of its algorithms possible, which allows for a very flexible use of either the entire framework, or only parts of it. Thus, good extendability is given and interfaces to other code bases are easy to realise. The GPU algorithms are implemented using CUDA and associated libraries.
For our finite element implementation, we build on the \texttt{deal.ii} library \cite{dealII95}, combined with the \texttt{SUNDIALS} solver suite \cite{hindmarsh2005sundials, gardner2022sundials} and \texttt{Boost} for efficient time-stepping. Jacobians are calculated using automatic differentiation, implemented with the \texttt{autodiff} library \cite{autodiff}. Autodifferentiation is a technique popular in machine learning and hydrodynamics, among others, where derivatives of computations are evaluated at machine precision by using standard rules of differentiation within the computer calculus. 

This work provides a general overview of the entire framework. Besides the information given here, it is also useful to work through the set of tutorials and examples provided with \DiFfRG. These tutorials, that provide in-depth step-by-step explanations, can be accessed through the documentation, which is automatically built with \DiFfRG.

In \Cref{sec:theory} we briefly introduce the fRG and some central concepts that will be referred to later.
In \Cref{sec:examples}, we describe and show some results from three illustrative benchmark applications of the \DiFfRG framework: A finite temperature O(N) model in LPA, SU(N) Yang-Mills theory in a vertex expansion, and a Quark-Meson model in LPA'.
In \Cref{sec:install} we describe the installation of the \DiFfRG framework and the first steps to setup a bare-bones application.
In \Cref{sec:design} the structure of \DiFfRG is explained. Specifically, we discuss the entire process of creating a \DiFfRG application from starting with the Mathematica setup of deriving flow equations and automatically generating the code to putting the system of equations together in C++ and running the flow integration.

\section{Theoretical background}
\label{sec:theory}

\subsection{The functional Renormalisation Group}
\label{sec:fRG}
\noindent
A recent comprehensive review on the functional renormalisation group is with applications ranging from condensed matter systems to quantum gravity is provided by \cite{Dupuis:2020fhh}, further reviews dedicated to specific research areas and works can be found there. Here we only sketch the most important derivations and properties that are relevant for the current numerical context. 

The standard fRG setup provides a flow from the microscopic theory at a large momentum scale $k = \Lambda_\textrm{UV}$ to the full macroscopic theory at $k=0$. At a given momentum cutoff scale the flow incorporates fluctuations at momenta $p^2\approx k^2$, and all momentum scales are included by successively integrating the flow. Moreover, the general setup already accommodates composite degrees of freedom, obtained by coupling the source $J_\phi$ not to the fundamental fields $\hat\varphi$, but to general composite operators $\hat\phi(\hat\varphi)$ that may or may not include the fundamental field, see \cite{Pawlowski:2005xe, Ihssen:2024ihp}.  A well-known example is a scalar theory with $\hat\phi=(\,\hat\varphi(x)\,,\, \hat \varphi(x)\hat\varphi(y)\,)$, where the second entry is nothing but the full two-point function, this leads to a variant of the two-particle irreducible effective action. The general composite field may also depend on the cutoff scale, which allows for a flowing change of the degrees of freedom, leading to flowing fields. The starting point of the derivation of the generalised functional flow equation  \cite{Pawlowski:2005xe} is the generating functional $Z_k[J_\phi]$ of the theory with a quadratic regulator insertion $R_k$, 
\begin{align}
	Z_{\phi,k}[J_\phi] = \exp\left\{\frac12 \frac{\delta}{\delta J_\phi^a}R^{ab}_k\frac{\delta}{\delta J_\phi^b}\right\}Z_\phi[J_\phi]\,, 
\label{eq:Zphi}
\end{align}
where $Z_\phi[J_\phi]$ is the full generating functional of the theory with currents $J_\phi$ that couple to the (potentially) composite field $\hat\phi$ and the index $a$ sums over space-time, Lorentz and internal indices as well as species of fields, for more notational details see e.g.~\cite{Pawlowski:2005xe, Pawlowski:2021tkk}. The generating functional $Z_k$ in \labelcref{eq:Zphi} depends on the RG-scale $k$ and $R^{ab}_k$ is a matrix in field space. By lowering $k$, the momentum fluctuations are successively included  the generating functional $Z_k[J]$. Note also that while we explain the setup with a momentum cutoff scale $k$, the setup also incorporates general RG-flows or reparametrisations of the theory. Further cases considered in the literature are cutoffs in thermal fluctuations, cutoffs in the amplitude of the field, and cutoffs that introduces a maximal central time in a non-equilibrium time evolution, for these and more examples see \cite{Dupuis:2020fhh} and literature therein. 

The functional flow equation is usually formulated in terms of the one-particle effective action, the modified Legendre transform of $\ln Z[J_\phi]$, 
\begin{align}
	\Gamma_{\phi,k}[\phi] = \sup_{J_\phi} \left( J^a_\phi \phi_a - \ln Z_{\phi,k}[J_\phi]\right) - \frac12 \int \phi_a R^{ab}_k \phi_b\,.  
\end{align}
The RG-scale derivative of the effective action is directly related to that of $\ln Z_{\phi,k}[J_\phi]$. We formulate it for a general theory with \cite{Pawlowski:2005xe}, 
\begin{align}
  \left( k\partial_k  + \dot\Phi^a[\Phi] \frac{\delta}{\delta \Phi^a} \right) \Gamma_k[\Phi] = \frac12\textrm{Tr}\left[\frac{1}{\Gamma^{(2)}_k[\Phi] + R_k}\,\left( k\partial_k + 2 \frac{\delta  }{\delta \Phi}  \dot \Phi[\Phi] \right) R_k\right]
  \,. 
\label{eq:GenfRG}
\end{align}
\Cref{eq:GenfRG} depends on the super field $\Phi$, that collects all degrees of freedom including possible composite fields, and the flowing field transformation $\dot\Phi_k[\Phi]$ is at our disposal. The trace sums over all indices including space-time, often it is also abbreviated as $\textrm{STr}$, a super trace. The expression $\Gamma^{(2)}_k$ denotes the second derivative of the effective action w.r.t.~the fields $\Phi$, see \labelcref{eq:FlowGamman} 

The generalised flow equation \labelcref{eq:GenfRG} includes emergent composites \cite{Gies:2001nw, Pawlowski:2005xe, Floerchinger:2009uf}, and it is the Legendre transform of the Wegner equation \cite{Wegner:1974sla} for the Wilsonian effective action. \Cref{eq:GenfRG} boils down to the Wetterich equation \cite{Wetterich:1992yh} for $\Phi=\varphi$ with the fundamental mean fields $\varphi$, which implies $\dot\Phi_k[\Phi]=0$. Its analogue with $\dot\Phi[\Phi]=0$ in the Wegner equation is the Polchinski equation \cite{Polchinski:1983gv}. For a respective discussion see \cite{Ihssen:2022xjv}.   

\Cref{eq:GenfRG} comprises a general RG-flow of the QFT at hand, including general reparametrisations of the theory via flowing fields $\dot\Phi$. By taking derivatives, flow equations for all correlation functions of the theory can be obtained. These flow equations have a structure such that
\begin{align}
  k\partial_k \Gamma_k^{(n)} = \textrm{Flow}(\Gamma_k^{(2)},\dotsc, \Gamma_k^{(n+2)}, \dot\Phi)\,, \qquad \textrm{with} \qquad \Gamma_k^{(n)} =\frac{\delta^n \Gamma_k}{\delta\Phi^n}\,, 
  \label{eq:FlowGamman} 
\end{align}
for the 1PI correlation functions $\Gamma^{(n)}$. Thus, \labelcref{eq:GenfRG} generates an infinite tower of flow equations. This tower has to be truncated for numerical applications and we explain this procedure at the example of the two most commonly used systematic approximation schemes: \\[-2ex]

The first of these systematic expansion schemes is the \textit{vertex expansion}, in which the flow equations are ordered in terms of the order $n$ of the correlation functions. Then, the tower of flow equations is terminated at some finite order $n=N_\textrm{max}$ with $\Gamma^{(n> N_\textrm{max})}\equiv 0$.
In this approximation, one considers only a finite number of correlation functions which are integrated along the RG-trajectory.

Another systematic expansion scheme, which has shown its great potential in studies of systems with spontaneous symmetry breaking is the \textit{derivative expansion}, where the effective action is expanded in powers of momenta $p^2/m^2_\textrm{gap}$ where $m^2_\textrm{gap}$ is the mass gap of the theory in the presence of the cutoff $k$. In such an expansion scheme, one keeps the full field dependence at each expansion order, increasing only the order of the momenta considered. The momentum cutoff fRG is tailor made for this as, roughly speaking, it increases the mass gap $m^2_\textrm{gap}$ by $k^2$, leading to a well-controlled expansion. Starting with the classical dispersion, the zeroth order of the derivative expansion is called the local potential approximation (LPA). The first order takes into account any term of the shapes $1/2 \int Z(\rho)\, (\partial \varphi^a)^2$ and $1/2 \int Y(\rho) \,(\varphi^a\partial\varphi^a)^2$. Here $\rho= (\varphi^a )^2/2$ and $a,b$ simply label internal indices, an example being a field $\varphi^a$ with $a,b=1,...,N$ in an $O(N)$ theory. 

For complex systems and if striving for quantitative accuracy, combinations of these two schemes and further ones are used. 
A recent detailed general analysis of the systematic error control of expansion schemes in functional RG equations is provided in \cite{Ihssen:2024miv}, based on the example of QCD. In this context it is worth emphasising that the modular form of the fRG, mainly due to its one-loop exact form, facilitates systematic error estimates, as it allows to directly use the error estimates and stability of results in subsystems. This is captured in the \LEGO principle, see the discussion in \cite{Ihssen:2024miv}.
Finally, we stress the optimisation potential of \textit{physics-informed} RG flows (PIRGs) put forward in \cite{Ihssen:2024ihp}. In these flows one is fully exploiting the possibility of general reparametrisations induced by the flowing fields $\dot\Phi[\Phi]$ for devising rapidly converging expansion schemes or simply optimising the numerical effort. The combination of the numerical framework \DiFfRG presented here and the optimisation potential of PIRGs should allow for many fRG applications beyond the current state of the art, as well as significantly lowering the entry threshold to these fRG applications.

Both vertex expansions (which do not have an additional field dependence) and fully field-dependent quantities are directly supported by \DiFfRG. Their implementation is explained first separately for the two expansion approaches and then their combination in \Cref{sec:design2}.
Once a truncation is chosen, one obtains a system of differential equations for a set of chosen RG-scale dependent objects:
\begin{align}
	k\partial_k \lambda_{k,i}(\dotsc) = \textrm{Flow}(\lambda_i)(\lambda_{k,1},\,\dots)\qquad i = 1,\dotsc
\end{align}
The flows $\textrm{Flow}(\lambda_i)$ are obtained by performing a projection of $k\partial_k\Gamma_k[\Phi]$ onto $k\partial_k\lambda_i$ and then carrying out the full trace, i.e.~tracing the group structure of the system and performing integrations where possible. The functions $\textrm{Flow}(\lambda_i)$, or integration kernels thereof if momentum-integrals have not yet been performed, are the building blocks of the system of equations which is then solved with \DiFfRG. We explain how to obtain these flows in an automated way and how to set up their integration in \Cref{sec:design}.

Lastly, we comment quickly on our definition of RG-time. Numerically, time evolution will always go \textit{forward} in time. Therefore, if $t$ as used in \DiFfRG is interpreted as an RG-time, its definition is given by
\begin{align}
	t = t_+ = \log\left({\Lambda}/{k}\right)\,.
\end{align}
Note that this is different from the commonly used definition of the RG-time in the literature,
\begin{align}
	t_- = \log\left({k}/{\Lambda}\right) = -t_+\,,
\end{align}
and thus
\begin{align}
	k\partial_k = \partial_{t_-} = - \partial_{t_+}\,.
\end{align}
For reasons of clarity, we distinguish these definitions from here on both here and within the documentation of the \DiFfRG  framework. If no subscript is given, we mean the numerical $t = t_+$.

\subsection{Finite element methods}
\label{sec:FEM}
\noindent
A stand-out feature of \DiFfRG is the ability to quickly and easily define fully field-dependent functions and integrate their flow equations. To make the description and usage of this part of the code clearer, we give in this section a very brief overview how FE methods work in general and why they are useful for fRG applications. This section is recommended to read in order to understand how and why to use the field-dependent flows of \DiFfRG with FE methods.

In contrast to finite-difference grid methods or spectral methods, finite elements allow for higher-order local precision while providing increased numerical stability. Due to their geometrical flexibility, they are also better suited for applications in higher dimensions, which is encountered in a QFT context when a potential of two or more field-invariants is required, e.g. $V(\rho_1, \rho_2)$, where $\rho_1$ and $\rho_2$ are associated with two different particle species that may mix.

FE methods, especially discontinuous ones, allow furthermore for stable and precise capturing of non-analytic features of functions. This has been exploited in \cite{Grossi:2021ksl}, where shocks in field-space have been observed in the large-N Quark-Meson model at high densities and accurately captured using a discontinuous Galerkin (DG) method. 
DG methods are especially well-suited to convection-dominated problems, allowing for very high numerical precision. We will show below that fRG flow equations can be cast into convection-diffusion equations.

Going with the hydrodynamic analogue, we call field discretisations \textit{spatial discretisations}.
The mesh of such a discretisation is given by dividing the computational domain (i.e.~field space) $\Omega = \cup_k D_k$ into cells $D_k$. We define a function space $V_h$, whose elements all have support either only in one cell (discontinuous Galerkin) or in a small set of neighboring cells (continuous Galerkin). In \DiFfRG these are by default Legendre polynomials up to some finite order that can be chosen by the user.

All quantities are then expanded in terms of these basis functions $\varphi_i \in V_h$. This means that any FE function $u$ is represented as a vector of coefficients $u^{(i)}$ in a global decomposition
\begin{align}
	u_h(x) = \sum_i u^{(i)} \varphi_i(x)\,,\qquad x \in \Omega\,.
\end{align}
The subscript $h$ is used to distinguish any numerically approximated quantity, such as $u_h$, from the exact solution $u$. Note that $u = (u_1,u_2,\dots,u_n)$ may in general be vector-valued, and thus $u^{(i)} = (u_1^{(i)},u_2^{(i)},\dots,u_n^{(i)})$ may be too.
We make a very general Ansatz and assume that the flow equation of $u$ in field space has the shape
\begin{align}
	m_i(\partial_{t}u, u, t, x) + \partial_{x_j}F_{ij}(u,t,x,\dots) + s_i(u,t,x,\dots)  = 0\,,\quad i=1\dots n \,.
	\label{eq:FEMeq}
\end{align}
Possible additional arguments, indicated by $\dots$, can stand in for derivatives of $u$ or other quantities, depending on the precise discretisation scheme chosen.
Some standard nomenclature should be also explained here:
\begin{itemize}
	\item $m_i$ is called the mass function. In its simplest form, which is also the most often used, $m_i = \partial_t u_i$. We allow for relatively general mass functions to make certain transformations of the flow equation possible, e.g. an exponential formulation where one solves for $e^{u + k^2}$ as has been investigated in \cite{Ihssen:2023qaq}.
	Furthermore, $m_i$ can in some of its components not depend on $\partial_t u$ and thus represent an ODE instead of a PDE. Therefore, \labelcref{eq:FEMeq} allows for mixtures of at least one PDE and any number of ODEs.

	\item $F_{ij}$ is called the flux term. If it does not depend on higher derivatives of $u$, the flux term is purely convective and represents wave propagation through field space. Dependence on the first derivative leads to convective contributions to the flow of $u$.

	\item $s_i$ is called the source term, where all terms are stored that do not fit the criteria for the flux term.
\end{itemize}
The Ansatz \labelcref{eq:FEMeq} can be motivated by an often encountered example, investigated e.g. in \cite{Grossi:2019urj, Koenigstein:2021rxj, Koenigstein:2021syz, Ihssen:2022xkr, Ihssen:2023qaq, Ihssen:2023xlp}: 
The flow equation of the effective potential $V(\rho)$ of an O(N) or a Yukawa model, with the spatial variable $x=\rho=\phi^2/2$ being the O(N) invariant built from the bosonic fields. It does only depend on its first and second derivatives in the local potential approximation (LPA):
\begin{align}
	\partial_t V(\rho) = - F(\partial_\rho V, \partial_\rho^2V, \rho, t)\,.
\end{align}
One can introduce the variable $u(\rho) = \partial_\rho V(\rho)$ and solve only for $u$ instead of $V$ without loss of information. We investigate this example further in \Cref{ex:ONfiniteT}.
The hydrodynamic equation \labelcref{eq:FEMeq} has to be discretised twofold: Its time-dependence and its field dependence. We will discretise here only the spatial part and discuss the time-discretisation in \Cref{sec:design2}.

The spatial discretisation is achieved by requiring \labelcref{eq:FEMeq} to hold true in its weak form. This means that the equation holds true if multiplied with all test functions $\varphi_i(x)$ from some set $U_h$, and integrated over the entire domain.
Here the name \textit{Galerkin} comes into play: in a Galerkin method, test and trial function spaces are chosen to be equal, $V_h = U_h$.
Explicitly, this procedure leads to the set of equations
\begin{align}
	R_{ij} = &\int_{\Omega} m_i(\partial_{t}u, u, x) \varphi_j(x) + \int_{\Omega} s_i(u,t,x,\dots) \varphi_j(x) \notag\\[2ex]
	&- \int_{\Omega} F_{ik}(u,t,x,\dots) \partial_{x_k}\varphi_j(x)
	+ \int_{\partial\Omega} \hat{\boldsymbol{n}}_k \hat F_{ik}(u,t,x,\dots) \varphi_j(x)\overset{!}{=}\, 0\,,\quad\forall\,\varphi_j \in V_h\,.
	\label{eq:FEresidual}
\end{align}
Here, $\hat{\boldsymbol{n}}$ is the outward pointing normal vector on the boundary $\partial\Omega$ of the domain $\Omega$ within which we define our system. In an fRG context, this is usually a part of the field space.

Here, we have also introduced $\hat F$, the flux on the boundary. It can be used to set von Neumann boundary conditions at $\partial \Omega$, which in a flux formulation are given by "mirroring" or inflow/outflow conditions. In a discontinuous Galerkin formulation additional boundary terms pop up due to the discontinuity at every cell border. This is discussed in more detail in \Cref{app:DG}.

\section{Example applications}
\label{sec:examples}
\noindent
All examples are  present in the folder \texttt{Examples/} and can be built using the scripts \texttt{build.sh}, present in every single subfolder, corresponding to the below examples.

\subsection{Finite temperature O(N) model}
\label{ex:ONfiniteT}
\noindent
The full Mathematica and C++ code for this example is also provided in  \texttt{Examples/ONfiniteT}. This example is meant to show the setup and performance of the available finite element methods. Therefore, there are multiple applications defined here for the continuous Galerkin (CG), direct discontinuous Galerkin (dDG) and local discontinuous Galerkin (LDG) finite element methods.

We consider an O(N) model in LPA, with the Ansatz for the effective action given by
\begin{align}
	\Gamma_k[\phi] = \int_x\left( \frac12 (\partial_\mu \phi)^2 + V(\rho) \right)\,,
\end{align}
where we have introduced the O(N) symmetric invariant $\rho= {\phi^2}/{2}$ and $\phi = (\phi_1,\,\dotsc,\phi_N)$. The integration $\int_x = \int_0^\beta d\tau\int d^{d-1}x$ splits into the integral over imaginary time $\tau$, with $\beta= 1/T$, and the spatial part. The regulator is chosen to be spatial, i.e. $R_{\phi,k}(p) = R_{\phi,k}(\boldsymbol{p})$, so that Matsubara sums can be performed analytically.

\begin{figure}[t]
	\centering
	\begin{minipage}[b]{0.48\linewidth}
		\vspace{-1.8cm}
		\includegraphics[width=1\linewidth]{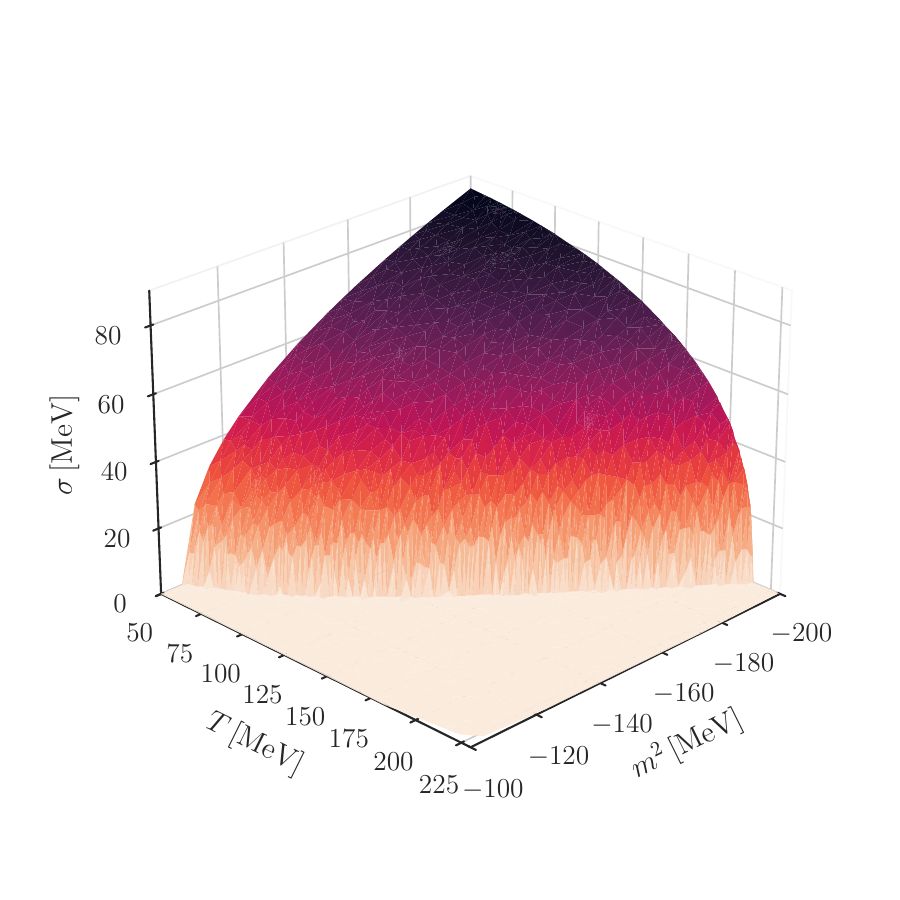}
		\vspace{-0.7cm}
		\caption{The phase diagram of the O(N) theory as described in \Cref{ex:ONfiniteT}. We show the order parameter $\sigma$ depending on temperature $T$ and initial mass parameter $m^2$. \hspace*{\fill}}
		\label{fig:ONFiniteTPD}
	\end{minipage}
	\hspace{0.02\linewidth}
	\begin{minipage}[b]{0.48\linewidth}
		\includegraphics[width=0.95\textwidth]{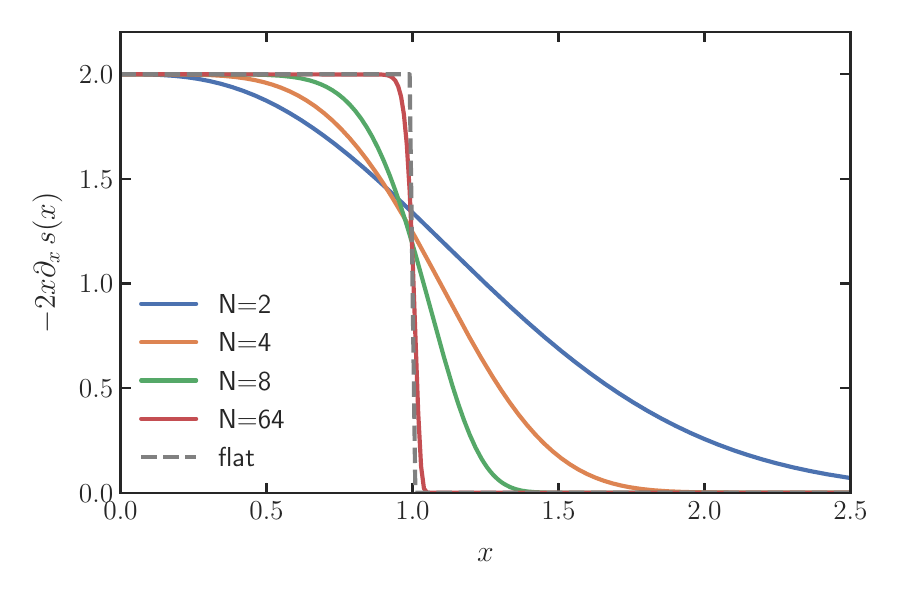}
		\caption{The derivative $-2x\partial_x s(x)$ of the polynomial exponential regulator shape function given in \labelcref{eq:RegPolyExp}. Note that $k\partial_k = -2x\partial_x$ if $x = \boldsymbol{p}^2/k^2$. We compare to the flat regulator, given by $f(x) = (1/x-1)\Theta(1-x)$.
			\hspace*{\fill}}\label{fig:polyExpReg}
		\vspace{-9pt}
	\end{minipage}
\end{figure}

As momentum integrals are performed numerically, we can use a smooth shape function. We choose $R_{\phi,k} = s_N\left(\frac{\boldsymbol{p}^2}{k^2}\right)$ with $s_N(x)$ given by
\begin{align}\label{eq:RegPolyExp}
	s_N(x) = k^2 e^{-\sum_{i=1}^{N} \frac{x^i}{i}}\,,
\end{align}
and we call the associated regulator the polynomial exponential regulator.
In all following examples, we will use the same shape function $s_8(x)$ for the regulators. We show the RG-time derivative of the  shape function in \Cref{fig:polyExpReg}.

From the Ansatz, one could easily derive the flow equation for $V(\rho)$ on paper. However, in the corresponding Mathematica notebook \texttt{ON.nb}, we perform the derivation using \texttt{QMeS} and \texttt{FormTracer}, do the Matsubara sum and export the resulting flow equation code directly to C++.
Once one has obtained $\text{Flow}(V)$, note that $\partial_\rho V = m_\pi^2$, and one can implement the following PDE using a FE method:
\begin{align}\label{eq:ONfiniteTCD}
	\partial_t m_\pi^2(\rho) + \partial_\rho \text{Flow}(V) = 0
\end{align}
The field invariant $\rho$ is chosen as the spatial coordinate, the analogue to $x$ in \labelcref{eq:FEMeq} and $m_\pi^2 = u_1$ is the (only) FE function.
\labelcref{eq:ONfiniteTCD} is a convection-diffusion equation fitting the general Ansatz \labelcref{eq:FEMeq}, as $\text{Flow}(V)$ depends only on $m_\pi^2(\rho)$ and $\partial_\rho m_\pi^2(\rho)$. To implement it in the numerical model, we can identify $F_{11}$ as introduced in \labelcref{eq:FEMeq} with $\text{Flow}(V)$. The resulting numerical model is thus very compact and easy to understand.

At the initial scale $\Lambda_\textrm{UV} = \qty{650}{\MeV}$ we use the initial condition $m_\pi^2(\rho) = m^2 + \frac\lambda2  \rho$, where $\lambda=71.6$.
A python notebook in the folder, \texttt{phasediagram.ipynb} can be used to plot the $m^2$--$T$ phase diagram of the O(N) theory and we show the result in \Cref{fig:ONFiniteTPD}.

The simulation parameters, i.e. the used group size $N$, the initial UV scale, temperature (in units of GeV) and initial potential can be adjusted by modifying the \texttt{parameter.json} file:
\begin{lstlisting}[language=C++, style=myStyle]
	{
		"physical": {
			"Lambda": 0.65,
			"N": 2.0,
			"m2": -0.2,
			"lambda": 71.6,
			"T": 5e-2
		},
		...
	}
\end{lstlisting}
Alternatively, parameters can be overridden directly from the shell when running the program:
\begin{lstlisting}[language=Bash]
	$ ./CG -sd /physical/T=0.05 -sd /physical/m2=-0.2 
\end{lstlisting}
The above call sets the two parameters \cpp{"T"} and \cpp{"m2"} to custom values, ignoring the information in \texttt{parameter.json}. For an explanation of the syntax, simply call
\begin{lstlisting}[language=Bash]
	$ ./CG --help
\end{lstlisting}

Additionally, in this example adaptive mesh refinement is implemented in a very straight-forward way; the refinement criterion for CG and dDG is chosen as the size of the local Hessian, i.e. $\partial_\rho^2 m_\pi^2$, by implementing the method
\begin{lstlisting}[language=C++, style=myStyle]
template <int dim, typename NumberType, typename Solution>
void cell_indicator(NumberType &indicator, const Point<dim> &x,
                    const Solution &sol) const
{
	const auto &fe_hessians = get<"fe_hessians">(sol);
	indicator = fe_hessians[idxf("m2")][0][0];
}
\end{lstlisting}
in the numerical model, the central object in a \DiFfRG application.
We use the \cpp{HAdaptivity} class (passed to the time-stepper) to perform the refinement.
One can turn on the adaptive refinement by modifying the appropriate section of the parameter file \texttt{parameter.json}:
\begin{lstlisting}[language=C++, style=myStyle]
{
	...
	"discretization": {
		...
		"adaptivity": {
			"start_adapt_at": 0E0,
			"adapt_dt": 2E-1,
			"level": 6,
			"refine_percent": 1E-1,
			"coarsen_percent": 2E-1
		}
	},
	...
}
\end{lstlisting}
If the parameter \cpp{"level"} is set to a number greater than $0$, adaptive mesh refinement is used. \cpp{"level"} sets the maximum number of refinements on a single cell. Then, after a time \cpp{"start_adapt_at"}, the refinement indicator is evaluated at an interval of \cpp{"adapt_dt"}. Every time, a fraction \cpp{"refine_percent"} of all cells are refined and the fraction \cpp{"coarsen_percent"} of cells are coarsened (if they have been refined before), in accordance with the local size of the refinement indicator.

\subsection{Four-Fermi flows}
\label{ex:fourFermi}
\noindent
The full Mathematica and C++ code for this example is also provided in  \texttt{Examples/FourFermi}. This example is meant as a short showcase for the Mathematica functionality of the framework and we present here the main ideas of how the Mathematica setup works. 
A thorough explanation of the code is given in \Cref{app:fourFermi}.

We embed this example in a very simplified QCD setup - the system consists of only gluons and quarks. We use vacuum data for the glue and quark propagators, the quark-gluon coupling $g_{A\bar qq,k}$, and assume that the quark-mass $M_q(p) = 0$. The data is taken from \cite{Ihssen:2024miv} and the diagrammatic rules can be also found in the appendix thereof. We regulate everything using 3D-regulators and the same polynomial exponential regulator as in \Cref{ex:ONfiniteT}.

On top of this data, we calculate the flows of a set of four-Fermi couplings, associated with a ten-component four-Fermi basis.
We utilise the following Fierz-complete four-Fermi basis \cite{Ihssen:2024miv}, adapted from \cite{Cyrol:2017ewj}:

\begin{minipage}{0.48\textwidth}
	\begin{alignat}{2}
		& \mathcal{L}_{\scriptscriptstyle\bar q\bar qqq}^{\scriptscriptstyle(\sigma-\pi)}       &  & = (\bar q\tau_0  q)^2 + (\bar q i \gamma_5 \boldsymbol{\tau} q )^2
		\,,\notag                                                                                                                                                                             \\[1ex]
		& \mathcal{L}_{\scriptscriptstyle\bar q\bar qqq}^{\scriptscriptstyle(\eta')}            &  & = (\bar q \boldsymbol{\tau} q )^2 + (\bar q i \gamma_5 \tau_0 q )^2
		\,,\notag                                                                                                                                                                             \\[1ex]
		& \mathcal{L}_{\scriptscriptstyle\bar q\bar qqq}^{\scriptscriptstyle(S+P)_+}            &  & = (\bar{q} \tau_0 q)^2 + (\bar{q}\gamma_5 \boldsymbol{\tau} q)^2
		\notag                                                                                                                                                                                \\&&&\hspace{4ex} - (\bar q \gamma_5 \tau_0 q )^2 - (\bar q \boldsymbol{\tau} q )^2
		\,,\notag                                                                                                                                                                             \\[1ex]
		& \mathcal{L}_{\scriptscriptstyle\bar q\bar qqq}^{\scriptscriptstyle(S-P)_-}            &  & = (\bar{q} \tau_0 q)^2 - (\bar{q}i \gamma_5 \boldsymbol{\tau} q)^2
		\notag                                                                                                                                                                                \\&&&\hspace{4ex} + (\bar q i \gamma_5 \tau_0 q )^2 - (\bar q \boldsymbol{\tau} q )^2
		\,,\notag                                                                                                                                                                             \\[1ex]
		& \mathcal{L}_{\scriptscriptstyle\bar q\bar qqq}^{\scriptscriptstyle(S+P)_+^\text{adj}} &  & = (\bar{q} \tau_0 T^a q)^2 + (\bar{q}\gamma_5 \boldsymbol{\tau} T^a q)^2
		\notag                                                                                                                                                                                \\&&&\hspace{4ex} + (\bar q \gamma_5 \tau_0 T^a q )^2 + (\bar q \boldsymbol{\tau} T^a q )^2
		\,,\notag
	\end{alignat}
\end{minipage}
\begin{minipage}{0.48\textwidth}\vspace{-1ex}
	\begin{alignat}{2}
		& \mathcal{L}_{\scriptscriptstyle\bar q\bar qqq}^{\scriptscriptstyle(S+P)_-^\text{adj}} &  & = (\bar{q} \tau_0 T^a q)^2 + (\bar{q}i \gamma_5 \boldsymbol{\tau} T^a q)^2
		\notag                                                                                                                                                                                \\&&&\hspace{4ex} - (\bar q i \gamma_5 \tau_0 T^a q )^2 - (\bar q \boldsymbol{\tau} T^a q )^2
		\,,\notag                                                                                                                                                                             \\[1ex]
		& \mathcal{L}_{\scriptscriptstyle\bar q\bar qqq}^{\scriptscriptstyle(S-P)_-^\text{adj}} &  & = (\bar{q} \tau_0 T^a q)^2 - (\bar{q}i \gamma_5 \boldsymbol{\tau} T^a q)^2
		\notag                                                                                                                                                                                \\&&&\hspace{4ex} + (\bar q i \gamma_5 \tau_0 T^a q )^2 - (\bar q \boldsymbol{\tau} T^a q )^2
		\,,\notag                                                                                                                                                                             \\[1ex]	
		& \mathcal{L}_{\scriptscriptstyle\bar q\bar qqq}^{\scriptscriptstyle(V+A)}              &  & = (\bar{q}\gamma_\mu \tau_0 q)^2 + (\bar{q}\, i \gamma_\mu\gamma_5\tau_0\,q)^2
		\,,\notag                                                                                                                                                                             \\[1ex]
		& \mathcal{L}_{\scriptscriptstyle\bar q\bar qqq}^{\scriptscriptstyle(V-A)}              &  & = (\bar{q}\gamma_\mu \tau_0 q)^2 - (\bar{q}\, i \gamma_\mu\gamma_5\tau_0\,q)^2
		\,,\notag                                                                                                                                                                             \\[1ex]
		& \mathcal{L}_{\scriptscriptstyle\bar q\bar qqq}^{\scriptscriptstyle(V-A)^\text{adj}}   &  & = (\bar{q}\gamma_\mu \tau_0 \boldsymbol{\tau} q)^2 - (\bar{q}\, i \gamma_\mu\gamma_5\tau_0 \boldsymbol{\tau}\, q)^2
		\,\notag \\\,
	\end{alignat}
\end{minipage}\vspace{2ex}
Three sets of diagram types are thus present in this example:
\begin{equation}
	\includegraphics[width=0.14\linewidth,valign=c]{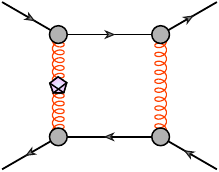}%
	\qquad%
	\includegraphics[width=0.14\linewidth,valign=c]{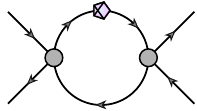}%
	\qquad%
	\includegraphics[width=0.14\linewidth,valign=c]{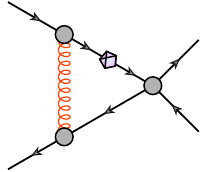}%
	\vspace{-6pt}
\end{equation}
Here, the regulator symbol has been chosen as in \cite{Ihssen:2024miv} and indicates regulators together with general field reparametrisation terms.
The first diagram initially seeds the four-fermi couplings, whereas the others feed the generated couplings into themselves and fully mix them.
One can adjust whether the two latter diagrams are included or not by switching the parameter \cpp{"feedback"} in the \texttt{.json} file:
\begin{lstlisting}[language=C++, style=myStyle]
{
	"physical": {
		"Lambda": 20.0,
		"T": 0.01,
		"muq": 0.34,
		"feedback": false
	},
	...
}
\end{lstlisting}
Furthermore, the temperature and chemical potential at which the flows are evaluated can be adjusted by \cpp{"T"} and \cpp{"muq"}. Note that the input data is not being corrected for thermal and density effects - the resulting four-Fermi couplings are thus at best qualitative indications of how the system reacts to different values of $T$ and $\mu_q$.

\begin{figure}[t]
	\centering
	\begin{subfigure}{0.48\linewidth}
		\includegraphics[width=1\linewidth]{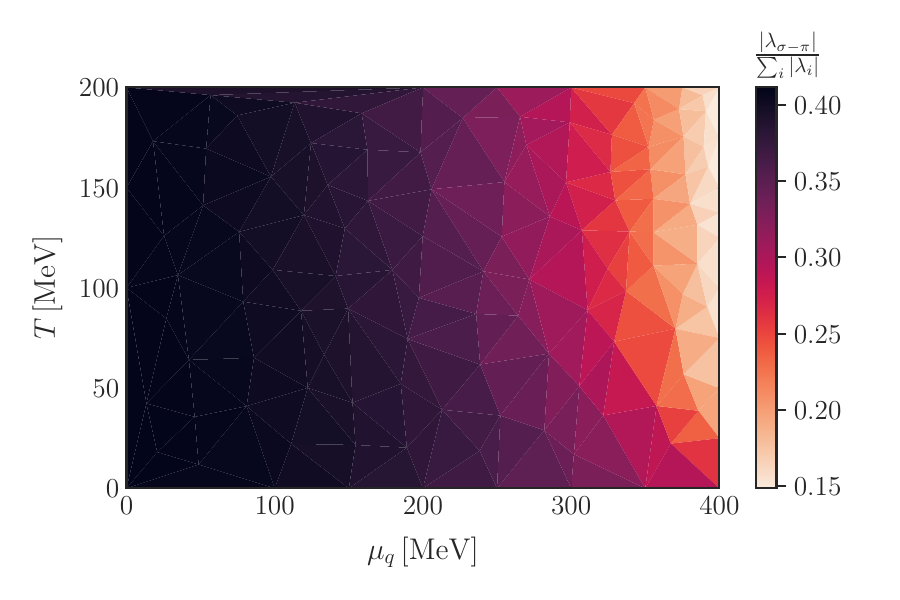}
		\caption{Relative size of the sigma-pion channel $\lambda_{\sigma-\pi}$.}
	\end{subfigure}%
	\hspace{0.02\linewidth}
	\begin{subfigure}{0.48\linewidth}
		\includegraphics[width=1\linewidth]{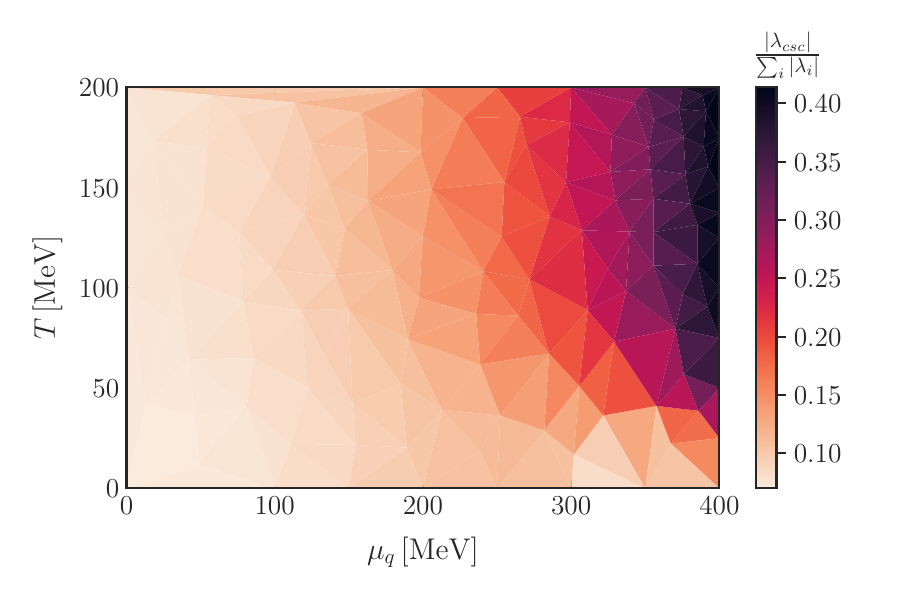}
		\caption{Relative size of the csc channel $\lambda_{csc}$.}
	\end{subfigure}%
	\caption{We show the relative sizes of the sigma-pion and the color-super-conducting (csc) channel in the $T$-$\mu_q$ phase diagram using the setup from \Cref{ex:fourFermi}. All points are run from an initial scale $\Lambda_\textrm{UV}=\qty{20}{\GeV}$ down to $k = \qty{0.5}{\GeV}$, where the flow is stopped. Going lower in $k$ leads to divergences in the couplings. These are expected as the four-Fermi channels turn resonant, signifying massless modes and possible condensates forming which require additional regularisation.
		\hspace*{\fill}}
	\label{fig:FourFermi}
\end{figure}

We show the relative sizes of the sigma-pion coupling and the color-super-conducting channel in \Cref{fig:FourFermi} for different $T$ and $\mu_q$, generating the couplings purely from the glue dynamics,  i.e. \cpp{"feedback"} is set to \cpp{false}. The color-superconducting channel $\lambda_{csc}$ is associated with diquarks and given by 
\begin{align}
	\mathcal{L}_{\mathrm{csc}} &= (\bar\psi \gamma_5 \mathcal{C} \epsilon_F \epsilon_A \bar\psi^T)(\psi^T \mathcal{C}\gamma_5 \epsilon_F \epsilon_A \psi)\,,
\end{align}
where $\mathcal{C}$ is the charge conjugation operator and $\epsilon_A$ and $\epsilon_F$ are fully antisymmetric tensors in the fundamental color and flavor representation and summation over $F$ and $A$ is implied.

%

\subsection{SU(N) Yang-Mills theory}
\label{ex:YangMills}
\noindent
The full Mathematica and C++ code for this example is also provided in  \texttt{Examples/YangMills}. We show here how a large, fully momentum-dependent vertex expansion can be implemented in \DiFfRG.
	
For this example, we consider 4-dimensional SU(N) Yang-Mills gauge theory in two truncations:
\begin{enumerate}
	\item The classical vertex structures are evaluated on the symmetric point. This code can be found in \texttt{Examples/YangMills/SP}.
	\item The classical three-point vertices are evaluated fully on a momentum grid consisting of the average momentum and two angles. The four-gluon vertex is calculated both on the symmetric point configuration and in the special configuration $\lambda_{A^4}(p,q,-p,-q)$, which is fed back into the gluon propagator flow. This code can be found in \texttt{Examples/YangMills/Full}.
\end{enumerate}
Our setup is a reproduction of \cite{Cyrol:2016tym}, so we do also explicitly include the tadpole-configuration of the four-gluon vertex in the flow of the glue propagator. We initialise the setup at an initial scale of $\Lambda_\textrm{UV} = \qty{100}{\GeV}$ and flow it down to $k \lesssim 10^{-4}\qty{}{\GeV}$. The (average) momentum grid is similarly chosen to be logarithmic from $10^{-4}\qty{}{\GeV}$ to $\qty{100}{GeV}$ with $96$ grid points. The three-dimensional dependence  of the three-point functions and the special four-gluon configuration is calculated using a $96\times8\times7$ grid for the average momentum $S_0$ and two angles $a,\,s$, in a parametrisation as introduced in \cite{Eichmann:2014xya}.
We do not describe here the full system of flow equations in detail, as it is quite large, and instead refer the reader to \cite{Cyrol:2016tym} for more details. 

\begin{figure}[t]
	\vspace{-40pt}
	\centering
	\begin{minipage}[b]{0.48\linewidth}
		\centering
		\includegraphics[width=0.95\linewidth]{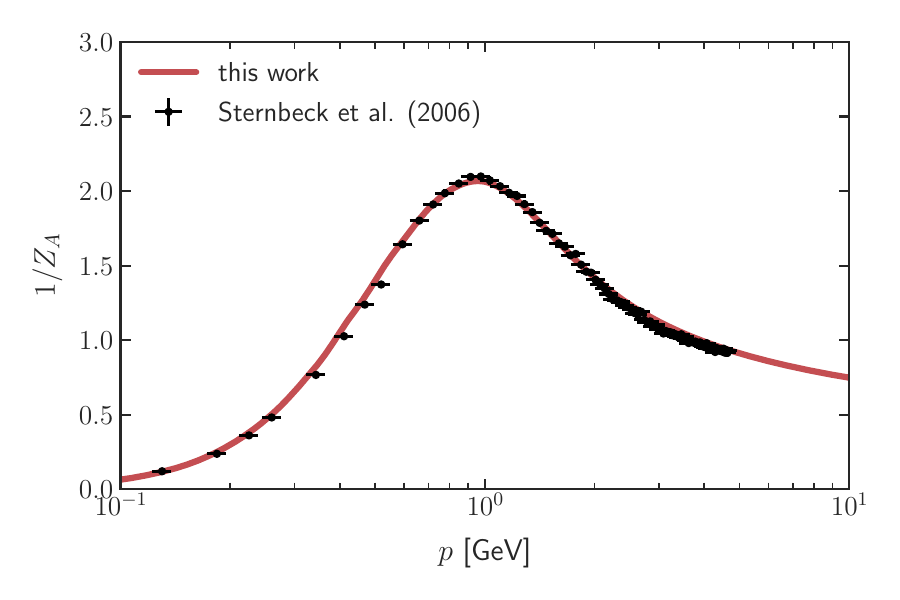}
		\caption{Gluon dressing function $1/Z_A$ in comparison to lattice results from \cite{Sternbeck:2006cg}. \hspace*{\fill}}
		\label{fig:YMLat}
	\end{minipage}%
	\hspace{0.02\linewidth}%
	\begin{minipage}[b]{0.48\linewidth}
		\centering
		\includegraphics[width=0.99\linewidth]{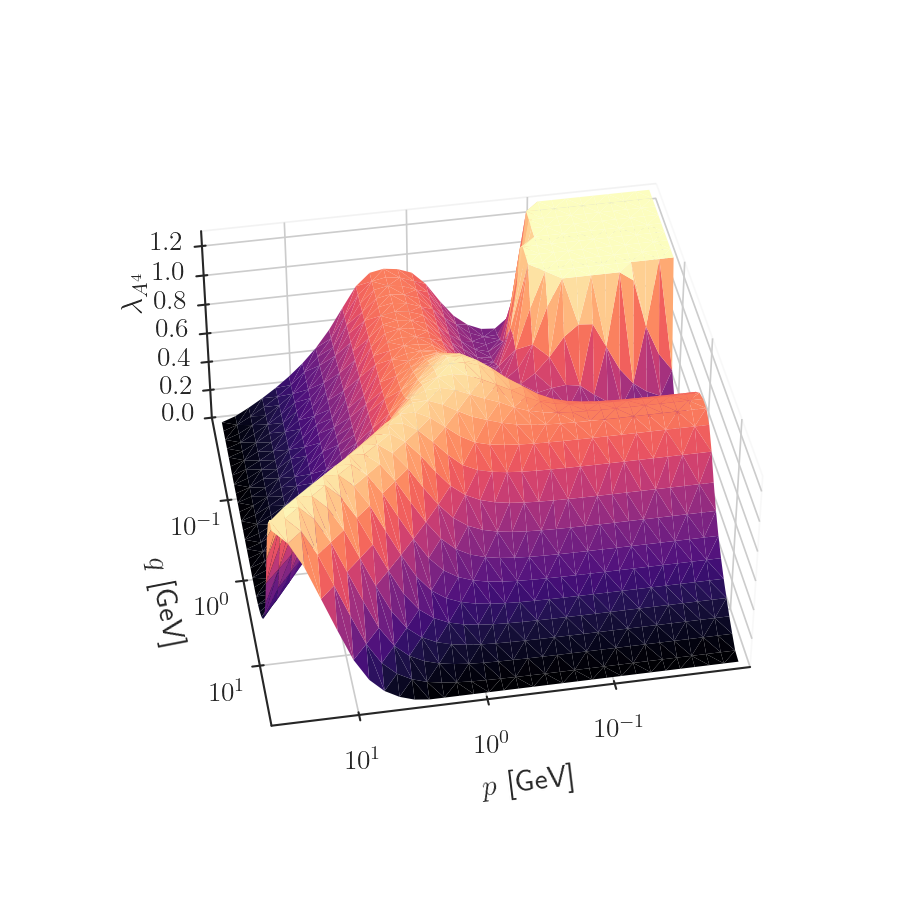}%
		\vspace{-25pt}%
		\caption{The tadpole configuration of the four-gluon dressing $\lambda_{A^4}(p,q,-p,-q)$ at the angular configuration $p\cdot q=0$. The shown coupling has been cut off for small $p$ and $q$ as it scales to very large values, see also \Cref{fig:YMDressings}. \hspace*{\fill}}
		\label{fig:YMA4tadpole}
		\vspace{-19pt}
	\end{minipage}%
	\vspace{20pt}
\end{figure}

\begin{figure}[h!]	
	\centering
	\begin{subfigure}[t]{0.45\textwidth}
		\includegraphics[width=0.99\linewidth]{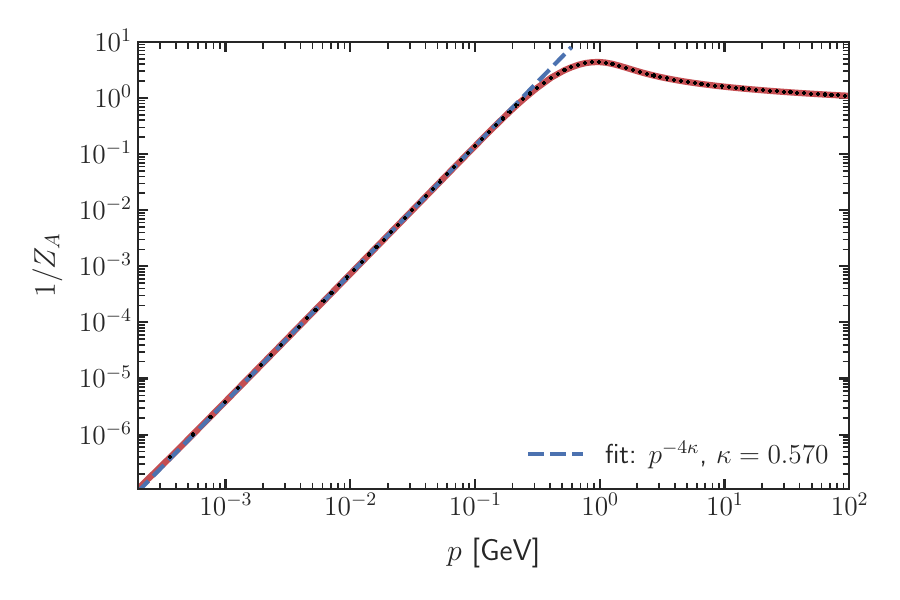}
		\caption{Gluon dressing}
	\end{subfigure}%
	\hspace{0.05\textheight}
	\begin{subfigure}[t]{0.45\textwidth}
		\includegraphics[width=0.99\linewidth]{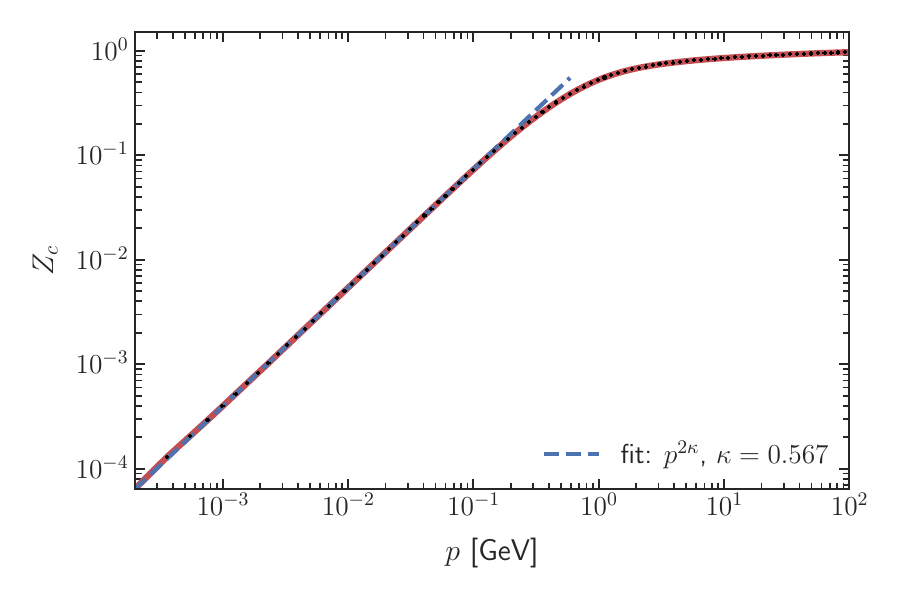}
		\caption{Ghost dressing}
	\end{subfigure}
	\caption{Propagator dressings of SU(3) Yang-Mills theory at the scaling solution. We fit a scaling Ansatz to the curves to obtain the exponent $\kappa$, which is consistent with \cite{Cyrol:2016tym}.
		\hspace*{\fill}}
	\label{fig:YangMillsProps}
	\vspace{10pt}
	\begin{subfigure}[t]{0.45\textwidth}
		\centering
		\includegraphics[width=0.8\linewidth]{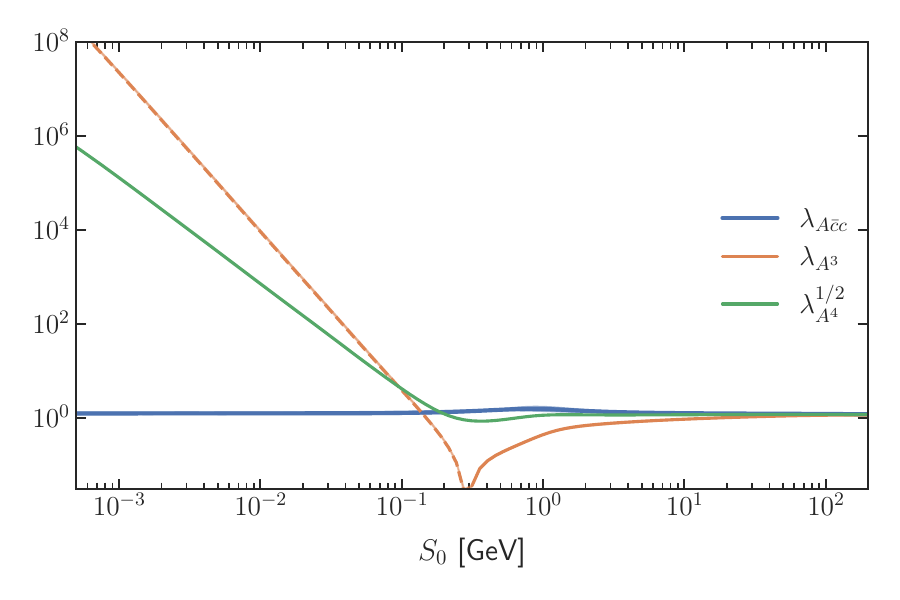}
		\caption{Average-momentum dependence of the vertex dressings on the symmetric point configuration. For the three-point functions, the shaded region indicates the angular dependence at fixed average momentum. The three-gluon coupling turns negative at $S_0 \approx \qty{0.3}{\GeV}$ and we show its absolute value below this with a dashed line. \hspace*{\fill}}
		\label{fig:YMDressings}
	\end{subfigure}%
	\hspace{0.05\textheight}
	\begin{subfigure}[t]{0.45\textwidth}
		\centering
		\includegraphics[width=0.8\linewidth]{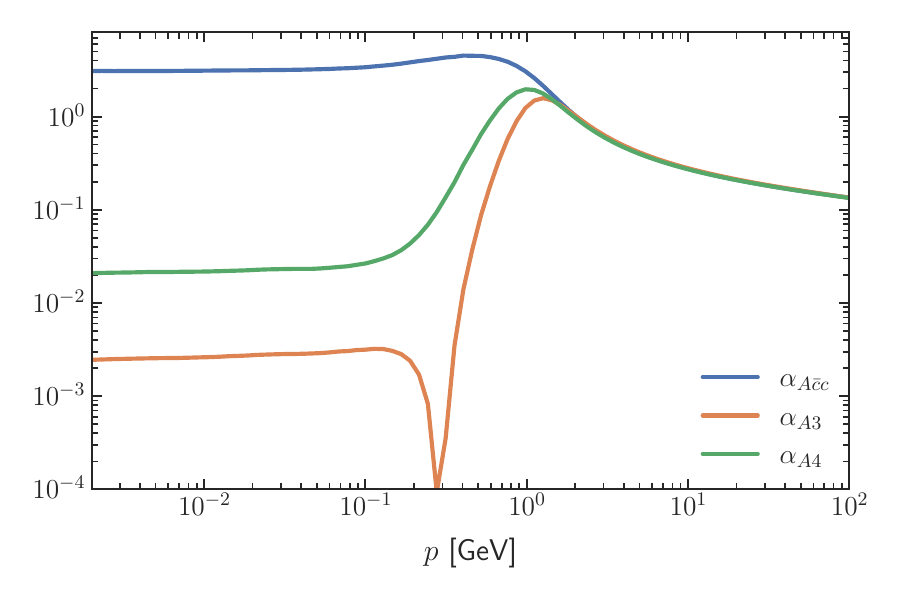}
		\caption{Average-momentum dependence of the renormalised couplings on the symmetric point configuration.\hspace*{\fill}}
	\end{subfigure}
	
	\vspace{5ex}
	
	\begin{subfigure}[t]{0.45\textwidth}
		\centering
		\includegraphics[width=0.8\linewidth]{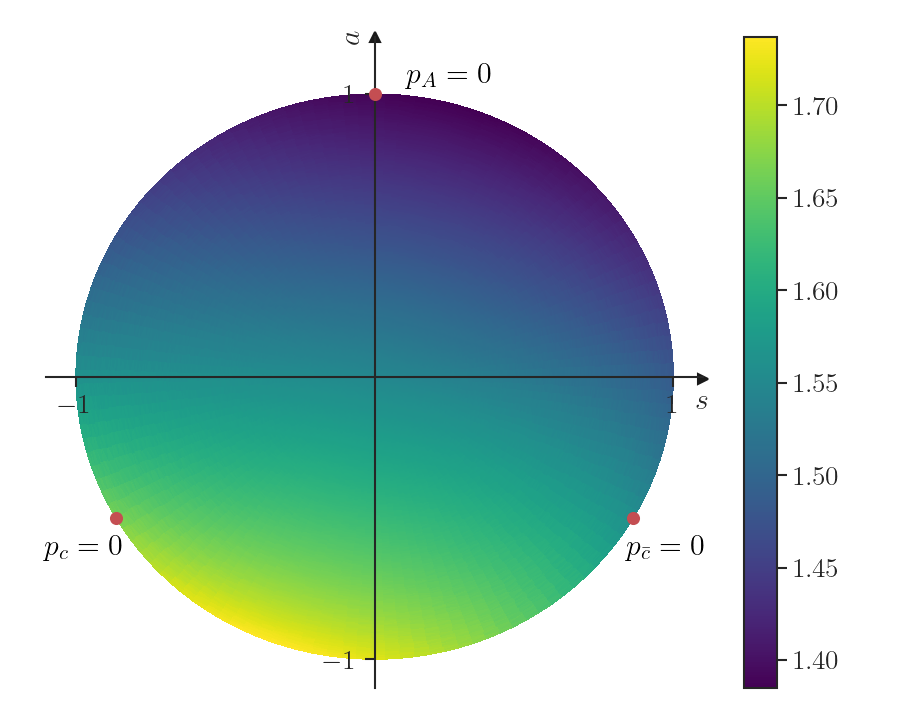}
		\caption{Angular dependence of the ghost-gluon dressing $\lambda_{A\bar{c}c}$ at an average momentum $S_0 = \qty{1}{\GeV}$. \hspace*{\fill}}
	\end{subfigure}%
	\hspace{0.05\textheight}
	\begin{subfigure}[t]{0.45\textwidth}
		\centering
		\includegraphics[width=0.8\linewidth]{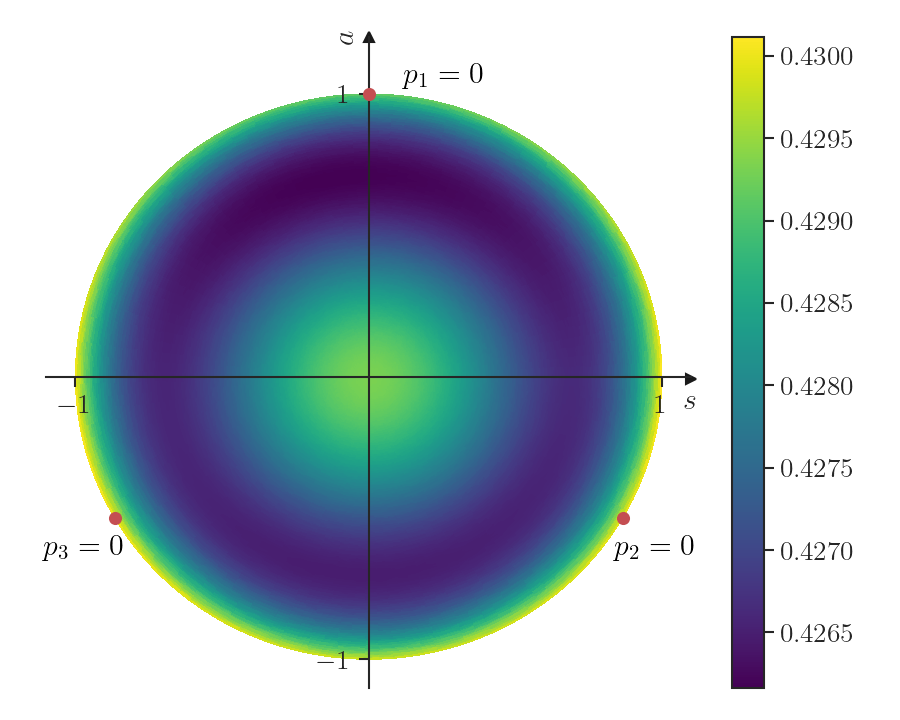}
		\caption{Angular dependence of the three-gluon dressing $\lambda_{A^3}$ at an average momentum $S_0 = \qty{1}{\GeV}$. The vertex has much less angular dependence than the ghost-gluon vertex. \hspace*{\fill}}
	\end{subfigure}
	\caption{Vertex dressings of SU(3) Yang-Mills theory in the scaling solution. In the lower figures, the soft limits located at the edge of the angular disk are pointed out. \hspace*{\fill}}
	\label{fig:YangMillsCouplings_Full}
	\vspace{5ex}
\end{figure}

For both of the truncations described above we include an ipython notebooks which can visualise the generated data. In \Cref{fig:YMLat} we compare our resulting gluon propagator to lattice results. We show in \Cref{fig:YMA4tadpole} the tadpole configuration of the four-gluon vertex. \Cref{fig:YangMillsProps} depicts the scaling propagators and in \Cref{fig:YangMillsCouplings_Full} we show the three-gluon and ghost-gluon dressings in the scaling limit for SU(3). For the latter, we show a slice at constant average momentum of $\qty{1}{\GeV}$ and use a parametrisation of angles as described in \cite{Eichmann:2014xya}. These figures can be directly reproduced by running the example and evaluating the notebook. Furthermore, with the supplied code the gluon mass gap can also be automatically tuned to scaling.


We show here also the advantage of utilizing the GPU for the flow equations: in \Cref{tab:YMPerf} we compare the two truncations of the SU(3) Yang-Mills model on the CPU and GPU. Even for this medium-sized setup, using a GPU can give a speedup of a factor of $4.6$.
\begin{table}[h!]
	\centering
	\begin{tabular}{c|c|c|c}
		Truncation & Time per evaluation (GPU) & Time per evaluation (CPU) & Speedup \\[0.5ex]
		\hline &&& \\[-1.5ex]
		Symmetric point & 45.25 ms & 118.6 ms & 2.6$\times$\\[2ex]
		Full & 1086 ms & 5041 ms & 4.6$\times$
	\end{tabular}
	\caption{Evaluation time for the flow equations of the two truncations of the Yang-Mills setup. The system has been evaluated once fully on the CPU and once on the GPU and the average time per evaluation measured. For all integrals, a quadrature rule of order $32\times8^n$ has been used, where $n$ is the number of integrated angles.
	The performance test has been performed on a Laptop with an AMD Ryzen 9 7945HX 16-core CPU and a NVIDIA RTX 4070 graphics card.
	\hspace*{\fill}}
	\label{tab:YMPerf}
\end{table}
%

\subsection{Quark-Meson model in LPA'}
\label{ex:QM}
\noindent
The full Mathematica and C++ code for this example is also provided in  \texttt{Examples/QuarkMesonLPAprime}.
In this example, we show how to use FE functions together with extractors and variables as previously explained in \Cref{sec:design2}. We utilise the CG assembler in this case.
	
\begin{figure}[t]
	\centering
	\begin{subfigure}{0.48\linewidth}
		\includegraphics[width=1\linewidth]{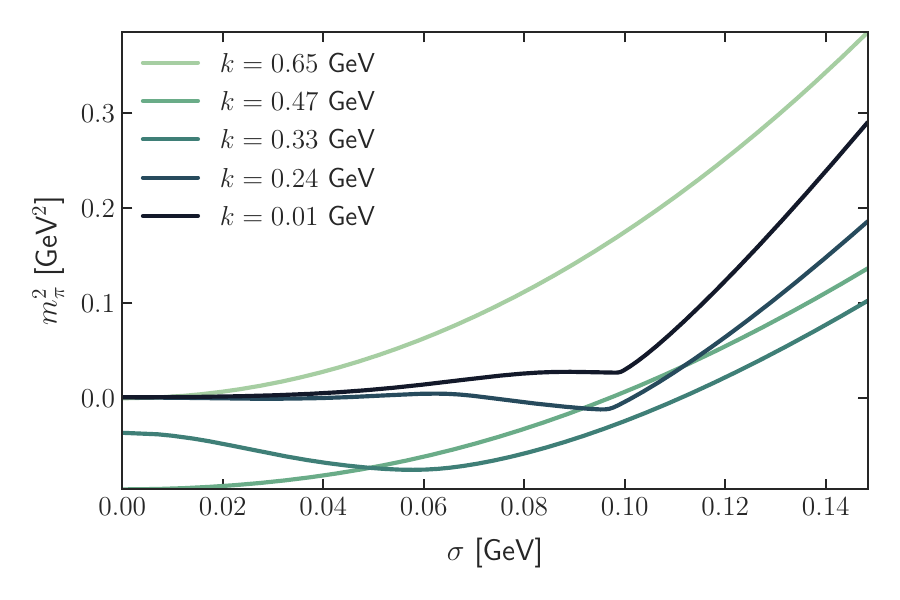}
		\caption{Evolution of $m_\pi^2(\sigma)$.}
	\end{subfigure}%
	\hspace{0.02\linewidth}
	\begin{subfigure}{0.48\linewidth}
		\includegraphics[width=1\linewidth]{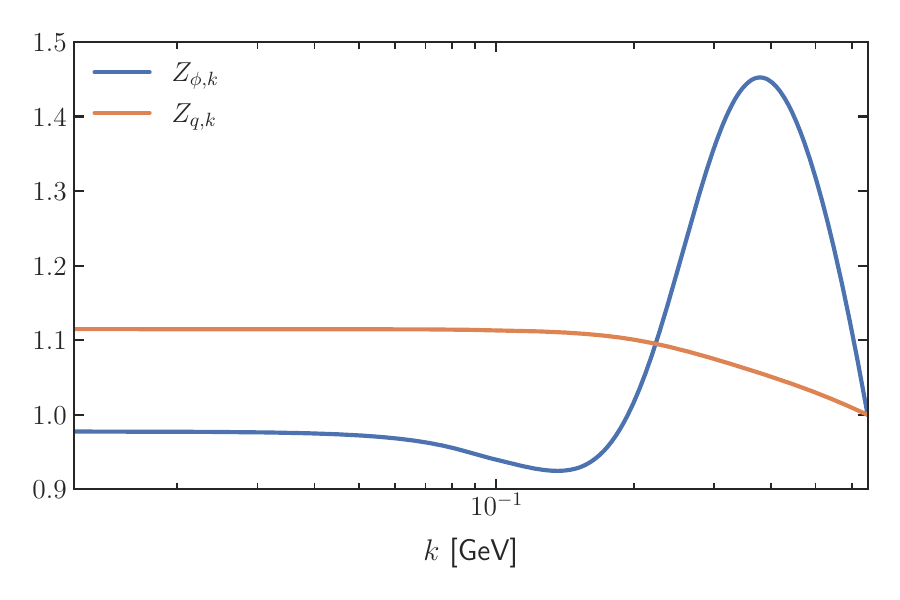}
		\caption{Evolution of the wave-functions $Z_\phi$ and $Z_q$.}
	\end{subfigure}%
	
	\begin{subfigure}{0.48\linewidth}
		\includegraphics[width=1\linewidth]{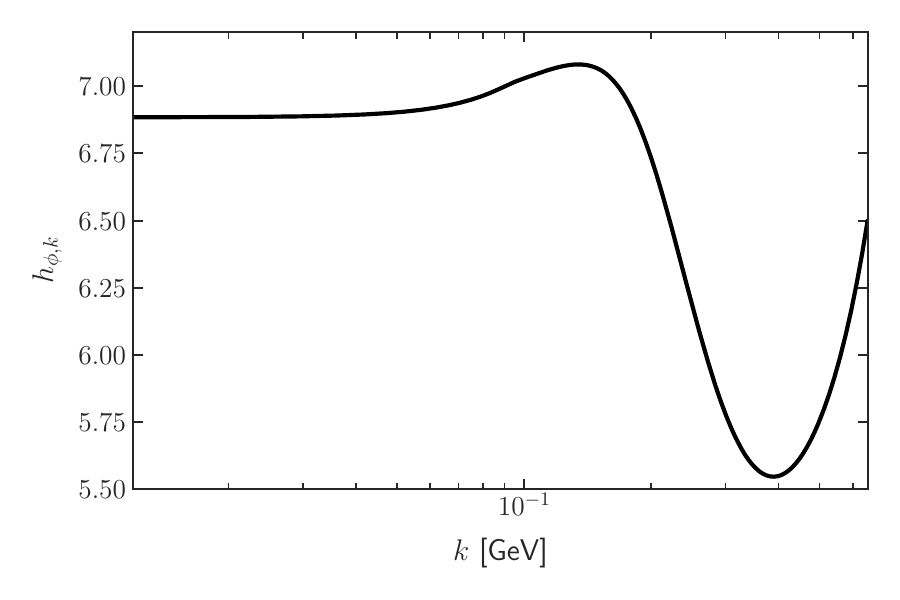}
		\caption{Evolution of the renormalised Yukawa coupling ${h_{\phi,k}}/{Z_{q,k}Z_{\phi,k}^{1/2}}$.}
	\end{subfigure}%
	\caption{We show the RG-evolution of all flowing quantities of the Quark-meson model given by the Ansatz \labelcref{eq:QMLPAp}. The depicted simulation has been run at a temperature of $T = \qty{0.01}{\GeV}$ and chemical potential $\mu_q = \qty{0.34}{\GeV}$, with initial conditions tuned to $m_\pi = \qty{140}{\MeV}$ and $m_q = \qty{350}{\MeV}$ in the vacuum. In (a), we show the field-space evolution of $m_\pi^2$, (b) and (c) show the running of the wave-functions and Yukawa coupling.
	\hspace*{\fill}}
	\label{fig:QM_data}
\end{figure}

The model utilised here is a standard Yukawa-type setup, the Quark-Meson model, a low energy effective model of QCD. The Ansatz for the effective action reads
\begin{align}\label{eq:QMLPAp}
	\Gamma_k[\phi] = \int_x\left\{ \bar{q}(Z_{q,k}\slashed \partial + \textrm{i}\mu\gamma_0)q + h_{\phi,k} \,\bar{q}\, \bigg( \frac{1}{\sqrt{2N_f}}\sigma + \textrm{i}\gamma_5 \boldsymbol{\pi}\boldsymbol{\tau}\bigg)\,q + \frac12 (Z_{\phi,k}\partial_\mu \phi)^2 + V(\rho) \right\}\,,
\end{align}
where $\boldsymbol{\tau}$ is the vector of all generators of the adjoint representation of the flavor group SU($N_f$). $\rho = \phi^2/2$ is again the chirally symmetric field-invariant formed by the mesons.
We use, for simplicity, an RG-invariant language in the example, i.e. we absorb the wave functions $Z_i$ fully into the fields and couplings. Therefore, only the $\eta_i = -\partial_t Z_i / Z_i$ enter the flow equations. This is also described in detail in \cite{Ihssen:2024miv}.
The flow equations are again easily derived from the above Ansatz. The flows for the $\eta_i$ are now slightly more complicated and their derivation is shown in the corresponding notebook, \texttt{QuarkMesonLPAprime.nb}. Furthermore, we also flow the Yukawa coupling $h_{\phi,k}$, although without additional rebosonisation of the (also generated) four-Fermi interactions. The setup is similar to \cite{Pawlowski:2014zaa}, with the difference that we do not take into account higher-order quark-meson scatterings, but instead calculate the full mesonic potential.
Similar to the previous example, the flow of the effective potential is formulated in terms of the renormalised pion mass function
\begin{align}\label{eq:QMFlow}
	\partial_t m_\pi^2(\rho) + \partial_\rho \text{Flow}(V) = 0
\end{align}
We show the evolution of the given model in \Cref{fig:QM_data}, which depicts the evolution of the described Quark-Meson setup at a temperature of $T = \qty{0.01}{\GeV}$ and chemical potential $\mu_q = \qty{0.34}{\GeV}$.
%

\section{Installation and usage}
\label{sec:install}

\subsection{Installation}
\noindent
To set up the \DiFfRG library, the following requirements need to be fulfilled:
\begin{itemize}
	\item \texttt{CMake} $\geq 3.26.4$ and \texttt{GNU Make}, which are used for the build system of \DiFfRG, \texttt{deal.ii} and other libraries.
	\item A compiler supporting at least the C++20 standard. The GCC compiler suite with with a version $\geq12$ is recommended, but  both the latest Clang and AppleClang will also work, although without GPU support.
	\item LAPACK and BLAS in some form, e.g. OpenBlas,
	\item the GNU Scientific Library GSL,
	\item oneTBB, which is used for CPU-side parallel computations,
	\item Doxygen and graphviz to build the documentation.
\end{itemize}
These programs are standard and usually packaged by any Linux distribution; HPC clusters usually provide all of the requirements, as they are integral to many HPC applications.

Additionally, if a valid \texttt{CUDA} toolkit is present, \DiFfRG and any \DiFfRG-applications will be automatically built with GPU-support\footnote{For installations with \texttt{CUDA}, possible difficulties due to versioning are discussed in the pertaining section of the documentation. The critical requirement is to have a \texttt{CUDA} version compatible with an installed compiler which also supports the C++20 dialect, see the \texttt{CUDA} installation guide: \url{https://docs.nvidia.com/cuda/cuda-installation-guide-linux/\#host-compiler-support-policy}}.
Furthermore, in the documentation we also provide a quick guide to  set up a reproducible container environment for use with \DiFfRG, useful e.g. on HPC clusters.

For a standard installation, one has to clone the official \DiFfRG repository to a given directory and run the build script:
\begin{lstlisting}[language=Bash]
$ git clone https://github.com/satfra/DiFfRG.git
$ cd DiFfRG
$ bash -i build.sh -j8 -f -i /opt/DiFfRG
\end{lstlisting}
This will build and install the library to \texttt{/opt/DiFfRG}. For a local installation, e.g. to \texttt{\$HOME/.local/share/DiFfRG}, just change the path in the above command.
The options for the script are also explained by using the \texttt{--help} message:
\begin{itemize}
	\item \texttt{-f } Perform a full build and install of everything without confirmations.
	\item \texttt{-c } Use CUDA when building the DiFfRG library.
	\item \texttt{-i <directory> } Set the installation directory for the library.
	\item \texttt{-j <threads> } Set the number of threads passed to make and git fetch.
\end{itemize}
Thus, to build the library with CUDA, the \texttt{-c} flag has to be used, e.g.
\begin{lstlisting}[language=Bash]
$ bash -i  build.sh -j8 -cf -i $HOME/.local/share/DiFfRG
\end{lstlisting}
Depending on the machine, building the library together with all requirements takes from 10-40 minutes.
Additional configuration may be performed in the file \texttt{config} in the top folder of the repository, e.g. to force a specific compiler or CUDA architecture. The file is well documented and should be self-explanatory.

\subsection{Usage}
\noindent
For exhaustive examples and tutorials, the user is referred to the documentation, which is automatically built with the library and can be found at \texttt{Documentation.html} in the install directory or alternatively on \textit{github pages}\footnote{\url{https://satfra.github.io/DiFfRG/cpp/index.html}}. The full tutorial code is also available in the \texttt{Tutorials/} directory. The tutorials are fully described in the documentation and are a good starting point for new users. In particular, they describe step-by-step the setup of a \DiFfRG-application from the very start.

In addition to the tutorials, four exhaustive examples are provided with the library. In the subfolder \texttt{Examples/} of the cloned repository, the full codes for an O(N)-theory in LPA at finite temperature, a four-Fermi calculation, a Yang-Mills setup with fully momentum dependent classical vertices, and a Quark-Meson model in LPA' are available.
These are also briefly discussed in \Cref{ex:ONfiniteT}, \ref{ex:fourFermi}, \ref{ex:YangMills} and \ref{ex:QM} respectively.

In general, when setting up a new \DiFfRG application, one has to make use of the \texttt{CMake} build system to find and integrate the framework into the application. For example, a typical \texttt{CMakeLists.txt} for an application using the \DiFfRG framework may look like this:
\begin{lstlisting}[language=CMake]
	cmake_minimum_required(VERSION 3.26.4)
	project(myApp)
	
	find_package(DiFfRG REQUIRED HINTS /opt/DiFfRG)
	
	add_executable(myApp main.cc)
	setup_application(myApp)
	add_flows(myApp flows)
\end{lstlisting}
In the above, CMake searches first for the \DiFfRG framework at the location \texttt{/opt/DiFfRG}, then creates an executable and calls \cmake{setup_application(<target>)} to make it a \DiFfRG-application, i.e. link against the library and add the include directories. After that, the auto-generated code for the flow equations is included in the program with the command \cmake{add_flows(<target> <folder>)}.

From the directory where the \texttt{CMakeLists.txt} is located, the application can be then set up and built by running the commands
\begin{lstlisting}[language=Bash]
	$ mkdir build
	$ cd build
	$ cmake -DCMAKE_BUILD_TYPE=Release ../
\end{lstlisting}
Here \texttt{/opt/DiFfRG} has to be replaced with the chosen install directory of DiFfRG.

\begin{figure*}[t]
	\centering
	\begin{tikzpicture}
		\node [roundbox] at (0,0.9) (start) {Derive symbolic functional equations: \textit{QMeS-Derivation} \cite{Pawlowski:2021tkk}};
		\node [roundbox] at (0,0) (step1) {Insert correlation functions};
		\draw [thick,->,color=black!50] (start) -- (step1);
		\node [roundbox] at (0,-0.9) (step2a) {Project onto dressings of specific tensor structures};
		\draw [thick,->,color=black!50] (step1) -- (step2a);
		\node [roundbox] at (0,-1.8) (step2b) {Perform traces over Lorentz, Dirac and other group indices: \textit{FormTracer} \cite{Cyrol:2016zqb}};
		\draw [thick,->,color=black!50] (step2a) -- (step2b);
		\node [roundbox] at (0,-2.7) (step2) {Auto-generate C++ integration kernels for the flow equations};
		
		\draw [decorate,decoration={brace,amplitude=10pt},xshift=-4pt,yshift=0pt,color=black!50]
		(5.9,1.3) -- (5.9,-3.1) node [black,midway,xshift=2.75cm]
		{\footnotesize \DiFfRG Mathematica package};
		
		\draw [thick,->,color=black!50] (step2b) -- (step2);
		\node [roundbox] at (0,-3.6) (step3a) {Set up the PDE system as a \texttt{numerical model}};
		\draw [thick,->,color=black!50] (step2) -- (step3a);
		\node [roundbox] at (0,-4.5) (step3) {Choose assembler, time-stepper, set up program logic};
		\draw [thick,->,color=black!50] (step3a) -- (step3);
		
		\draw [decorate,decoration={brace,amplitude=10pt},xshift=-4pt,yshift=0pt,color=black!50]
		(5.9,-3.3) -- (5.9,-4.8) node [black,midway,xshift=2.5cm]
		{\footnotesize \DiFfRG C++ application};
		
		\node [roundbox, fill=red!5,fill opacity=0.01,text width=11cm, color=black!60,line width=0.5mm] at (-0.,-6.97) (cbox) {\vspace{3.5cm}};
		
		\node [process] at (-1.5,-5.9) (step4) {$k =\Lambda$};
		\node [integration] at (-1.5,-6.8) (step5) {$k \to k - \Delta k$};
		\node [process] at (-1.5,-7.8) (step6) {$k \to 0$};
		\draw [thick,->,color=black!50] (step3) -- (cbox);
		\draw [thick,->,color=black!50] (step4) -- (step5);
		\draw [thick,->,color=black!50] (step5) -- (step6);
		
		\draw [decorate,decoration={brace,amplitude=10pt},xshift=-4pt,yshift=0pt,color=black!50]
		(5.9,-5.0) -- (5.9,-8.8) node [black,midway,xshift=1.8cm]
		{\footnotesize {Computation}};

		\node [roundbox, fill=white,text width=4.5cm] at (3.0,-5.9) (pdep) {Numerically integrate momentum loops and evolve p-dependent vertices};
		\node [roundbox, fill=white,text width=4.5cm] at (3.0,-7.8) (phidep) {Numerically integrate momentum loops and evolve field dependent quantites using \textit{deal.II} \cite{dealII95}};
		\node [roundbox, fill=white,text width=1.8cm] at (-4.3,-6.8) (output) {Data output};
		\draw [thick,<->,color=black!50] (step5) -- (pdep);
		\draw [thick,->,color=black!50]  (step5) -- (output);
		\draw [thick,<->,color=black!50] (step5) -- (phidep);
		\draw [thick,<->,color=black!50] (pdep) -- (phidep);
	\end{tikzpicture}
	\vspace{1mm}
	\caption{A typical workflow for the setup and evaluation of an fRG model using the \DiFfRG package. The three major parts of the workflow are discussed in detail in \Cref{sec:design}. \hspace*{\fill}}
	\label{fig:design}
\end{figure*}
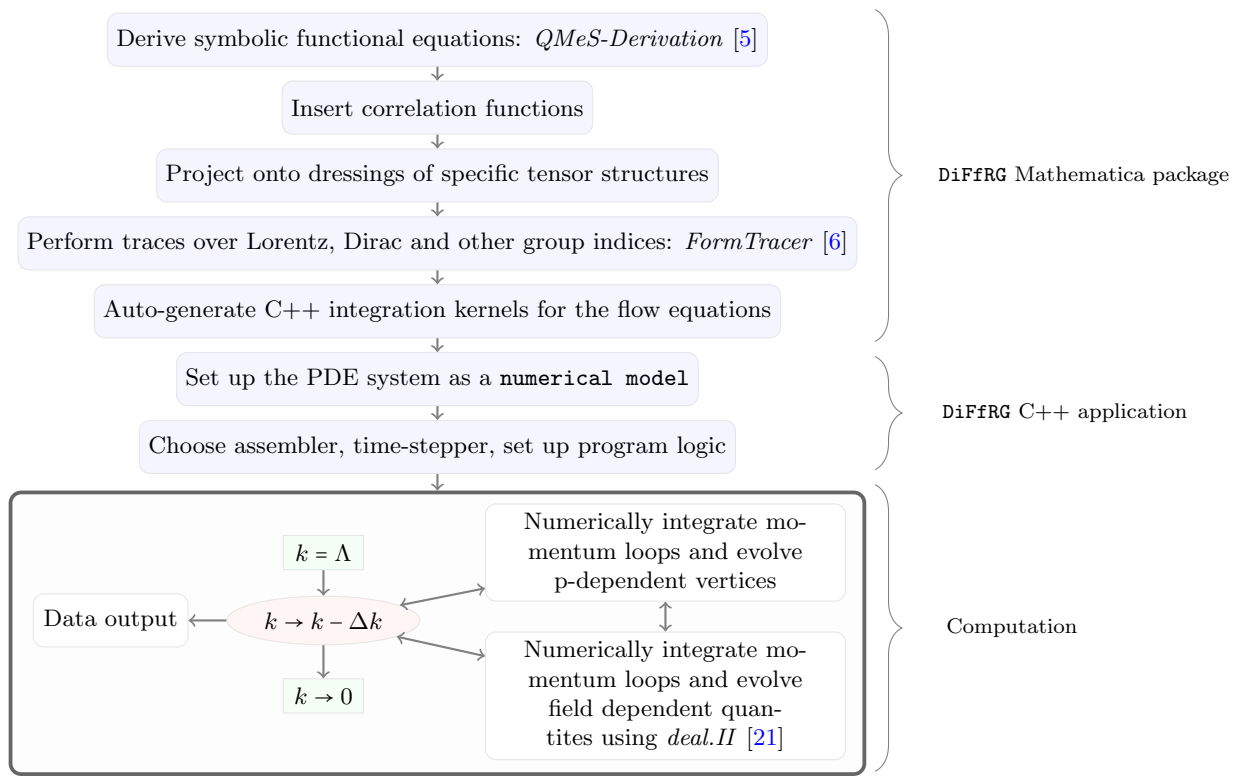
%

\section{Design and features}
\label{sec:design}

In this section we discuss the overall structure and central concepts of the framework. We refrain from discussing the code in detail, rather giving a general overview of how \DiFfRG works and how it may be used. For a thorough explanation of the entire code, the user is referred to the documentation and the tutorials provided with the library.

Besides the main functionality of \DiFfRG, a large set of convenience functions have been defined to quickly create standard scenarios of fRG setups. For example, usual threshold functions as obtained for the flat or Litim regulator \cite{Litim:2000ci} are present at \texttt{DiFfRG/physics/}, as well as interpolation classes and other tools like minimisation, root finding and data import algorithms at \texttt{DiFfRG/common/}.

In \Cref{fig:design}, we give an overview for the process of setting up and running an fRG simulation using \DiFfRG. In the following subsections we discuss the three central stages of creating a simulation and how they interplay with each other.
First, we explain the derivation of flow equations and automatic code-generation using \texttt{Mathematica} in \Cref{sec:design1}. Then, the C++ setup is shown in \Cref{sec:design2}. Finally, in \Cref{sec:design3} some details are given on the explicit computation and numerical algorithms used therein.

\subsection{Mathematica setup}
\label{sec:design1}
\noindent
The Mathematica sub-package of \DiFfRG supplies all tools necessary to define a truncation of some effective action and generate the flow equations.
These are then parsed to auto-generated C++ code, which in turn can be directly integrated in simulations. Thus, most of the analytic derivation necessary to set up a system of flow equations is fully automated.
Note that in principle, this is optional: All C++ facilities can also be used directly and flow equations entered either by hand or from other programs. Furthermore, one may also only use the code-export features of the Mathematica package and nothing else.

For the derivation of the abstract flow equations we rely on the \texttt{QMeS-derivation} Mathematica package \cite{Pawlowski:2021tkk}. In its current, updated version, it also allows for grouping of identical diagrams, where additional symmetries of vertices can be taken into account.

In the next step, the diagrammatic rules of the theory are inserted into the flow equations. After full contraction with projection operators, all tensor structures, i.e. Lorentz, Dirac and group tensors, are fully traced out with the help of \texttt{FormTracer} \cite{Cyrol:2016zqb}. We include the \texttt{TensorBases} Mathematica package \cite{TensorBases} in the workflow for easy handling of tensor structures both for the diagrammatic rules and the projection operators.
Furthermore, we extend some tracing capabilities of \texttt{FormTracer}:
\DiFfRG implements an algorithm to trace out (even) pairs of charge conjugation operators. The tracing algorithm directly implements the action of the charge conjugation operator on gamma matrices. As an example, one can trace now
{
	\footnotesize
	\begin{mmaCell}{Code}
	\mmaDef{Get}["DiFfRG`"]
	\end{mmaCell}
	\begin{mmaCell}{Code}
	ChargeConj[d1,d2]gamma5[d2,d3]ChargeConj[d3,d4]gamma5[d4,d1]//\mmaDef{ExtendedFormTrace}
	\end{mmaCell}
	\begin{mmaCell}{Output}
	-4
	\end{mmaCell}
	\begin{mmaCell}{Code}
	gamma5[d1,d4]gamma5[d4,d1]//\mmaDef{ExtendedFormTrace}
	\end{mmaCell}
	\begin{mmaCell}{Output}
	4
	\end{mmaCell}
	\par
}\noindent
Fully traced expressions are translated to C++ code in the form of \textit{integration kernels} which can be evaluated by any application built with the \DiFfRG package. Doing so requires one to first specify all \textit{kernels}, i.e. flow equations that shall be exported to code, and a list of \textit{kernel parameters}, which represent all C++ arguments passed to the integration routines.
As an example, for the flow of the effective potential of the O(N)-model in \Cref{ex:ONfiniteT}, one first sets up the integration kernel for the effective potential:
{
\footnotesize
\begin{mmaCell}{Code}
	kernelV=<|
	    "Path"->"V",
	    "Name"->"V",
	    "Type"->"Quadrature",
	    "Angles"->0,
	    "d"->4,
	    "AD"->True,
	    "ctype"->"double",
	    "Device"->"CPU"
	  |>;
	kernelParameters= kernelParameterList={
	    <|"Name"->"N","Type"->"Constant","AD"->False|>,
	    <|"Name"->"rhoPhi","Type"->"Constant","AD"->False|>,
	    <|"Name"->"m2Pion","Type"->"Variable","AD"->True|>,
	    <|"Name"->"d2V","Type"->"Variable","AD"->True|>
	  };
	MakeFlowClass["ON",{kernelV}];
\end{mmaCell}
\noindent\hspace{-7pt}
}
The first declaration describes the kind of integration kernel to define, whereas the second gives a list of parameters to be passed to the integration routine - the specific meanings of the keys and possible values are given in \Cref{app:mathematica}.
Then, after the flow for the potential has been obtained, it can be converted to C++ code using
{%
	\footnotesize
	\begin{mmaCell}{Code}
	\mmaDef{MakeKernel}[kernelV,kernelParameterList,VLoop];
	\end{mmaCell}
	\noindent\hspace{-7pt}
}
where it is assumed that \mathem{VLoop} is the (already derived) equation for the flow of the effective potential.
This command will create the files \texttt{flows/V/V.kernel}, \texttt{flows/V/V.cu}, \texttt{flows/V/V.cc} and \texttt{flows/V/V.hh} which contain the full routines to calculate the effective potential.
For the full O(N) example, see also \Cref{ex:ONfiniteT} and the directory \texttt {Examples/ON\_finiteT}, where the full Mathematica and C++ code for this model is provided.

The exported C++ code is automatically placed in a sub-folder \texttt{flows} of the working directory, although this behavior can be changed using \mathem{SetFlowDirectory[dir_String]}. The user can include and build the entire set of flow equations using the \texttt{CMake}-function \cmake{add_flows(<target> <folder>)} in the \texttt{CMakeLists.txt} of the application:
\begin{lstlisting}[language=CMake, firstnumber=6]
add_executable(myApp main.cc)
setup_application(myApp)
add_flows(myApp flows)
\end{lstlisting}
We provide an explanations and references for the Mathematica sub-package in \Cref{app:mathematica}. \Cref{app:fourFermi} discusses the four-fermi example of \Cref{ex:fourFermi} and may be also instructive for the usage of the Mathematica sub-package of \DiFfRG.

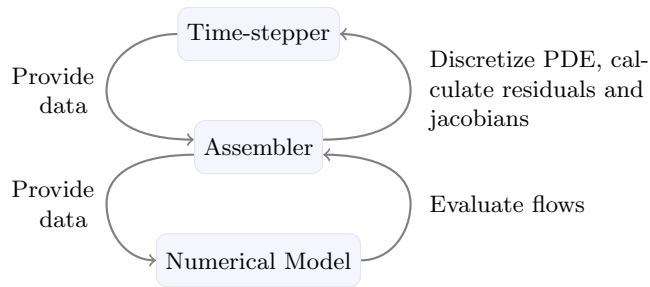
\begin{figure}[t]
	\centering
	\begin{tikzpicture}
		\node [roundbox] at (0,0.) (timestepper) {Time-stepper};
		\node [roundbox] at (0,-1.5) (assembler) {Assembler};
		\node [roundbox] at (0,-3.) (model) {Numerical Model};

		\draw [thick,->,color=black!50] (timestepper) to [out=180,in=90] ($(timestepper)+(-2,-0.75)$) to [out=-90,in=180] ($(assembler)+(-0.85,0.1)$);
		\draw [thick,->,color=black!50] ($(assembler)+(-0.85,-0.1)$) to [out=180,in=90] ($(assembler)+(-2,-0.75)$) to [out=-90,in=180] (model);

		\draw [thick,->,color=black!50] (model) to [out=0,in=-90] ($(model)+(2,0.75)$) to [out=90,in=0] ($(assembler)+(0.85,-0.1)$);
		\draw [thick,->,color=black!50] ($(assembler)+(0.85,0.1)$) to [out=0,in=-90] ($(assembler)+(2,0.75)$) to [out=90,in=0] (timestepper);

		\node [draw=none, fill=none,text width=1cm, align=right] at ($(timestepper)+(-2.75,-0.75)$)() {Provide data};
		\node [draw=none, fill=none,text width=1cm, align=right] at ($(assembler)+(-2.75,-0.75)$)() {Provide data};

		\node [draw=none, fill=none,text width=3cm, align=left] at ($(model)+(3.75,0.75)$)() {Evaluate flows};
		\node [draw=none, fill=none,text width=3cm, align=left] at ($(assembler)+(3.75,0.75)$)() {Discretize PDE, calculate residuals and jacobians};
	\end{tikzpicture}
	\vspace{1mm}
	\caption{Structure of a \DiFfRG\, simulation. The chosen \textit{Time-stepper} manages the overall data flow and requests residuals and jacobians from the \textit{Assembler}, which are then used to advance the RG-flow. The assembler constructs these by calling the methods of a used-defined \textit{Numerical model}, where the actual flow equations are evaluated. \hspace*{\fill}}
	\label{fig:numericsStructure}
\end{figure}

\subsection{C++ setup}
\label{sec:design2}
\noindent
A \DiFfRG simulation consists in general of three parts, which are shown in \Cref{fig:numericsStructure} and have to be combined by the user in a problem-appropriate manner:
\begin{itemize}
	\item The time-stepper, which manages the overall data flow and evolves the system of PDEs between RG-scales,
	\item the assembler, which constructs the numeric representation of the system of equations,
	\item and the numerical model, where the actual flow equations are implemented and the structure of the equations is specified.
\end{itemize}
In the following we briefly discuss these three aspects and the algorithms \DiFfRG makes available for the corresponding tasks.

\subsubsection{Time-stepping}
\noindent
All time-steppers are accessible by including the file \texttt{DiFfRG/timestepping/timestepping.hh} and are templated on the type of the field, the type of system matrix and the dimension of the field space.
The system matrix type specifies the structure of the field discretisation and how the degrees of freedom (i.e. basis function coefficients) are coupled through the PDE. For most FEM systems this will be some sparse matrix type, as degrees of freedom within the discretisation are only locally connected along neighboring cells.
These two arguments do not have to be specified if no spatial discretisation is given - then, they are per default set to the necessary values. Furthermore, these types can be directly taken from a chosen discretisation, e.g. \cpp{CG::Discretization::SparseMatrixType}.
If the time-stepper is implicit, additionally the type of linear solver used for the implicit part has to be specified. As an example, one may find a specification like
\begin{lstlisting}[language=C++, style=myStyle]
#include <DiFfRG/timestepping/timestepping.hh>

using VectorType = dealii::Vector<double>;
using SparseMatrixType = dealii::SparseMatrix<double>;
constexpr uint dim = 2;
using TimeStepper = DiFfRG::TimeStepperSUNDIALS_IDA<
													VectorType, SparseMatrixType, dim, DiFfRG::UMFPack>;
\end{lstlisting}
where we use the \texttt{SUNDIALS IDA} time-stepper with the \texttt{UMFPack} linear solver and we assumed that the types \lstinline[language=C++,style=myStyle]|VectorType| and \lstinline[language=C++,style=myStyle]|SparseMatrixType| have been specified before - here, we took a two-dimensional example.

\DiFfRG provides three classes of time-steppers:
\begin{enumerate}
	\item Implicit time-steppers, which are necessary to integrate theories with spontaneous symmetry breaking. They take information from earlier times but also the to-be calculated result itself into account, which leads to the name "implicit".
	      \begin{itemize}
		      \item \texttt{SUNDIALS\_IDA}, the differential-algebraic solver from the \texttt{SUNDIALS}-suite \cite{hindmarsh2005sundials,gardner2022sundials}. In the time direction, this solver is a variable order BDF time-stepper. This type of implicit stepper has been found to be most favourable for convexity restoration in spontaneaously broken theories within an fRG approach in \cite{Ihssen:2023qaq} and is also the most performant stepper for field-discretisations in \DiFfRG.
	      \end{itemize}
	\item Explicit time-steppers which only rely on information from earlier steps, favored for larger vertex expansions with slower dynamics in $k$-scale:
	      \begin{itemize}
		      \item \texttt{Boost\_ABM} is the Adams-Bashforth-Moulton multistep algorithm from the \texttt{Boost} library. It is most useful for very large and slow momentum-dependent setups, where the evaluation of a single residual is very costly.
		      \item \texttt{Boost\_RK45} and \texttt{Boost\_RK78} are respectively 5th and 8th order Runge-Kutta methods from the \texttt{Boost} library, useful for high-precision calculations, but less performant than \texttt{Boost\_ABM}.
	      \end{itemize}
	\item ImEx (mixed implicit / explicit) time-steppers, useful for mixing setups with both fully field dependent quantities and vertex expansions:
	      \begin{itemize}
		      \item \texttt{SUNDIALS\_Arkode}, the variable-order ImEx time-stepper from the \texttt{SUNDIALS}-suite \cite{hindmarsh2005sundials,gardner2022sundials}. It is an alternative to  \texttt{SUNDIALS\_IDA}, but is less performant in almost all fRG-contexts.
		      \item \texttt{SUNDIALS\_IDA\_Arkode},
		            \texttt{SUNDIALS\_IDA\_Boost\_ABM},
		            \texttt{SUNDIALS\_IDA\_Boost\_RK45} and\\ \texttt{SUNDIALS\_IDA\_Boost\_RK78} are combinations of the  \texttt{SUNDIALS\_IDA} time-stepper for the field discretisation and an explicit stepper for the variable discretisation.
	      \end{itemize}
\end{enumerate}
For linear solvers, we supply currently two algorithms:
\begin{itemize}
	\item \cpp{DiFfRG::UMFPack} \cite{10.1145/992200.992206} is a direct sparse solver utilizing a multifrontal LU factorisation for inversion of sparse matrices. It is preferred for small discretisations, i.e. one-dimensional systems due to its precision and relative speed.
	\item \cpp{DiFfRG::GMRES} is an iterative solver, giving an interface to the GMRES solver from \textit{deal.ii}. As it does not directly invert the full jacobian, it can be much more efficient than \cpp{DiFfRG::UMFPack} for large systems and is preferred in spatial dimensions greater than one.
\end{itemize}
%

\subsubsection{Equation setup and assembly}
\begin{figure}[t]
	\centering
	\begin{tikzpicture}
		\node [roundbox] at (4,-1.) (fefunctions) {FE functions};
		\node [roundbox] at (0,-1.) (extractors) {Extractors};
		\node [roundbox] at (-4,-1.) (variables) {Variables};

		\draw [thick,->,color=black!50] (fefunctions) to [out=135,in=0] ($(fefunctions)+(-2.,1.0)$) to [out=180,in=45] (extractors);
		\draw [thick,->,color=black!50] (extractors) to [out=135,in=0] ($(extractors)+(-2.,1.0)$) to [out=180,in=45] (variables);
		\draw [thick,->,color=black!50] (variables) to [out=-45,in=180] ($(variables)+(2.,-1.0)$) to [out=0,in=-135] (extractors);
		\draw [thick,->,color=black!50] (extractors) to [out=-45,in=180] ($(extractors)+(2.,-1.0)$) to [out=0,in=-135] (fefunctions);

		\draw [thick,->,color=black!50] (variables) to [out=-75,in=180] ($(variables)+(4,-1.75)$) to [out=0,in=-105] (fefunctions);

		\node [draw=none, fill=none,text width=5cm, align=center] at  ($(fefunctions)+(-2.,1.25)$)() {Extract data at EoM};
		\node [draw=none, fill=none,text width=3cm, align=center] at  ($(extractors)+(-2.,1.25)$)() {Provide data};

		\node [draw=none, fill=none,text width=3cm, align=center] at  ($(fefunctions)+(-2.,-1.25)$)() {Provide data};
		\node [draw=none, fill=none,text width=3cm, align=center] at  ($(extractors)+(-2.,-1.25)$)() {Provide data};
		\node [draw=none, fill=none,text width=3cm, align=center] at   ($(variables)+(4,-2.05)$)() {Provide data};
	\end{tikzpicture}
	\vspace{1mm}
	\caption{Relationship and data sharing between Variables, Extractors and FE functions. \hspace*{\fill}}
	\label{fig:dataRels}
\end{figure}
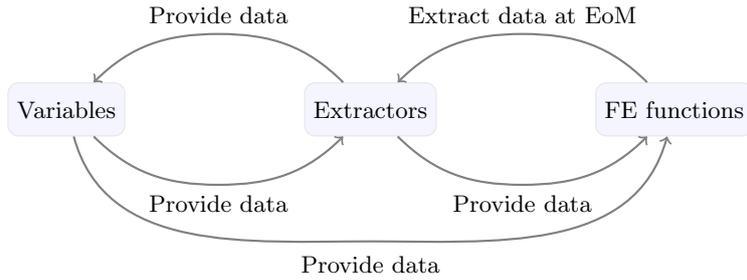
\noindent
Variables, functions and other flowing quantities are treated within \DiFfRG by splitting them into two subsectors: \textit{variables} and \textit{FE functions}.
\begin{itemize}
	\item \textit{Variables} hold arbitrary data including (multi-dimensional) momentum grids, finite difference grids for field space discretisation, Taylor expansions, etc. For example, one would describe a momentum-dependent $n$-point function usually as a grid in the variable space.
	\item \textit{FE functions} are functions defined on finite element basis functions, used for field space discretisations as explained in \Cref{sec:FEM}.
\end{itemize}
A \DiFfRG setup may employ both subsectors, e.g. variables to calculate full momentum dependences on the equation of motion and FE functions for a full effective potential. In such a case, the FE function sector is able to communicate information to the variables sector through \textit{extractors}:
\begin{itemize}
	\item \textit{Extractors} read out or calculate data from the FE functions at the EoM and pass this data to the variables, consituting the link between the two subsectors. Additionally, FE functions can also use the full variable and extractor data for their calculations, which is for example necessary for an LPA' setup.
	If extractors are present, \DiFfRG will solve a given scale-dependent EoM
	\begin{align}
		\frac{\delta \Gamma_k}{\delta \Phi}\Big\vert_{x=x_{\text{EoM},k}} = 0
	\end{align}
	at each time-step and evaluate the extractors at $x_{\text{EoM},k}$.
\end{itemize}
The relationship between the two subsectors and the extractors as discussed is also depicted in \Cref{fig:dataRels}. 
The structure of the overall system informs an appropriate choice of \textit{Discretisation} and \textit{Assembler}.
As mentioned before, the assembler will evaluate certain methods of a numerical model to construct the system of equations. The full set of methods the user can specify in a numerical model is given in \Cref{app:model}.
The preferred way to define any numerical model is by inheriting from the \cpp{DiFfRG::AbstractModel} class, which implements all methods with their default behavior as explained in \Cref{app:model}. 
One can also derive the model class from other classes providing further standard implementations, e.g.
\begin{lstlisting}[language=C++, style=myStyle]
class ON_finiteT :
	public def::AbstractModel<ON_finiteT, Components>,
	public def::LLFFlux<ON_finiteT>,        // use a LLF numflux for DG
	public def::FlowBoundaries<ON_finiteT>, // Inflow/Outflow boundaries
	public def::AD<ON_finiteT>              // define all jacobians per AD
\end{lstlisting}
Here, the \cpp{Components} template parameter is a struct which describes the structure of the discretisation i.e. the number of FE functions, variables and extractors. This is described in more detail in \Cref{sec:componentsDesc}.

\subsubsection*{Variable Assembler}\label{sec:varAssembler}
\noindent
If no field-space discretisation is present, an assembler only using variables can be used:
\begin{lstlisting}[language=C++, style=myStyle]
#include <DiFfRG/discretisation/discretisation.hh>
...
using Assembler = DiFfRG::Variables::Assembler<Model>;
\end{lstlisting}
This assembler creates a system of equations of the very general shape
\begin{align}\label{eq:assVar}
	\partial_{t} v_n + \mathcal{R}_{v,n}(v,t) & = 0
	\,,
\end{align}
where $v = \{v_1, v_2, \dotsc\}$ is the vector of all variables in the system.
The assembler will call methods within the provided \textit{numerical model} \cpp{Model} to construct this system of equations. Explicitly, the variable assembler calls the method
\begin{lstlisting}[language=C++, style=myStyle]
template<typename Vector, typename Solution>
void dt_variables(Vector &residual,	const Solution &data) const;
\end{lstlisting}
which is expected to write $\mathcal{R}_{v,n}(v,t)$ into the variable \cpp{residual}.
Thus, if the flow $k\partial_k \lambda = \partial_{t_-}\lambda = \textrm{Flow}(\lambda)$ of some coupling $\lambda$ is to be integrated, one would implement
\begin{align}\label{eq:flowVar}
	\mathcal{R}_{v,n}(v,t) = \textrm{Flow}(\lambda)\,.
\end{align}
If an implicit method has been chosen for the variables, the method
\begin{lstlisting}[language=C++, style=myStyle]
template<uint to, typename NT, typename Solution>
void jacobian_variables(dealii::FullMatrix<NT> &jacobian,
                        const Solution &data) const;
\end{lstlisting}
will also be called to obtain the the jacobian of the variables.
Note that the standard form of the flow equations can be directly implemented into the numerical models, without the need for additional minus signs due to the definition of RG-time. One can convince oneself easily that this is true, as
\begin{align}
	\partial_{t_-} = k\partial_k = -\partial_{t}\quad \Rightarrow \quad\partial_t \lambda = -k\partial_k \lambda = - \textrm{Flow}(\lambda)\,,
\end{align}
in agreement with \labelcref{eq:assVar} and \labelcref{eq:flowVar}. The choice of RG-time has also been discussed above in \Cref{sec:fRG}.

\subsubsection*{FEM Assemblers}
\noindent
If, on the other hand, a FEM field discretisation is requested, four assemblers are currently available:
\begin{itemize}
	\setlength\itemsep{-0.8em}
	\item \begin{lstlisting}[language=C++, style=myStyle, numbers=none]
DiFfRG::CG::Assembler<typename Discretization, typename Model>;
\end{lstlisting}

	\item \begin{lstlisting}[language=C++, style=myStyle, numbers=none]
DiFfRG::DG::Assembler<typename Discretization, typename Model>;
\end{lstlisting}

	\item \begin{lstlisting}[language=C++, style=myStyle, numbers=none]
DiFfRG::dDG::Assembler<typename Discretization, typename Model>;
\end{lstlisting}

	\item \begin{lstlisting}[language=C++, style=myStyle, numbers=none]
DiFfRG::LDG::Assembler<typename Discretization, typename Model>;
\end{lstlisting}
\end{itemize}
These correspond, respectively, to the continuous Galerkin (CG), discontinuous Galerkin (DG), direct discontinuous Galerkin (dDG) and local discontinuous Galerkin (LDG) finite-element discretisations, which are detailed in \Cref{app:FEM} and \ref{app:assembler}. The first template parameter of these assemblers is a valid discretisation class type, i.e. a class which describes the mesh and the finite element basis. We will discuss this in a moment.
For a documentation of the assemblers, see also \Cref{app:assembler}.

In the following, we label the vector of all FE functions as $u = \{u_1, u_2, \dotsc\}$ and the set of extractors as $e = \{e_1, e_2, \dotsc\}$. 
One can implement any set of flow equations in field-space whose shape fits
\begin{align}\notag
	m_i(\partial_{t_+}u, u, t, x) + \partial_{x_j}F_{ij}(u,e,v,t,x,\dots) + s_i(u,v,e,t,x,\dots) & = 0 \\
	\partial_{t_+} v_n + \mathcal{R}_{v,n}(v,e,t)                                             & = 0
	\,,
	\label{eq:FEMsystem}
\end{align}
which is the PDE all of the FEM assemblers set up, see also \Cref{sec:FEM}. Here, we have also indicated the variable/extractor systems which are always solved alongside the FEM system if there is at least one variable/extractor present.

Depending on the assembler, additional information may be passed to the flux $F_{ij}$ and source $s_i$ methods, e.g. field-space derivatives.
We stress that the entire numerical procedure as detailed below is automatically performed by the \DiFfRG framework. The only thing the user needs to do is to choose a set of algorithms and then explicitly implement at least one of the methods $F_{ij}$ or $s_i$ (and possibly $m_i$).

In the O(N) example code, the three assemblers that also provide field derivatives (i.e. CG, dDG, LDG) are all showcased, see \Cref{ex:ONfiniteT} and the code itself at \texttt{Examples/ONfiniteT}. With an assembler that implements a FE discretisation, at least three methods of the numerical model are called to construct the first equation:
\begin{itemize}
	\item The method \cpp{mass} calculates the mass function of the system, i.e. $m_i(\partial_{t_+}u, u, x, t)$ in \labelcref{eq:FEMsystem}. Usually, $m_i(\partial_{t_+}u, u, x) = \partial_{t_+}u_i$ and this is the default behavior if this function is not explicitly re-implemented. The method is expected to write its result into \cpp{m_i}:
\begin{lstlisting}[language=C++, style=myStyle, numbers=none]
template<int dim, typename NumberType, typename Vector,
         typename Vector_dot, size_t n_fe_functions>
void mass(std::array<NumberType, n_fe_functions> &m_i,
          const dealii::Point<dim> &x,
          const Vector &u_i,
          const Vector_dot &dt_u_i) const;
\end{lstlisting}
	\item The function \cpp{flux} calculates the flux function of the system, i.e. $F_{ij}(u,e,v,t,x,\dots)$ in \labelcref{eq:FEMsystem}. The contribution of the flux to the residual is calculated by means of a partial integration against the test functions of the finite element discretisation, as explained in \Cref{sec:FEM}:
	\begin{align}
		- \int_{D_k} F_{ik}(u,t,x,\dots) \partial_{x_k}\varphi_j(x)
		+ \int_{\partial D_k} \hat{\boldsymbol{n}}_k \hat F_{ik}(u,t,x,\dots) \varphi_j(x)
		\label{eq:fluxCell}
	\end{align}
	In \labelcref{eq:fluxCell} we have restricted the partial integration onto a single cell. The second term cancels everywhere except on the boundary for continuous methods, but remains finite for DG methods. Thus, the flux function will be evaluated inside the mesh cells and (possibly) on the boundary faces - for DG methods it is additionally needed on the interfaces of cells. If not explicitly re-implemented, the default behavior of this method is to set $F_{ij} = 0$.  The method is expected to write its result into \cpp{F_ij}:
\begin{lstlisting}[language=C++, style=myStyle, numbers=none]
template<int dim, typename NumberType, typename Solutions,
         size_t n_fe_functions>
void flux(std::array<Tensor<1, dim, NumberType>, n_fe_functions> &F_ij,
          const dealii::Point<dim> &x,
          const Solutions &sol) const;
\end{lstlisting}
	\item The function \cpp{source} calculates the source function of the system, i.e. $s_i(u,v,e,t,x,\dots)$ in \labelcref{eq:FEMsystem}. This function is called on all cells of the mesh and directly integrated against test functions. The default implementation of the source term returns zero. The method is expected to write its result into \cpp{s_i}:
\begin{lstlisting}[language=C++, style=myStyle, numbers=none]
template<int dim, typename NumberType, typename Solutions,
         size_t n_fe_functions>
void source(std::array<NumberType, n_fe_functions> &s_i,
            const dealii::Point<dim> &x,
            const Solutions &sol) const;
\end{lstlisting}
\end{itemize}
In principle, most equations can be straight-forwardly implemented by writing only custom \cpp{flux} and/or \cpp{source} functions. One further necessary step is the inclusion of a boundary condition which is also discussed below.
In the following, we briefly explain a few details of what the Assemblers explicitly do and how to implement boundary conditions and Jacobians.

The choice of the basis functions $\varphi_j(x)$ is made by the \texttt{Discretisation} one uses. \DiFfRG implements the following discretisations:
\begin{itemize}
	\item \lstinline[language=C++,style=myStyle]|DiFfRG::CG::Discretisation<Components, NumberType, Mesh>|
	\item \lstinline[language=C++,style=myStyle]|DiFfRG::DG::Discretisation<Components, NumberType, Mesh>|
	\item \lstinline[language=C++,style=myStyle]|DiFfRG::LDG::Discretisation<Components, NumberType, Mesh>|
\end{itemize}
For CG the basis functions are continuous piecewise polynomials, with support on a small neighbourhood of cells. On the other hand, for the three DG assemblers, the basis functions are cell-wise discontinuous.
The discretisation classes are templated on the structure of the system, given by its \cpp{Components} description, the \cpp{NumberType}, e.g. \cpp{double} that should be used for the system and a description of the cell-mesh \cpp{Mesh}.

Additionally, the boundary term in \labelcref{eq:FEMsystem} needs to be specified, i.e. some kind of boundary condition has to be given. The specification of Dirichlet boundary conditions is explained in \Cref{app:model}.

Often, a flux formulation can be found for derivative expansions in the fRG; in O(N)-models and related setups this has been shown e.g. in \cite{Koenigstein:2021rxj, Koenigstein:2021syz, Grossi:2019urj, Ihssen:2022xkr, Ihssen:2023qaq, Ihssen:2023xlp}. In such a formulation, it is more appropriate to specify a Von-Neumann type boundary condition, which is naturally given by a direct extrapolation at the upper boundary. At the lower boundary, a similar condition can be applied, which is fully justified if one resolves $V(\sigma)$, which is $\mathbb{Z}_2$ symmetric around $\phi=(\sigma,\boldsymbol{\pi}) = 0$. A similar argument can be made if resolving $m^2(\rho) = \partial_{\rho}V(\rho)$ with $\rho= \phi^2 / 2$ by noting that $\partial_{\rho}V(\rho) = \frac{1}{\sigma}\partial_\sigma V(\sigma)$, leading to the appropriate Von-Neumann boundary condition for $m^2(\rho)$ to be the mirroring at $\rho = 0$.

In a \DiFfRG-setup these kinds of boundary conditions can be set by defining the model function
\begin{lstlisting}[language=C++, style=myStyle]
template <int dim, typename NumberType, typename Solutions, size_t n_fe_functions>
void boundary_numflux(
    std::array<Tensor<1, dim, NumberType>, n_fe_functions> &F_ij,
    const Tensor<1, dim> & normal,
    const Point<dim> &p,
    const Solutions &sol) const
{
  flux(F_ij, p, sol)
}
\end{lstlisting}
In this example, we straight-forwardly call the flux function on the inside of the boundary, effectively imposing $F_{ij}^+ = F_{ij}^-$ by means of $u^+ = u^-$. Here, a superscript $^+$ signifies the value to be extrapolated at the outside of the boundary cell, whereas $^-$ is the value inside. This procedure corresponds to the direct extrapolation of derivatives over the boundary. The above, rather standard boundary condition can also be implemented by using the base class
\begin{lstlisting}[language=C++, style=myStyle, numbers=none]
class ON_finiteT : public def::FlowBoundaries<ON_finiteT>, ... // Inflow/Outflow
\end{lstlisting}
which simply implements the method given above.

Finally, due to its inherently discontinuous nature, any DG setup necessitates the definition of boundary terms for every single cell, effectively specifying some kind of Riemann solver over the intersections. This is described in more detail in \Cref{app:model}, but once again this does not have to be done by hand - as an example, the Lax-Friedrichs numerical flux can be directly used like
\begin{lstlisting}[language=C++, style=myStyle, numbers=none]
class ON_finiteT : public def::LLFFlux<ON_finiteT>, ... // use a LLF numflux
\end{lstlisting}
We remark here that the LDG assembler requires a more complex setup, which is deferred to \Cref{app:LDG}.

There is another important ingredient for FE-simulations of fully field dependent quantities. As the main application of such methods are theories where spontaneous symmetry breaking takes place, the resulting PDEs are highly stiff, see \cite{Ihssen:2023qaq}. Explicit time-stepping becomes exponentially expensive - therefore, implicit time-stepping is usually much more efficient, resulting in a numerical speedup of multiple orders of magnitude for small systems. The trade-off is that implicit steppers require the calculation of the jacobians of the PDE, i.e. the derivative of \labelcref{eq:FEresidual},
\begin{align}
	J_{ij} = \frac{\partial R_i(u)}{\partial u_j}\,.
\end{align}
Deriving the explicit form of $J_{ij}$ can be very tedious, especially for large flow equations. Therefore, we provide the automatic evaluation of $J_{ij}$ by means of (forward) auto-differentiation. The implementation is based on the \texttt{autodiff} \cite{autodiff} library and can be used very easily by inheriting the numerical model from the \lstinline[language=C++,style=myStyle]|DiFfRG::def::AD| class:
\begin{lstlisting}[language=C++, style=myStyle, numbers=none]
class ON_finiteT : public def::AD<ON_finiteT>, ...
\end{lstlisting}
Usage of the \cpp{AD} class requires one to write all methods going into the construction of $R_i$ with arbitrary types in mind: If a residual is calculated, the \cpp{Solution} tuple is passed in populated with data of the type \cpp{double} and expects another \cpp{double} as output, whereas any evaluation for $J_{ij}$ passes in \cpp{autodiff::real} and also expects \cpp{autodiff::real} as output. Thus, using the \cpp{auto} keyword is highly useful here. As an example, one can easily write the flux of a Burger's equation as
\begin{lstlisting}[language=C++, style=myStyle]
template<int dim, typename NumberType, typename Solutions, size_t n_fe_functions>
void flux(std::array<Tensor<1, dim, NumberType>, n_fe_functions> &F_ij,
          const dealii::Point<dim> &x, const Solutions &sol) const
{
  const auto &fe_functions = get<"fe_functions">(sol);
  const auto u = fe_functions[idxf("u")];
  F_ij[idxf("u")][0] = 0.5 * u * u;
}
\end{lstlisting}
which accepts general types for both inputs and outputs.

If one wishes to not define jacobians, e.g. when using an explicit time-stepper, one can use the class \lstinline[language=C++,style=myStyle]|DiFfRG::def::NoJacobians|:
\begin{lstlisting}[language=C++, style=myStyle, numbers=none]
class ON_finiteT : public def::NoJacobians, ...
\end{lstlisting}
If jacobians are needed for the FE part of the calculation, but not for the variables, one may use the class \lstinline[language=C++,style=myStyle]|DiFfRG::def::FE_AD|:
\begin{lstlisting}[language=C++, style=myStyle, numbers=none]
class ON_finiteT : public def::FE_AD<ON_finiteT>, ...
\end{lstlisting}
In mixed FE-variable systems with an ImEx stepper this is highly convenient, as it removes the requirement of writing the variable and extractor methods to be type-agnostic.

\subsubsection{Component description}
\label{sec:componentsDesc}
\noindent
The \cpp{Components} struct, which is passed to the \lstinline[language=C++,style=myStyle]|DiFfRG::AbstractModel| base class template as shown above, is used to describe the structure of the discretisation and thus the entire system. For an O(N) model, this may look like
\begin{lstlisting}[language=C++, style=myStyle, numbers=none]
using FEFunctionDesc = FEFunctionDescriptor<Scalar<"u">>;
using Components = ComponentDescriptor<FEFunctionDesc>;
\end{lstlisting}
where the \cpp{FEFunctionDescriptor} describes the FE field discretisation. In this case, it is a single scalar field, which is described by the \cpp{Scalar} template. The \cpp{ComponentDescriptor} then describes the entire system, which in this case only contains the FE subsystem.

A more advanced example may look like
\begin{lstlisting}[language=C++, style=myStyle]
constexpr uint p_grid_size = 32;
constexpr uint angle_size = 7;
// Full field-dependent propagator
using FEFunctionDesc = FEFunctionDescriptor<
  FunctionND<"Gamma2", p_grid_size>
  >;
// Momentum-dependent full three-point and symmetric-point four-point vertices
using VariableDesc = VariableDescriptor<
  FunctionND<"Gamma3", p_grid_size, angle_size, angle_size>,
  FunctionND<"Gamma4", p_grid_size>
  >;
// Extract the propagator at the EoM to compute the three-point vertex
using ExtractorDesc = ExtractorDescriptor<
  FunctionND<"Gamma2", p_grid_size, angle_size, angle_size>
  >;
// Put the system together
using Components = ComponentDescriptor<
  FEFunctionDesc, VariableDesc, ExtractorDesc
  >;
constexpr auto idxf = FEFunctionDesc{};
constexpr auto idxv = VariableDesc{};
constexpr auto idxe = ExtractorDesc{};
\end{lstlisting}
We can use the index maps \cpp{idxf}, \cpp{idxv} and \cpp{idxe} to access indices of the subsystems in the code. These should always be declared to be compile-time objects by using \cpp{constexpr} for optimal performance. For example, to access the curvature mass, one may do the following:
\begin{lstlisting}[language=C++, style=myStyle]
const double m2 = variables[idxf("Gamma2")];
\end{lstlisting}
where the lookup is optimised to happen at compile-time.

Finally, for the LDG assembler, the component description is explained in \Cref{app:LDG}.

\subsubsection{Local data passing}
\label{sec:tupleStruct}
\noindent
To simplify the interface of the \textit{numerical model}, the assembler passes local data, i.e. the current state of the system, to the methods of the model packed into tuples. Usually this is the last argument to the method, e.g. in the flux function the parameter \cpp{sol},
\begin{lstlisting}[language=C++, style=myStyle]
template<int dim, typename NumberType, typename Solutions,
	size_t n_fe_functions>
void flux(std::array<Tensor<1, dim, NumberType>, n_fe_functions> &F_ij,
	const dealii::Point<dim> &x,
	const Solutions &sol) const
{
\end{lstlisting}
The last argument, \cpp{sol}, contains all the data passed on from the assembler. It can be accessed either index-wise, using \cpp{std::get}, or preferably using \cpp{DiFfRG::get}, passing the name of the required object:
\begin{lstlisting}[language=C++, style=myStyle]
const auto &fe_functions = DiFfRG::get<"fe_functions">(sol);
const auto &fe_derivatives = DiFfRG::get<"fe_derivatives">(sol);
const auto &fe_hessians = DiFfRG::get<"fe_hessians">(sol);
const auto &variables = DiFfRG::get<"variables">(sol);
const auto &extractors = DiFfRG::get<"extractors">(sol);
\end{lstlisting}
The objects \cpp{fe_functions, variables, extractors} are all vectors of data, either \cpp{std::vector}, \cpp{std::array} or \cpp{dealii::Vector}, with the layout specified in the \cpp{Components} class. Their values can be accessed as usual with the \cpp{[]}-operator.
\cpp{fe_derivatives} is a vector (one for each FE function) of order-1 tensors and \cpp{fe_hessians} is a vector of order-2 tensors, which can be accessed using multiple \cpp{[]}-operators.

The methods in the \textit{numerical model} usually get the following data structures passed in, in accordance with the data dependencies shown in \Cref{fig:dataRels}:
\begin{itemize}
	\item The variable subsystem, i.e. \cpp{dt\_variables} and its jacobian function, get only the \cpp{"variables"} and \cpp{"extractors"}.
	\item The FE functions, i.e. \cpp{flux} and \cpp{source} get the full set of objects, i.e. \lstinline[language=C++,style=myStyle]|"fe_functions", "variables", "extractors"| and also, assembler-dependent further ones:
	\begin{itemize}
		\item With the \texttt{dDG} and \texttt{CG} assemblers, additionally,  \cpp{"fe_derivatives"} and \cpp{"fe_hessians"} are passed. These represent all first derivatives of all FE functions and the full hessian matrices of all FE functions, respectively:
		\begin{align}
			&\text{\cpp{get<"fe_functions">(sol)}:}& &u(x) \notag\\[1ex]
			&\text{\cpp{get<"fe_derivatives">(sol)[i]}:}& &\partial_{x_i}u(x) \notag\\[1ex]			
			&\text{\cpp{get<"fe_hessians">(sol)[i][j]}:}& &\partial_{x_i}\partial_{x_j}u(x)\notag
		\end{align}
		\item Using the \texttt{LDG} assembler, the LDG subsystems as explained in \Cref{app:LDG} are passed, with the names \cpp{"LDG1"}, \cpp{"LDG2"}, \dots.
	\end{itemize}
	\item The initial condition functions and the mass do not get their data passed in using tuples, but get directly the FE functions (and their time derivatives, in the case of the mass) passed as vectors. It is not allowed to use the other data parts there, as this would make the jacobian construction significantly more complex. The same applies to the \texttt{LDG} \cpp{flux} and \cpp{source} functions, which have only access to the LDG-subsystem one level higher, as explained in \Cref{app:LDG}.
	\item Extractors get the entire set of objects at the EoM as explained above, except for the extractors themselves.
\end{itemize}
%

\subsubsection{Global data storage}
\noindent
To interface with the time-steppers, a class reimplementing the abstract \cpp{DiFfRG::AbstractFlowingVariables} should be passed.
Such a class holds the initial condition of a simulation in a \cpp{dealii::BlockVector} with two blocks: the first block holds FE functions, the second one variables. 
We implement this abstract interface in the 
\begin{lstlisting}[language=C++,style=myStyle]
template<typename Discretization>
DiFfRG::FlowingVariables;
\end{lstlisting}
class, which uses the \cpp{initial_condition} and \cpp{initial_conditon_variables} methods of a given method to interpolate the data at the intial RG scale:
\begin{lstlisting}[language=C++,style=myStyle]
FlowingVariables initial_condition(discretization);
initial_condition.interpolate(model);
\end{lstlisting}
%

\subsubsection{Full setup}
\noindent
As an example, we one can set up the logic of a full \DiFfRG-application in the following way:
\begin{lstlisting}[language=C++, style=myStyle]
#include <common/configuration_helper.hh>
#include <common/utils.hh>
#include <discretisation/discretisation.hh>
#include <timestepping/timestepping.hh>
#include "model.hh"

using namespace DiFfRG;

// Choices for types of algorithms
using Model = ON_finiteT;
constexpr uint dim = Model::dim;
using Discretisation = CG::Discretisation<Model::Components, double, RectangularMesh<dim>>;
using VectorType = typename Discretisation::VectorType;
using SparseMatrixType = typename Discretisation::SparseMatrixType;
using Assembler = CG::Assembler<Discretisation, Model>;
using TimeStepper = TimeStepperSUNDIALS_IDA<VectorType, SparseMatrixType, dim, UMFPack>;

int main(int argc, char *argv[])
{
	// get all needed parameters and parse from the CLI
	ConfigurationHelper config_helper(argc, argv);
	const auto json = config_helper.get_json();

	// Define the objects needed to run the simulation
	Model model(json);
	RectangularMesh<dim> mesh(json);
	Discretisation discretisation(mesh, json);
	Assembler assembler(discretisation, model, json);
	DataOutput<dim, VectorType> data_out(json);
	HAdaptivity mesh_adaptor(assembler, json);
	TimeStepper time_stepper(json, &assembler, &data_out, &mesh_adaptor);

	// Set up the initial condition
	FlowingVariables initial_condition(discretisation);
	initial_condition.interpolate(model);

	// Now we start the timestepping
	Timer timer;
	try {
		time_stepper.run(&initial_condition, 0.,
		  json.get_double("/timestepping/final_time"));
	} catch (std::exception &e) {
		spdlog::get("log")->error("Simulation finished with exception {}", e.what());
		return -1;
	}
	auto time = timer.wall_time();
	
	// Let the assembler write its performance information to the log
	assembler.log("log");
	spdlog::get("log")->info("Simulation finished after " + time_format(time));
	
	return 0;
}
\end{lstlisting}
This is the main file of the continuous Galerkin implementation of the O(N) finite temperature model described in \Cref{ex:ONfiniteT}.

\subsection{Computation}
\label{sec:design3}

\subsubsection{Regulator choice}
\noindent
\DiFfRG implements currently a list of standard and non-standard regulators, which can be found in the file \texttt{DiFfRG/physics/regulators.hh}.

If one uses the standard code generation of the Mathematica subpackage, a specific regulator can be chosen by modifying the code to generate the overall flow equation class:
{
	\footnotesize
	\begin{mmaCell}{Code}
		MakeFlowClass["ONfiniteT", kernels, "Regulator"->"RationalExpRegulator"]
	\end{mmaCell}
	\noindent\hspace{-7pt}
}
The default regulator is given by the \cpp{"PolynomialExpRegulator"}, which is a smooth regulator with a short tail, optimised for convexity-restoration problems.
The above Mathematica command will generate the code
\begin{lstlisting}[language=C++, style=myStyle]
#define __REGULATOR__ ::DiFfRG::PolynomialExpRegulator<>
\end{lstlisting}
in the file \texttt{flows/def.hh}, which is used by all flows generated to determine the used regulator class.

The standard implementation of this simply makes the functions
\begin{lstlisting}[language=C++,style=myStyle,numbers=none]
template<typename NT1, typename NT2> auto RF(NT1 k2, NT2 p2);
template<typename NT1, typename NT2> auto RB(NT1 k2, NT2 p2);
template<typename NT1, typename NT2> auto RFdot(NT1 k2, NT2 p2);
template<typename NT1, typename NT2> auto RBdot(NT1 k2, NT2 p2);
\end{lstlisting}
available within any integration kernel, where \cpp{NT1} and \cpp{NT2} are floating-point number types. One can change this behavior using the \mathem{SetKernelDefinitions[definitionsCode]}. In this context, also refer to the standard code which can be shown by using \mathem{ShowKernelDefinitions[]}.

If desired, one may also modify the regulator options, although this is slightly more technical. To do so, one has to use the further optional parameter \texttt{"RegulatorOptionCode"}:
{
	\footnotesize
	\begin{mmaCell}{Code}
		regOpts = "struct RegOpts {
			static constexpr int order = 10;
			constexpr static double c = 5.;
			constexpr static double b0 = 0.;
		};";
		MakeFlowClass["QCD", kernels, "Regulator" -> "RationalExpRegulator",
			"RegulatorOptionCode" -> {"RegOpts", regOpts}];

	\end{mmaCell}
	\noindent\hspace{-7pt}
}
which will lead to the code
\begin{lstlisting}[language=C++, style=myStyle]
struct RegOpts {
	static constexpr int order = 10;
	constexpr static double c = 5.;
	constexpr static double b0 = 0.;
};
#define __REGULATOR__ ::DiFfRG::RationalExpRegulator<RegOpts>
\end{lstlisting}
The necessary structure of the regulator options can be read off from the appropriate default structs in the file \texttt{DiFfRG/physics/regulators.hh}.

\subsubsection{Momentum integrations and flow generation}
\label{sec:momIntegrals}
\noindent
The flow kernels are optimised for use both on GPU and CPU during their generation by the Mathematica package. We do not rely on the code optimisation provided by \texttt{FORM}, as it concentrates on minimisation of mathematical operations during the evaluation of loops at the cost of large buffers necessary to hold intermediate results. While on modern CPUs the cost of these buffers is negligible due to their rather large hardware buffer sizes, local memory on GPUs cannot accommodate them and the compiler usually reserves the memory within slow, global GPU memory.

Therefore, we optimise conservatively for mathematical operations, but maximally for memory retrieval: If interpolator objects, which fetch data from global memory are called multiple times, this is optimised out by the code generation. The user can also define calls to other functions, e.g. regulators, to be optimised out by the code generation using
{
	\footnotesize
	\begin{mmaCell}{Code}
		AddCodeOptimizeFunctions[RBdot[__], RFdot[__], RB[__], RF[__]]
	\end{mmaCell}
	\noindent\hspace{-7pt}
}
The above line is automatically called within the \DiFfRG mathematica package. Here, we request regulator functions to be pre-evaluated, as some of the regulators included in \DiFfRG can become quite complex.

One can choose which kind of device to use for the integration routines.
Switching between devices is as simple as changing the choice in the kernel definitions:
{
	\footnotesize
	\begin{mmaCell}{Code}
	kernelAqbq1 = <|
			"Path"->"Aqbq",
			"Name"->"Aqbq",
			"Type"->"Quadrature",
			"Angles"->2,
			"d"->4,
			"AD"->False,
			"ctype"->"double",
			"Device"->"GPU"
	  |>;
		
	\end{mmaCell}
	\noindent\hspace{-7pt}
}
Here, we have chosen to run the flow equation on the GPU. By changing the \texttt{"Device"} option to \texttt{"CPU"}, one can use again the CPU algorithm.
\DiFfRG implements a large range of integration routines, all of which can be found in \\ \texttt{DiFfRG/physics/integration/} and \texttt{DiFfRG/physics/integration\_finiteT/}. Most of the integrators there implement multi-dimensional quadrature rules of arbitrary order. However, for some integrators we also provide an interface to a GPU quasi-Monte-Carlo integration library \cite{QMCgit}.
Valid choices for \texttt{"Type"} are:
\begin{itemize}
	\item \texttt{"Quadrature"} will use an appropriate quadrature which treats all space dimensions equally, i.e. in the vacuum.
	\item \texttt{"Quadratureq0"} and \texttt{"Quadraturex0"} will both lead to integration routines which treat the time direction in terms of a Matsubara sum. The former will perform an integration within a fixed, bounded imaginary time domain, whereas the latter will scale the domain with $k$. The second option is of course much more efficient if one regulates also the imaginary-time direction, e.g. with 4-dimensional regulators.
	\item \texttt{"QMC"} will use a bounded, adaptive quasi-Monte-Carlo method. This currently only works for vacuum computations.
\end{itemize}
As long as angular dependences are not strongly peaked, quadrature rules provide both very good performance and good numerical accuracy. Thus, numerical effort is closely linked to the choice of interaction tensor basis, see \cite{TensorBases}. In standard computations of structures such as a full vacuum quark-gluon vertex, we found the integration error of a medium-sized quadrature rule ($32\times8\times8$), measured against a high-precision adaptive integration, to be of the order $10^{-4}$.

\subsubsection{Data export}
\label{sec:DataOut}
\noindent
\DiFfRG allows for maximal flexibility when exporting data from the simulation. The full set of FE functions, their time derivatives and the error of the non-linear solver are all saved to disk at every save step. This data is saved to disk in the \texttt{VTK} unstructured grid format \footnote{see also the documentation \url{https://vtk.org/doc/nightly/html/classvtkUnstructuredGrid.html}} in \texttt{.vtu} files, separately for each time step. For easy access, a \texttt{.pvd} time-series file is always saved, so that the user may simply open the entire time-dependent data set in a software such as \texttt{ParaView}\footnote{\url{https://www.paraview.org/}}.

Additionally, the full set of variables, extractors and FE functions on the EoM is passed to a method \lstinline[language=C++,style=myStyle]|readouts(...)| in the numerical model, where the user may save and post-process the data however they wish.
For convenience, \DiFfRG provides a set of standard \texttt{.csv} data export classes, persistent throughout the RG-integration, so that data can be written in a continuous fashion to multiple \texttt{.csv} files.
For example, in the Yang-Mills example code, we save the running gluon mass parameter and momentum grids for the propagators as
\begin{lstlisting}[language=C++, style=myStyle]
  template <int dim, typename DataOut, typename Solutions>
  void readouts(DataOut &output, const Point<dim> &, const Solutions &sol) const
  {
  	// "unpack" the variables from the solution tuple
    const auto &variables = get<"variables">(sol);
    
    // get the gluon mass parameter
    const auto m2A = variables[idxv("m2A")];

    // access/create a continuous csv file, one row per time step
    auto &out_file = output.csv_file("data.csv");
    // set the initial UV scale
    out_file.set_Lambda(Lambda);
    // write the gluon mass to the file
    out_file.value("m2A", m2A);
	
    // access propagator dressings
    const auto *ZA = &variables.data()[idxv("ZA")];
    const auto *Zc = &variables.data()[idxv("Zc")];
    // create the data to be written and a header
    std::vector<std::vector<double>> Zs_data(p_grid.size(),
                                             std::vector<double>(5, 0.));
    const std::vector<std::string> Zs_header =
      std::vector<std::string>{"k [GeV]", "p [GeV]", "ZAbar", "ZA", "Zc"};
    for (uint i = 0; i < p_grid_size; ++i) {
      Zs_data[i][0] = k;
      Zs_data[i][1] = p_grid[i];
      Zs_data[i][2] = ZA[i];
      // this is the gluon propagator dressing without splitting the mass
      Zs_data[i][3] = (ZA[i] * powr<2>(p_grid[i]) + m2A) / powr<2>(p_grid[i]);
      Zs_data[i][4] = Zc[i];
    }

    // create a custom csv file, attach a large data set at each time step
    if (is_close(t, 0.))
      output.dump_to_csv("Zs.csv", Zs_data, false, Zs_header);
    else
      output.dump_to_csv("Zs.csv", Zs_data, true);
  }
\end{lstlisting}
In the future, \DiFfRG may also implement an interface to the HDF5 data format for a more structured approach to data management.

\subsubsection{Mesh adaptivity}
\noindent
Currently, \DiFfRG provides a mechanism to execute h-adaptive refinement, based on indicator functions. The indicator functions are integrated on the cell faces and interiors, respectively and their signatures read as in the following:
\begin{lstlisting}[language=C++, style=myStyle]
  template <int dim, typename NumberType,
            typename Solutions_s, typename Solutions_n>
  void face_indicator(std::array<NumberType, 2> &indicator,
                      const Tensor<1, dim> &normal,
                      const Point<dim> &p,
                      const Solutions_s &sol_s, const Solutions_n &sol_n) const;

  template <int dim, typename NumberType, typename Solution>
  void cell_indicator(NumberType &indicator, const Point<dim> &p,
                      const Solution &sol) const;
\end{lstlisting}
Both methods are expected to write the resulting indicator to their first argument.
The CG, DG, dDG assemblers provide FE functions, their direct derivatives and hessians to the indicators. The LDG assembler only provides only the FE functions, but also all LDG subsystems in the \cpp{Solution} type. Furthermore, the CG assembler does not evalute the \cpp{face\_indicator} method, as the discretisation is continuous across cell boundaries.
We showcase the adaptivity routines in the O(N) example below in \Cref{ex:ONfiniteT}.

\section{Summary and outlook}
\label{sec:outlook}

The \DiFfRG framework is still evolving and there are several straightforward extensions in preparation:
\begin{itemize}
	\item Especially for higher-dimensional potentials, as well as large systems of correlation functions, using large scale MPI and multi-GPU parallelisation would be very advantageous. Due to the modular structure of \DiFfRG this can be achieved without large changes to the code structure.
	\item Currently, \DiFfRG only implements h-adaptivity which is the most straight-forward way to refine a mesh during a computation. As the \texttt{deal.ii} library natively supports hp-adaptivity, we hope to soon add this to the framework. Furthermore, there is current work on r-adaptivity, which may be very well suited for fRG, relying on deforming the mesh, moving cells where they are most needed, rather than adding to the structure by refinement.
	\item Currently, only flow equations are automatically translated to C++ code. However, automatically generating the code is feasible and will make it even simpler to use the framework. With such facilities, we also expect simple python or Mathematica interfaces to be possible in the near future, without the need to write C++ code (at least for less complicated setups).
	\item Numerical tracing of tensor structures is, especially for theories with complicated tensor structures like QCD, a near-future aim. \texttt{FORM} is highly efficient for tracing large flow equations, but also has its limits. Very large equations due to tracing are not numerically tractable - e.g. equations with a code size above 1MB take several days to be compiled into GPU code. Therefore, a (partial) numerical tracing instead of a purely symbolic one may make these flows both more feasible and also tractable in less time.
  	\item Expanding and improving the currently present integration algorithms is a work-in-progress. Especially minimally invasive adaptivity for Matsubara sums is currently being worked on.
\end{itemize}
Furthermore, although currently it is most suited for continuous systems, extensions to lattice-based fRG setups are easily possible and envisaged for the future.

\section*{Acknowledgments}

We thank Andreas Geissel, Friederike Ihssen, Keiwan Jamaly and Nicolas Wink for discussion and collaboration on related projects.
This work is done within the fQCD collaboration \cite{fQCD} and we thank its members for discussions and collaborations  on related projects.
This work is funded by the Deutsche Forschungsgemeinschaft (DFG, German Research Foundation) under Germany’s Excellence Strategy EXC 2181/1 - 390900948 (the Heidelberg STRUCTURES Excellence Cluster) and the Collaborative Research Centre SFB 1225 (ISOQUANT).
FRS acknowledges funding by the GSI Helmholtzzentrum f\"ur Schwerionenforschung.


\appendix
\gdef\thesection{\Alph{section}}
\makeatletter
\renewcommand\@seccntformat[1]{\appendixname\ \csname the#1\endcsname.\hspace{0.5em}}
\makeatother
\begingroup
\allowdisplaybreaks

\section{FEM discretisations}
\label{app:FEM}

\subsection{Continuous Galerkin (CG)}
\label{app:CG}
\noindent
For the CG discretisation, the basis functions are chosen such that they peak at some vertex of the given mesh $\{D_k\}$ and fall off into the neighboring cells, such that they become zero at the nearest-neighbor vertices. Thus, the support of such a basis function is always given by all cells containing the associated vertex.

The associated assembler \cpp{DiFfRG::CG::Assembler} can be only used with the discretisation \cpp{DiFfRG::CG::Discretisation}.

\subsection{Discontinuous Galerkin (DG)}
\label{app:DG}
\noindent
For a review on discontinuous Galerkin methods, see \cite{GeneralShu} and references therein. Here, we give a general overview on the way the method works.
In \Cref{sec:FEM} we have introduced a global decomposition of some FE function $u$ into local basis functions. In the context of DG methods it is useful to write a local, cell-wise decomposition of $u$: within the $k$-th cell in the domain we have
\begin{align}
	u_{h}(x)\Big\vert_{x\in D_k} = \sum_i u_{h,k}^{(i)} \, \phi_i(x)\quad \phi_i \in V_h(D_k)\,,
\end{align}
where we now require that all cells $D_k$ share the same basis function space $V_h(D_k) = V_{h,\text{local}}$.
Due to the locality of the function spaces $V_{h}(D_k)$, the equations can be solved separately within each cell. 
The cell-wise discontinuity of this approach still has to be stabilised. Hence, one usually couples the cell-wise equations on the interfaces with all nearest neighbour cells using a so-called \textit{numerical flux}.

From here on we use the usual DG notation and define for any variable $r$ at some cell interface at position $x$ the inner and outer limits
\begin{align}
	r^-(x) = \lim\limits_{y\rightarrow x, \, y<x}r(y),\qquad
	r^+(x) = \lim\limits_{y\rightarrow x, \, y>x}r(y).
\end{align}
Furthermore, we define the shorthand expressions for averages and jumps at a cell interface,
\begin{align}
	[\![ r ]\!] := \hat{n} \, (r^- - r^+), \qquad
	\{\!\!\{ r \}\!\!\} := \frac{r^- + r^+}{2},
\end{align}
where $\hat{n}$ is the outward-facing normal vector orthogonal to the cell border. As $u_h$ is double-valued at cell interfaces, the above quantities are important for treating cell-boundary terms. Suppose the system to discretise reads
\begin{align}\label{eq:LDG_cont_system}
	\partial_t u + \partial_x \mathcal{F}(u,x) + s(u,x) = 0\,.
\end{align}
Integrating \labelcref{eq:LDG_cont_system} against the cell-wise test functions $\phi_i$, one obtains the system of equations to be solved:
\begin{align}\label{eq:DGresidual}
	M_{ij}\partial_t u_{h,k}^{(j)}(t)
	- \int_{D_k} \mathcal{F}(u_h,x) \partial_x \phi_i
	+ \int_{\partial D_k}
	\widehat{\mathcal{F}}(u^-_h,u^+_h,x) \cdot\hat{n}_k\, \phi_i + \int_{D_k} s(u_h,x)\, \varphi_i = 0\,.
\end{align}
Here, we have introduced $\hat{n}_k$, which is the cell-wise outward-facing normal vector. The mass matrix $ M_{ij}$ is defined as
\begin{align}
	M_{ij} = \int_{D_k} \phi_i\, \phi_j \,.
\end{align}
Importantly, by means of a partial integration, we have introduced the \textit{numerical flux} $\widehat{\mathcal{F}}$, connecting solution variables across cell boundaries.
For the flux of $F$, i.e.~$\widehat{F}$, a common choice is the standard Lax-Friedrichs flux,
\begin{align}\label{eq:LLFflux}
	\widehat{F} = \{\!\!\{F\}\!\!\} + \frac{c}{2}\,[\![u]\!]\,,
\end{align}
where $c\,(x)$ is the local wave-speed, given by the largest eigenvalue of $\frac{\partial F}{\partial u}$. One can show that the LLF flux is one of many numerical fluxes which allow for stable convection and a unique solution, see also the DG review \cite{GeneralShu}.
Note that although \labelcref{eq:DGresidual} is given on a single cell, the solutions of neighbouring cells are coupled by the presence of the numerical flux and thus not independent of each other. The sparsity of a standard DG implementation is therefore given by a nearest-neighbour coupling matrix.

The associated assembler \cpp{DiFfRG::DG::Assembler} can be only used with the discretisation \cpp{DiFfRG::DG::Discretisation}. We additionall define the assembler \cpp{DiFfRG::dDG::Assembler} which also computes derivatives and hessians of the FE functions as described in \Cref{sec:design2}. Note that this may be less stable and introduce numerical errors in comparison to the LDG assembler.
For an example of an explicit usage of \cpp{DiFfRG::dDG::Assembler}, see the DG example for the O(N) finite T example, \Cref{ex:ONfiniteT}.
The involved methods are described in \Cref{sec:design2} and documented in \Cref{app:model}.

\subsection{Local Discontinuous Galerkin (LDG)}
\label{app:LDG}
\noindent
Suppose we wish to evolve a field-dependent, vectorial quantity $\boldsymbol{u}(x) = (u_1(x), u_2(x), \dotsc)$. Schematically, the flow of $\boldsymbol{u}$ can be brought into the shape
\begin{align} \label{eq:orig_diffeq}
	\partial_t u_i + \partial_{x_j} F_{ij}(t, x, u, \dotsc) = s_i(t, x, {u}, \dotsc)\,,
\end{align}
which structurally resembles a convection-diffusion equation, where $t$ has been chosen like above as $t=-\log k/\Lambda$. For example, in any bosonic O(N)-theory, or also for a Yukawa theory, the flow equation of the boson mass $m_B^2$ can be brought into the above form. Formally, this is a second-order PDE with a non-linear flow.

To overcome difficulties with the diffusive contributions to the flow, one can utilise a so-called local discontinuous Galerkin Method (LDG). For example, in \cite{Ihssen:2023xlp}, an LDG method as derived in \cite{FENG201668} has been applied to the Quark-Meson model with a fully field-dependent Yukawa coupling.

In general, the LDG method consists of decomposing a higher-order equation into a set of first-order ODEs and PDEs, which can all be discretised using the standard DG method.
This is usually done by introducing stationary equations for new variable vectors ${g}^{(n)}$ accompanying the instationary equation(s),
\begin{align}\label{eq:LDG_eq}
	\partial_t u_i + \partial_x F_{ij}(t, x, u, g^{(1)}, \dotsc) &= s_i(t, x, u, g^{(1)}, \dotsc)\,,\qquad &&i=1\dots n\,, \notag\\[1ex]
	g^{(1)}_i &= \partial_{x_j} G^{(1)}_{f,ij}(u) + G^{(1)}_{s,i}({u})\,,\qquad && i=1\dots n_1\,, \notag\\[1ex]
	g^{(2)}_i &= \partial_{x_j} G^{(2)}_{f,ij}(g^{(1)}) + G^{(2)}_{s,i}(g^{(1)})\,,\qquad && i=1\dots n_2\,, \notag\\[1ex]
	\dotsc
\end{align}
The higher-order objects ${g}^{(n)}$, set up in this manner, are most straightforwardly identified as derivative-like objects. E.g. one could simply project directly $g^{(1)} = \partial_x^+ u$ using a DG method, where $\partial_x^+$ is the derivative from the right side. The right-side derivative can be then implemented by an appropriate choice of numerical flux.
In \DiFfRG we call the ODEs for ${g}^{(n)}$ \textit{LDG subsystems}. They can be specified in the Components after all the other parts of the equation-system,
\begin{lstlisting}[language=C++, style=myStyle]
using FEFunctionDesc = FEFunctionDescriptor<Scalar<"u">>;
using LDG1FunctionDesc = FEFunctionDescriptor<Scalar<"du">>;
using LDG2FunctionDesc = FEFunctionDescriptor<Scalar<"ddu">>;
using Components = ComponentDescriptor<
	FEFunctionDesc,
	VariableDescriptor<>, ExtractorDescriptor<>, // Empty descriptors
	LDG1FunctionDesc, LDG2FunctionDesc
	>;
constexpr auto idxf = FEFunctionDesc{};
constexpr auto idxl1 = LDG1FunctionDesc{};
constexpr auto idxl2 = LDG2FunctionDesc{};
\end{lstlisting}
The equations in \labelcref{eq:LDG_eq} can be then implemented with the two functions
\begin{lstlisting}[language=C++, style=myStyle]
template<uint dependent, int dim, typename NumberType, typename Vector, 
	       size_t n_fe_functions_dep>
void ldg_flux(std::array<Tensor<1, dim, NumberType>, n_fe_functions_dep> &G_f,
              const Point<dim> &x, const Vector &g_nm1) const;
\end{lstlisting}
and
\begin{lstlisting}[language=C++, style=myStyle]
template<uint dependent, int dim, typename NumberType, typename Vector, 
	       size_t n_fe_functions_dep>
void ldg_source(std::array<NumberType, n_fe_functions_dep> &G_s,
                const Point<dim> &x, const Vector &g_nm1) const;
\end{lstlisting}
which correspond to the above $G^{(n)}_f$ and $G^{(n)}_s$ in \labelcref{eq:LDG_eq}. Both get a vector \cpp{g_nm1} passed, which contains only the data of $g^{(n-1)}$, where $g^{(0)}:=u$.
As an example, the above setup could implement first and second derivatives by means of
\begin{lstlisting}[language=C++, style=myStyle]
template<uint submodel, typename NT, typename Variables>
void ldg_flux(
	std::array<Tensor<1, dim, NT>, Components::count_fe_functions(submodel)> &G_f, 
	const Point<dim> & /*x*/, const Variables &g_nm1) const
{
	if constexpr (submodel == 1) {
		G_f[idxl1("du")]][0] = g_nm1[idxf("u")];
	} else if constexpr (submodel == 2) {
		G_f[idxl2("ddu")][0] = g_nm1[idxl1("du")];
	}
}
\end{lstlisting}
Furthermore, one has to choose numerical fluxes for the LDG system. If one only wants to take derivatives from the left or the right, the class \cpp{LDGUpDownFluxes} can be used to create a set of left and right derivative numfluxes in arbitrary dimensions.
Otherwise, one should reimplement the function \cpp{ldg_numflux} to provide the numerical fluxes, see \Cref{app:model}.
As an example, a variant of the O(N) model in \Cref{ex:ONfiniteT} uses the LDG method to calculate the derivatives of $m_\pi^2 = \partial_\rho V(\rho)$.

\section{Assemblers and FE Discretisations}
\label{app:assembler}
\noindent
Here, we briefly list and detail the usage of the different available Assemblers in \DiFfRG. All assemblers require a given discretisation, which defines the test function space $V_h$ used for the FE discretisation. The space $V_h$ is of course dependent on the given mesh.
Currently, we pre-define one kind of mesh with rectangular cells:


\begin{mdframed}
	\cpp{DiFfRG::RectangularMesh}: A standard, rectangular mesh.
	\begin{lstlisting}[language=C++,style=myStyle,numbers=none]
template <uint dim> class RectangularMesh
\end{lstlisting}
	\vspace{0.2cm}\noindent\textbf{Template parameters}:\vspace{-0.2cm}
	\begin{itemize}
		\item \textbf{dim} Dimensionality $d$ of the created FE mesh. It has the constraint $1\leq d \leq 3$.
	\end{itemize}
	The constructor of this class reads
	\begin{lstlisting}[language=C++,style=myStyle,numbers=none]
RectangularMesh::RectangularMesh(const JSONValue &json)
\end{lstlisting}
	The resulting mesh is read out from the configuration tree in the \cpp{json} argument. The following values are used:
	\begin{itemize}
		\item \texttt{discretization/grid/x\_grid}: Must be a string specifying the mesh cells along the x-axis using a python-like slice notation. As an example, the following value may be given for \texttt{x\_grid}:
		\begin{lstlisting}[language=c++,style=myStyle,numbers=none]
		"x_grid": "0:2.5e-5:3.5e-3, 3.5e-3:2.5e-4:8e-3"
\end{lstlisting}%
		This leads to a mesh with cells of the size of $2.5\cdot10^{-5}$ from $0$ to $3.5\cdot10^{-3}$ and cells with a size of $2.5\cdot10^{-4}$ from $3.5\cdot10^{-3}$ to $8\cdot10^{-3}$.
		\item \texttt{discretization/grid/y\_grid}: This is only required, if $d=2$. It works the same way as \texttt{x\_grid}, specifying the size and position of cells along the y-axis.
		\item \texttt{discretization/grid/z\_grid}: This is only required, if $d=3$. It works the same way as \texttt{x\_grid}, specifying the size and position of cells along the z-axis.
		\item \texttt{discretization/grid/refine}: Must be an integer. Every mesh cell, after being defined through the previous arguments, will be regularly subdivided into $(2^d)^\textrm{refine}$ sub-cells.
	\end{itemize}
	\vspace{0.2cm}\noindent\textbf{Parameters}:\vspace{-0.2cm}
	\begin{itemize}
		\item \textbf{json} The configuration tree holding simulation parameters. This class reads from the \\ \texttt{discretization} subsection of the configuration as described above.
	\end{itemize}
\end{mdframed}

%
\noindent
\DiFfRG provides the following discretisations:
\begin{itemize}
	\item \begin{lstlisting}[language=C++,style=myStyle,numbers=none]
template<typename Components, typename NumberType, typename Mesh> 
class DiFfRG::CG::Discretization;
\end{lstlisting}
	\item \begin{lstlisting}[language=C++,style=myStyle,numbers=none]
template<typename Components, typename NumberType, typename Mesh> 
class DiFfRG::DG::Discretization;
\end{lstlisting}
	\item \begin{lstlisting}[language=C++,style=myStyle,numbers=none]
template<typename Components, typename NumberType, typename Mesh> 
class DiFfRG::LDG::Discretization;
\end{lstlisting}
\end{itemize}
The first template argument must be a component structure as explained in \Cref{sec:componentsDesc}, the second one either \cpp{double} or \cpp{float}, and the third any kind of mesh, i.e. a class that provides a method
\begin{lstlisting}[language=C++,style=myStyle,numbers=none]
dealii::Triangulation<dim> &get_triangulation();
\end{lstlisting}
All of the above discretisations have the same constructor signature:
\begin{lstlisting}
Discretization(Mesh &mesh, const JSONValue &json);
\end{lstlisting}
The first argument is an object of the \cpp{Mesh} type as also specified in the class template, whereas the \cpp{JSONValue} object holds configuration data which is used for defining the discretisation.
Finally, we describe briefly the four assemblers currently included with \DiFfRG:


\begin{mdframed}
	\cpp{DiFfRG::CG::Assembler}: The continuous Galerkin assembler.
	\begin{lstlisting}[language=C++,style=myStyle,numbers=none]
template<typename Discretization, typename Model>
class DiFfRG::CG::Assembler;
\end{lstlisting}
	The CG Assembler currently only accepts the \cpp{DiFfRG::CG::Discretization}. The model only has to implement the \cpp{source}, \cpp{flux} and \cpp{mass} functions, as well as standard functions all assemblers need.
\end{mdframed}


\begin{mdframed}
	\cpp{DiFfRG::DG::Assembler}: The discontinuous Galerkin assembler.
	\begin{lstlisting}[language=C++,style=myStyle,numbers=none]
template<typename Discretization, typename Model>
class DiFfRG::DG::Assembler;
\end{lstlisting}
	The DG Assembler currently only accepts the \cpp{DiFfRG::DG::Discretization}. The model has to implement the \cpp{source}, \cpp{flux} and \cpp{mass} functions, as well as standard functions all assemblers need. Additionally, a \cpp{numerical_flux} function as described in \Cref{app:model} has to be provided.
\end{mdframed}


\begin{mdframed}
	\cpp{DiFfRG::dDG::Assembler}: The direct discontinuous Galerkin assembler.
	\begin{lstlisting}[language=C++,style=myStyle,numbers=none]
template<typename Discretization, typename Model>
class DiFfRG::dDG::Assembler;
\end{lstlisting}
	The DG Assembler currently only accepts the \cpp{DiFfRG::dDG::Discretization}. The model has to implement the \cpp{source}, \cpp{flux} and \cpp{mass} functions, as well as standard functions all assemblers need. Additionally, a \cpp{numerical_flux} function as described in \Cref{app:model} has to be provided.
\end{mdframed}


\begin{mdframed}
	\cpp{DiFfRG::LDG::Assembler}: The local discontinuous Galerkin assembler.
	\begin{lstlisting}[language=C++,style=myStyle,numbers=none]
template<typename Discretization, typename Model>
class DiFfRG::LDG::Assembler;
\end{lstlisting}
	The LDG Assembler currently only accepts the \cpp{DiFfRG::LDG::Discretization}. The model has to implement the \cpp{source}, \cpp{flux} and \cpp{mass} functions, as well as standard functions all assemblers need. Additionally, a \cpp{numerical_flux} function as described in \Cref{app:model} has to be provided. For the LDG sub-systems, \cpp{ldg_flux} and \cpp{ldg_source} have to be given.
\end{mdframed}

%
\noindent
All assemblers have the same signature for the constructor:
\begin{lstlisting}[language=C++,style=myStyle,numbers=none]
Assembler(Discretization &discretization, Model &model, const JSONValue &json);
\end{lstlisting}
The assemblers take a discretisation object and a model object with types as specified in the class template. The \cpp{JSONValue} object holds configuration pertaining the assembly configuration.

Additionally, the variables assembler, which is described in \Cref{sec:varAssembler} can be used for systems without a spatial discretisation.

\section{Parameters}
\label{app:parameters}
\noindent
In this section, we very briefly explain the configuration options available through the JSON file/CLI interface.
The parameter file contains usually user-defined quantities in a "physical" subsection:
\begin{lstlisting}[language=C++,style=myStyle,numbers=none]
{
	"physical": {
		"Lambda" : 1.0
	},
\end{lstlisting}
The only object that always has to be defined in this section is the initial RG-scale \cpp{"Lambda"}.
\begin{lstlisting}[language=C++,style=myStyle,numbers=none]
	"discretization": {
		"fe_order": 3,
		"threads": 8,
		"batch_size": 64,
		"overintegration": 0,
		"output_subdivisions": 2,
		
		"EoM_abs_tol": 1e-10,
		"EoM_max_iter": 100,
		
		"grid": {
			"x_grid": "0:1e-2:1",
			"y_grid": "0:0.1:1",
			"z_grid": "0:0.1:1",
			"refine": 0
		},
\end{lstlisting}
The discretization section configures the FEM setup of our simulation:
\begin{itemize}
	\item \cpp{"fe_order"} sets the polynomial order of the finite element space on the mesh cells.
	\item \cpp{"threads"} sets the number of CPU threads used for assembly. Note, that other multithreading parts of \DiFfRG, such as momentum integration, automatically use all available CPU threads and ignore this parameter.
	\item \cpp{"batch_size"}: The assembly threads get batches of \cpp{"batch_size"} which they sequentially process. Playing around with \cpp{"threads"} and \cpp{"batch_size"} may give a small performance boost, but keeping \cpp{"threads"} around the number of physical cores and \cpp{"batch_size"} around 32-64 should be sufficient for almost optimal performance.
	\item \cpp{"overintegration"} can be set to a value $>0$ to increase the order of the quadratures used in assembly when constructing the of the PDE. It is seldom necessary to increase this beyond 0.
	\item \cpp{"output_subdivisions"} gives the precision with which the grids in the output data are written. This goes exponentially, so don't choose it too high.
	\item \cpp{"EoM_abs_tol"} sets the absolute precision within which a specified equation of motion is solved. DiFfRG can be instructed to solve the EoM at every time-step and perform additional computations at this point.
	\item \cpp{"EoM_max_iter"} sets the number of bisections used in determining the position of the EoM. If this is set to 0, searching for the EoM is skipped and the origin of the grid is always used as the EoM.
	\item \cpp{"x_grid"} this works in a python-like slice syntax and sets the grid structure used in a \cpp{"RectangularMesh"}. The parameter also supports locally different cell sizes: \cpp{"0:1e-4:1e-2, 1e-2:1e-3:1"} creates 100 cells between 0 and $10^{-2}$, and 100 cells between $10^{-2}$ and 1.
	\item \cpp{"y_grid"} and \cpp{"z_grid"} work identically, but are only used in 2D / 3D simulations.
	\item \cpp{"refine"} can be used to quickly increase the cell count of the specified grid by $2^{d\cdot\textrm{refine}}$, where $d$ is the dimension of the grid.
\end{itemize}
\begin{lstlisting}[language=C++,style=myStyle,numbers=none]
		"adaptivity": {
			"start_adapt_at": 0E0,
			"adapt_dt": 1E-1,
			"level": 0,
			"refine_percent": 1E-1,
			"coarsen_percent": 5E-2
		}
	},
\end{lstlisting}
The adaptivity section sets parameters for adaptive refinement of the discretization mesh. 
\begin{itemize}
	\item \cpp{"start_adapt_at"} sets the first RG-time at which mesh adaptivity is being used. 
	\item \cpp{"adapt_dt"} sets the RG-time intervals in which mesh adaptations occur.
	\item The \cpp{"level"} parameter gives the maximum number of bisections allowed for a single cell of the base mesh.
	\item \cpp{"refine_percent"} sets which fraction of all cells get refined during adaption of the mesh.
	\item \cpp{"coarsen_percent"} sets which fraction of all cells get coarsened during adaption of the mesh.
\end{itemize}
\begin{lstlisting}[language=C++,style=myStyle,numbers=none]
	"timestepping": {
		"final_time": 10.0,
		"output_dt": 1E-1,
		"explicit": {
			"dt": 1E-4,
			"minimal_dt": 1E-6,
			"maximal_dt": 1E-1,
			"abs_tol": 0.0,
			"rel_tol": 1E-3
		},
		"implicit": {
			"dt": 1E-4,
			"minimal_dt": 1E-6,
			"maximal_dt": 1E-1,
			"abs_tol": 1E-13,
			"rel_tol": 1E-7
		}
	},
\end{lstlisting}
The time-stepping section controls the behavior of the chosen time-stepper class:
\begin{itemize}
	\item \cpp{"final_time"} sets the end-time of the simulation. This is actually not an internally required parameter, but one can use it in the call to \cpp{timestepper.run(...);}
	\item \cpp{"output_dt"} sets the intervals within which \DiFfRG writes the solution to disk. Note, that the time-steppers do not necessarily step exactly onto \cpp{"output_dt"} intervals, but interpolate the solution to the output times.
	\item The \cpp{"explicit"} and \cpp{"implicit"} sections control the behavior of explicit and implicit time-steppers respectively. The distinction is important for the ImEx steppers, which utilise both explicit and implicit time-stepping.
	\item \cpp{"dt"} is the initial time-step for adaptive time-steppers and the fixed time-step for non-adaptive ones.
	\item \cpp{"minimal_dt"} and \cpp{"maximal_dt"} set the bounds for the time-step size - if an adpative stepper goes above or below, the step is adjusted to the bound, if the time-stepper gets stuck below \cpp{"mminimal_dt"}, time evolution is aborted.
	\item \cpp{"abs_tol"} and \cpp{"rel_tol"} set absolute and relative error tolerance for the time-stepping. While the FE parts of the system usually require a higher precision ($\lesssim 10^{-7}$ relative precision), especially in the presence of spontaneaous symmetry breaking, explicit tolerances can be chosen much more freely.
\end{itemize}
\begin{lstlisting}[language=C++,style=myStyle,numbers=none]
	"output": {
		"verbosity": 0,
		"folder": "./",
		"name": "output"
	}
	}
\end{lstlisting}
The \cpp{"output"} section defines:
\begin{itemize}
	\item \cpp{"verbosity"} sets how much information is being written to console while running. If at 0, no information is written at all, whereas at 1 the system gives updates at every time-step.
	\item \cpp{"folder"} sets the base folder where data is stored. This is useful to not clutter the current directory (\texttt{"./"}) with output, or if one runs a large amount of simulations.
	\item \cpp{"name"} sets the beginning of all filenames of the output, i.e. in this case all files created by the simulation start with \cpp{"output"}.
\end{itemize}
%

\section{Numerical model interface}
\label{app:model}
\noindent
The numerical model is the central object for any \DiFfRG simulation, where the full set of equations to be solved can be specified. Thus, we give a quick overview over all methods that can be re-implemented here.


\begin{mdframed}
\cpp{initial\_condition}: Set the initial condition for FE discretisations.
\begin{lstlisting}[language=C++,style=myStyle,numbers=none]
template <int dim, typename Vector> 
void initial_condition(const Point<dim> &x, Vector &u_i) const
\end{lstlisting}
This method has to be defined if a FEM assembler is used. It is evaluated point-wise on the entire computational mesh and the result interpolated to the chosen FE discretisation.

\vspace{0.2cm}\noindent\textbf{Parameters}:\vspace{-0.2cm}
\begin{itemize}
	\item \textbf{x} The point for which the initial condition should be given
	\item \textbf{u\_i} The value of the initial condition at \textbf{x}. This variable is vector-valued in all cases with as many components as has been specified by the \cpp{Components} struct. The method is expected to write into this variable.
\end{itemize}
\end{mdframed}


\begin{mdframed}
	\cpp{initial_condition_variables}: Set the initial condition for the variables.
	\begin{lstlisting}[language=C++,style=myStyle,numbers=none]
template <typename Vector> void initial_condition_variables(Vector &v_a) const
\end{lstlisting}
	Per default, this method does nothing, i.e. the variables remain zero-initialised.
	
	\vspace{0.2cm}\noindent\textbf{Parameters}:\vspace{-0.2cm}
	\begin{itemize}
		\item \textbf{v\_a} The vector of all variables. This method is expected to write the initial values of the variables into this argument. 
	\end{itemize}
\end{mdframed}


\begin{mdframed}
	\cpp{dt_variables}: Obtain the flow for the variables.
	\begin{lstlisting}[language=C++,style=myStyle,numbers=none]
template <typename Vector, typename Solution> 
void dt_variables(Vector &r_a, const Solution &sol) const
\end{lstlisting}
	Per default, this method does nothing and thus the variables have a flow of zero. The method is expected to write $\partial_{t_-} v$ into \textbf{r\_a}.
	
	\vspace{0.2cm}\noindent\textbf{Parameters}:\vspace{-0.2cm}
	\begin{itemize}
		\item \textbf{r\_a} The flow of all variables. This method is expected to write into \textbf{r\_a}.
		\item \textbf{sol} A tuple containing the current state of the system. It contains the following vectors, accessible using \cpp{DiFfRG::get<...>}:
		\begin{itemize}
			\item \cpp{"variables"}: The vector of all variables at the current RG-scale.
			\item \cpp{"extractors"}: If a FEM discretisation has been chosen, this gives the vector of all extractors from the EoM. 
		\end{itemize}
	\end{itemize}
\end{mdframed}


\begin{mdframed}
	\cpp{mass}: Obtain the mass function for FE methods.
	\begin{lstlisting}[language=C++,style=myStyle,numbers=none]
template <int dim, typename NumberType, typename Vector, 
          typename Vector_dot, size_t n_fe_functions>
void mass(std::array<NumberType, n_fe_functions> &m_i, const Point<dim> &x, 
          const Vector &u_i, const Vector_dot &dt_u_i) const
\end{lstlisting}
	As introduced in \Cref{sec:FEM}, this method is used to evaluate $m_i(\partial_{t}u, u, t, x)$ of the FE residual, given by
	\begin{align}\notag
		m_i(\partial_{t}u, u, t, x) + \partial_{x_j}F_{ij}(u,t,x,\dots) + s_i(u,t,x,\dots)  = 0\,,\quad i=1\dots n \,.
	\end{align}
	The default implementation sets $m_i(\partial_{t}u, u, t, x) = \partial_tu\,$.
	
	\vspace{0.2cm}\noindent\textbf{Parameters}:\vspace{-0.2cm}
	\begin{itemize}
		\item \textbf{m\_i} This is the variable the method is expected to write its result $m_i$ into.
		\item \textbf{x} The point $x\in\Omega$.
		\item \textbf{u\_i} The current value of $u_i(x)$, given as a vector of values.
		\item \textbf{dt\_u\_i} The current value of $\partial_t u_i(x)$, given as a vector of values.
	\end{itemize}
\end{mdframed}


\begin{mdframed}
	\cpp{flux}: Obtain the flux function for FE methods.
\begin{lstlisting}[language=C++,style=myStyle,numbers=none]
template <int dim, typename NumberType, typename Solutions, size_t n_fe_functions>
void flux(std::array<Tensor<1, dim, NumberType>, n_fe_functions> &F_ij, const Point<dim> &x,
const Solutions &sol) const
\end{lstlisting}
	As introduced in \Cref{sec:FEM}, this method is used to evaluate $F_{ij}(u,t,x,\dots)$ of the FE residual, given by
	\begin{align}\notag
		m_i(\partial_{t}u, u, t, x) + \partial_{x_j}F_{ij}(u,t,x,\dots) + s_i(u,t,x,\dots)  = 0\,,\quad i=1\dots n \,.
	\end{align}
	The default implementation sets $F_{ij}(u,t,x,\dots) = 0\,$.
	
	\vspace{0.2cm}\noindent\textbf{Parameters}:\vspace{-0.2cm}
	\begin{itemize}
		\item \textbf{F\_ij} This is the variable the method is expected to write its result $F_{ij}$ into.
		\item \textbf{x} The point $x\in\Omega$.
		\item \textbf{sol} A tuple of values, which can be accessed using \cpp{DiFfRG::get}. For a description, see \Cref{sec:tupleStruct}.
	\end{itemize}
\end{mdframed}


\begin{mdframed}
	\cpp{source}: Obtain the source function for FE methods.
\begin{lstlisting}[language=C++,style=myStyle,numbers=none]
template <int dim, typename NumberType, typename Solutions, size_t n_fe_functions>
void source(std::array<NumberType, n_fe_functions> &s_i, const Point<dim> &x, const Solutions &sol) const
	\end{lstlisting}
	As introduced in \Cref{sec:FEM}, this method is used to evaluate $s_{i}(u,t,x,\dots)$ of the FE residual, given by
	\begin{align}\notag
		m_i(\partial_{t}u, u, t, x) + \partial_{x_j}F_{ij}(u,t,x,\dots) + s_i(u,t,x,\dots)  = 0\,,\quad i=1\dots n \,.
	\end{align}
	The default implementation sets $s_{i}(u,t,x,\dots) = 0\,$.
	
	\vspace{0.2cm}\noindent\textbf{Parameters}:\vspace{-0.2cm}
	\begin{itemize}
		\item \textbf{s\_i} This is the variable the method is expected to write its result $s_{i}$ into.
		\item \textbf{x} The point $x\in\Omega$.
		\item \textbf{sol} A tuple of solution values, which can be accessed using \cpp{DiFfRG::get}. For a description, see \Cref{sec:tupleStruct}.
	\end{itemize}
\end{mdframed}


\begin{mdframed}
\cpp{boundary_numflux}: Obtain the source function for FE methods.
\begin{lstlisting}[language=C++,style=myStyle,numbers=none]
template <int dim, typename NumberType, typename Solutions, 
          size_t n_fe_functions>
void boundary_numflux(std::array<Tensor<1,dim,NumberType>,n_fe_functions> &F,
                      const Tensor<1,dim> &normal, const Point<dim> &x, 
                      const Solutions &sol) const
\end{lstlisting}
This method is used to evaluate $F_{ij}(u,t,x,\dots)$ at the boundary, which can be used to set a von-Neumann type boundary condition. No default implementation is given, but the default class \cpp{DiFfRG::def::FlowBoundaries<typename Model>} can be used to set standard inflow/outflow boundary conditions.

\vspace{0.2cm}\noindent\textbf{Parameters}:\vspace{-0.2cm}
\begin{itemize}
	\item \textbf{F} This is the variable the method is expected to write its result $F_{ij}$ into.
	\item \textbf{normal} The outward-facing normal on $\partial\Omega$.
	\item \textbf{x} The point $x\in\partial\Omega$.
	\item \textbf{sol} A tuple of solution values, which can be accessed using \cpp{DiFfRG::get}. For a description, see \Cref{sec:tupleStruct}.
\end{itemize}
\end{mdframed}


\begin{mdframed}
	\cpp{numflux}: Obtain the source function for FE methods.
	\begin{lstlisting}[language=C++,style=myStyle,numbers=none]
template <int dim, typename NumberType, typename Solutions_s, 
          typename Solutions_n, size_t n_fe_functions>
void numflux(std::array<Tensor<1,dim,NumberType>,n_fe_functions> &NF,
             const Tensor<1,dim> &normal, const Point<dim> &x, 
             const Solutions_s &sol_s, const Solutions_n &sol_n) const
\end{lstlisting}
	This method is only called if a DG-type assembler is used. It evaluates the numerical flux $\hat F_{ij}(u_h^-,u_h^+,x)$ between two cells, as defined in \labelcref{eq:DGresidual}. A typical flux is the LLF-flux, also introduced in \labelcref{eq:LLFflux}. 
	
	\vspace{0.2cm}\noindent\textbf{Parameters}:\vspace{-0.2cm}
	\begin{itemize}
		\item \textbf{NF} This is the variable the method is expected to write its result $\hat F_{ij}$ into.
		\item \textbf{normal} The outward-facing normal on $\partial\Omega$.
		\item \textbf{x} The point $x\in\Omega$.
		\item \textbf{sol\_s} A tuple of solution values inside the cell, which can be accessed using \cpp{DiFfRG::get}. For a description, see \Cref{sec:tupleStruct}.
		\item \textbf{sol\_n} A tuple of solution values from the neighboring cell.
	\end{itemize}
\end{mdframed}


\begin{mdframed}
	\cpp{differential_components}: Define which model components are PDEs and which are ODEs.
	\begin{lstlisting}[language=C++,style=myStyle,numbers=none]
template <uint dim>
std::vector<bool> differential_components() const
	\end{lstlisting}
	The standard definition tries to work out which of the $n$ equations in \labelcref{eq:FEMeq} contain a $\partial_t u_i$. This is relevant for the \texttt{SUNDIALS\_IDA} time-stepper, which can solve for a mixture of PDEs and ODEs.
	Usually, reimplementing this method is unnecessary as the automatic detection should almost always work. If it does not work, this can be a sign that the given system of equations is ill-posed.
	
	\vspace{0.2cm}\noindent\textbf{Returns}: A \cpp{std::vector<bool>} of $n$ boolean values, which indicate whether the equation contains a $\partial_t u_i$ or not.
\end{mdframed}


\begin{mdframed}
	\cpp{extract}: Obtains the extractor values from the equation of motion.
	\begin{lstlisting}[language=C++,style=myStyle,numbers=none]
template <int dim, typename Vector, typename Solutions>
void extract(Vector &e_a, const Point<dim> &x, const Solutions &sol) const
\end{lstlisting}
	This method is expected to calculate or read out the full set of specified extractors, see \Cref{sec:design2}. It is invoked only at the EoM and gets the solution vector passed as specified in \Cref{sec:tupleStruct}.
	
	\vspace{0.2cm}\noindent\textbf{Parameters}:\vspace{-0.2cm}
	\begin{itemize}
		\item \textbf{e\_a} The extractor values should be written to this vector. 
		\item \textbf{x} The EoM point at which \cpp{extract} is being invoked.
		\item \textbf{sol} The set of solution data at the EoM.
	\end{itemize}
\end{mdframed}


\begin{mdframed}
	\cpp{ldg_flux}: Calculate the LDG flux function $G_f$.
	\begin{lstlisting}[language=C++,style=myStyle,numbers=none]
template <uint dependent, int dim, typename NumberType, typename Vector, size_t n_fe_functions_dep>
void ldg_flux(std::array<Tensor<1, dim, NumberType>, n_fe_functions_dep> &G_f, const Point<dim> &x, const Vector &g_nm1) const
\end{lstlisting}
	This method evaluates $G_f$ as described in \Cref{app:LDG} for the subsystem of $g^{(n)}$, where $n$~=~\cpp{dependent}.
	
	\vspace{0.2cm}\noindent\textbf{Parameters}:\vspace{-0.2cm}
	\begin{itemize}
		\item \textbf{G\_f} The method is expected to write to this variable.
		\item \textbf{x} The point at which this method is called.
		\item \textbf{g\_nm1} The (vector-valued) FE solution $g^{(n-1)}$.
	\end{itemize}
\end{mdframed}


\begin{mdframed}
	\cpp{ldg_source}: Calculate the LDG source function $G_s$.
	\begin{lstlisting}[language=C++,style=myStyle,numbers=none]
template<uint dependent, int dim, typename NumberType, typename Vector, 
         size_t n_fe_functions_dep>
void ldg_source(std::array<NumberType, n_fe_functions_dep> &G_s, 
                const Point<dim> &x, const Vector &g_nm1) const
\end{lstlisting}
	This method evaluates $G_s$ as described in \Cref{app:LDG} for the subsystem of $g^{(n)}$, where $n$~=~\cpp{dependent}.
	
	\vspace{0.2cm}\noindent\textbf{Parameters}:\vspace{-0.2cm}
	\begin{itemize}
		\item \textbf{G\_s} The method is expected to write to this variable.
		\item \textbf{x} The point at which this method is called.
		\item \textbf{g\_nm1} The (vector-valued) FE solution $g^{(n-1)}$.
	\end{itemize}
\end{mdframed}


\begin{mdframed}
	\cpp{ldg_numflux}: Obtain the numflux for LDG subsystems
	\begin{lstlisting}[language=C++,style=myStyle,numbers=none]
template <uint dependent, int dim, typename NumberType,
          typename Solutions_s, typename Solutions_n
          size_t fe_functions_dep>
void ldg_numflux(std::array<Tensor<1,dim,NumberType>,fe_functions_dep> &NF,
                 const Tensor<1,dim> &normal, const Point<dim> &x, 
                 const Solutions_s &g_nm1_s, const Solutions_n &g_nm1_n) const
\end{lstlisting}
	This method evaluates the numflux $\hat{G}_f$ as described in \Cref{app:LDG} for the subsystem of $g^{(n)}$, where $n$~=~\cpp{dependent}.

	\vspace{0.2cm}\noindent\textbf{Parameters}:\vspace{-0.2cm}
	\begin{itemize}
		\item \textbf{NF} This is the variable the method is expected to write its result $\hat{G}_f$ into.
		\item \textbf{normal} The outward-facing normal on $\partial\Omega$.
		\item \textbf{x} The point $x\in\Omega$.
		\item \textbf{g\_nm1\_s} The (vector-valued) FE solution $g^{(n-1)}$ inside the cell.
		\item \textbf{g\_nm1\_n} The (vector-valued) FE solution $g^{(n-1)}$ outside the cell.
	\end{itemize}
\end{mdframed}


\begin{mdframed}
	\cpp{face_indicator}: Obtain an indicator on cell faces for adaptive refinement.
	\begin{lstlisting}[language=C++,style=myStyle,numbers=none]
template <int dim, typename NumberType, typename Solutions_s, typename Solutions_n>
void face_indicator(std::array<NumberType, 2> &indicator, 
    const Tensor<1, dim> &normal, const Point<dim> &x, 
    const Solutions_s &sol_s, const Solutions_n &sol_n) const
\end{lstlisting}
	This method is expected to write the indicator values into its first argument, which is then used to decide which cells get refined and which coarsened. Smaller values indicate coarsening, whereas larger values lead to refinement.
	
	\vspace{0.2cm}\noindent\textbf{Parameters}:\vspace{-0.2cm}
	\begin{itemize}
		\item \textbf{indicator} The method is expected to write into this argument. Its first value is the indicator inside the cell, the second value in the neighbouring cell.
		\item \textbf{normal} The outward-facing normal on $\partial\Omega$.
		\item \textbf{x} The point $x\in\Omega$.
		\item \textbf{sol\_s} A tuple of solution values inside the cell, which can be accessed using \cpp{DiFfRG::get}. For a description, see \Cref{sec:tupleStruct}.
		\item \textbf{sol\_n} A tuple of solution values from the neighboring cell.
	\end{itemize}
\end{mdframed}


\begin{mdframed}
	\cpp{cell_indicator}: Obtain a cell-wise indicator for adaptive refinement.
	\begin{lstlisting}[language=C++,style=myStyle,numbers=none]
template <int dim, typename NumberType, typename Solution>
void cell_indicator(NumberType & indicator, const Point<dim> &x, const Solution &sol) const
\end{lstlisting}
	This method is expected to write the indicator value into its first argument, which is then used to decide which cells get refined and which coarsened. Smaller values indicate coarsening, whereas larger values lead to refinement.
	
	\vspace{0.2cm}\noindent\textbf{Parameters}:\vspace{-0.2cm}
	\begin{itemize}
		\item \textbf{indicator} The method is expected to write into this argument.
		\item \textbf{x} The point $x\in\Omega$.
		\item \textbf{sol} A tuple of solution values, which can be accessed using \cpp{DiFfRG::get}. For a description, see \Cref{sec:tupleStruct}.
	\end{itemize}
\end{mdframed}


\begin{mdframed}
	\cpp{EoM}: Obtain the equation of motion.
	\begin{lstlisting}[language=C++,style=myStyle,numbers=none]
template <int dim, typename Vector> 
std::array<double, dim> EoM(const Point<dim> &x, const Vector &u) const
\end{lstlisting}
	This method is expected to evaluate the value of $\frac{\delta \Gamma_k}{\delta \Phi}(x)$, possibly normalized with the spacetime volume. Its result is used to solve the equation of motion
	\begin{align*}
		\frac{\delta \Gamma_k}{\delta \Phi}\Big\vert_{x=x_{\text{EoM},k}} = 0\,.
	\end{align*}
	For $x_{\text{EoM},k}$ to have a unique solution, this method is required to specify an equation with \cpp{dim} components. Per default, it returns \cpp{u[0]}.
	
	\vspace{0.2cm}\noindent\textbf{Parameters}:\vspace{-0.2cm}
	\begin{itemize}
		\item \textbf{x} The point $x \in \Omega$ where this method is currently evaluated.
		\item \textbf{u} The value $u(x)$ of the FE functions.
	\end{itemize}
\end{mdframed}


\begin{mdframed}
	\cpp{EoM_postprocess}: Post-process the EoM.
	\begin{lstlisting}[language=C++,style=myStyle,numbers=none]
template <int dim, typename Vector> 
Point<dim> EoM_postprocess(const Point<dim> &EoM, const Vector &u) const
\end{lstlisting}
	If wanted, this method can be used to either set the EoM to some specific point or modify its result. Per default, this method returns the EoM and does not change it.
	
	\vspace{0.2cm}\noindent\textbf{Parameters}:\vspace{-0.2cm}
	\begin{itemize}
		\item \textbf{EoM} The EoM as found by solving $\frac{\delta \Gamma_k}{\delta \Phi} = 0$.
		\item \textbf{u} The value of the FE functions at the EoM.
	\end{itemize}
	
	\vspace{0.2cm}\noindent\textbf{Returns}: The modified EoM value.
\end{mdframed}


\begin{mdframed}
	\cpp{readouts}: Write scalar data to .csv files.
	\begin{lstlisting}[language=C++,style=myStyle,numbers=none]
template <int dim, typename DataOut, typename Solutions>
void readouts(DataOut &dataOut, const Point<dim> &x, 
              const Solutions &sol) const
\end{lstlisting}
	The readouts function allows one to output information to disk. If a FEM discretisation is used, \cpp{readouts} is evaluated at the given equation of motion. This function is expected to utilised the passed \cpp{dataOut} object in order to create and write to .csv files.
	
	\vspace{0.2cm}\noindent\textbf{Parameters}:\vspace{-0.2cm}
	\begin{itemize}
		\item \textbf{dataOut} An object of type \cpp{DiFfRG::DataOutput<dim,VectorType>}. It can be used to write \texttt{.csv} files, see \Cref{sec:DataOut}.
		\item \textbf{x} The point $x \in \Omega$ where this method is currently evaluated.
		\item \textbf{sol} A tuple of solution values, which can be accessed using \cpp{DiFfRG::get}. For a description, see \Cref{sec:tupleStruct}.
	\end{itemize}
\end{mdframed}


\begin{mdframed}
	\cpp{readouts_multiple}: Give multiple EoMs to solve and use for data output.
	\begin{lstlisting}[language=C++,style=myStyle,numbers=none]
template <typename FUN, typename DataOut>
void readouts_multiple(FUN &helper, DataOut &dataOut) const
\end{lstlisting}
	This method can be used to find solutions to multiple equations of motion and read out values at all of them. As an example, this method can be used with a Quark-Meson model in the following way:
	\begin{lstlisting}[language=C++,style=myStyle,numbers=none]
	// chiral EoM
	helper(
	[&](const auto &x, const auto &u_i) {
		const auto sigma = std::sqrt(2. * x[0]);
		const auto m2Pion = u_i[idxf("m2Pion")];
		return m2Pion;
	},
	[&](auto &output, const auto &x, const auto &sol) { 
		this->readouts(output, x, sol, "data_chiral.csv", 0);
		});
	// physical EoM
	helper(
	[&](const auto &x, const auto &u_i) {
		const auto sigma = std::sqrt(2. * x[0]);
		const auto m2Pion = u_i[idxf("m2Pion")];
		return m2Pion - cSigma / (sigma + 1e-14);
	},
	[&](auto &output, const auto &x, const auto &sol) {
		this->readouts(output, x, sol, "data_running_EoM.csv", 1);
		});
\end{lstlisting}
	
	\vspace{0.2cm}\noindent\textbf{Parameters}:\vspace{-0.2cm}
	\begin{itemize}
		\item \textbf{helper} A method that should be called to find a specified EoM.
		\item \textbf{dataOut} An object of type \cpp{DiFfRG::DataOutput<dim,VectorType>}. It can be used to write \texttt{.csv} files, see \Cref{sec:DataOut}.
	\end{itemize}
\end{mdframed}


\begin{mdframed}
	\cpp{affine_constraints}: Set Dirichlet boundary conditions for the FE functions.
	\begin{lstlisting}[language=C++,style=myStyle,numbers=none]
template <int dim, typename Constraints>
void affine_constraints(Constraints &constraints, 
			const std::vector<IndexSet> &component_boundary_dofs,
			const std::vector<std::vector<Point<dim>>> &component_boundary_points)
\end{lstlisting}
	This method allows one to set Dirichlet boundary conditions for FE functions. These are set only when the assembler is (re-)initialised and thus usually not regularly updated at later times. Setting the boundary conditions is done by very directly interfacing to \cpp{deal.ii}, see also the corresponding documentation\footnote{The \cpp{deal.ii} documentation: \url{https://www.dealii.org/current/doxygen/deal.II/classAffineConstraints.html}}.
	
	As an example, if one would like to set the value of the first FE function to $u_1(x=0)=0$ at $x = 0$, one would do the following:
\begin{lstlisting}[language=C++,style=myStyle,numbers=none]
constexpr uint FE_idx = 0;
constexpr double value = 0;
// find out which of the points is the boundary x=0
if(is_close(component_boundary_points[FE_idx][0][0], 0.))
	constraints.add_constraint(
		component_boundary_dofs[FE_idx].nth_index_in_set(0), {}, value);
else if(is_close(component_boundary_points[FE_idx][1][0], 0.))
	constraints.add_constraint(
		component_boundary_dofs[FE_idx].nth_index_in_set(1), {}, value);
\end{lstlisting}

	\vspace{0.2cm}\noindent\textbf{Parameters}:\vspace{-0.2cm}
	\begin{itemize}
		\item \textbf{constraints} An object usually of type \cpp{dealii::AffineConstraints<double>}. It should be used to register any Dirichlet boundary conditions requested by the user.
		\item \textbf{component\_boundary\_dofs} The components of this \cpp{std::vector} correspond to each FE function defined in the \cpp{Components} struct. In each component, an \cpp{dealii::IndexSet} is stored, which holds the indices of the degrees of freedom at boundaries belonging to the respective FE function.
		\item \textbf{component\_boundary\_points} These are ordered just like the \textbf{component\_boundary\_dofs} and provide the corresponding coordinates of the boundary DoFs. This can be used to infer on which boundary a given DoF is located.
	\end{itemize}
\end{mdframed}


\section{Python interface}
\label{app:python}
\noindent
We provide a small collection of python functions and classes to process data created by \DiFfRG simulations. 
The python interface of \DiFfRG is usually not automatically installed. 
It can be installed in a virtual environment, or globally, by using the wheel (distribution package) contained with \DiFfRG. To do so, run
\begin{lstlisting}[language=Bash]
$ pip install python/dist/DiFfRG.whl
\end{lstlisting}
For the functionality of the python interface, refer to the python documentation\footnote{\url{https://satfra.github.io/DiFfRG/python/DiFfRG.html}} and the examples, which make use of it to load, post-process and visualize the data.

\section{Mathematica interface}
\label{app:mathematica}
\noindent
The Mathematica sub-package of \DiFfRG is automatically installed when the \bash{build.sh} or \bash{update_DiFfRG.sh} scripts are run.

\subsection{Kernel definitions}
\noindent
An integration kernel has to be specified using an association, which contains the following keys:
\begin{itemize}
	\item \mathem{"Path"} has to be set to a string containing a path relative to \texttt{flows/}. This is the folder where the flow equations for this integration kernel will be exported.
	\item \mathem{"Name"} has to be set to a string containing the name of the integration kernel.
	\item \mathem{"Type"} has to be set to either \mathem{"Quadrature"}, \mathem{"Quadratureq0"}, \mathem{"Quadraturex0"} or \mathem{"QMC"}, see also \Cref{sec:momIntegrals}.
	\item \mathem{"Angles"} must be an integer, giving how many angular integrations should be performed in the kernel. Depending on the choice, the user has to make sure angular variables are called correctly in the Mathematica code to be exported:
	\begin{itemize}
		\item For \mathem{"Angles"->1}, the angular variable will be called \cpp{cos1}.
		\item For \mathem{"d"->3} and \mathem{"Angles"->2}, the angular variables will be called \cpp{cos1} and \cpp{phi}.
		\item For \mathem{"d"->4} and \mathem{"Angles"->2}, the angular variables will be called \cpp{cos1} and \cpp{cos2}.
		\item For \mathem{"d"->4} and \mathem{"Angles"->3}, the angular variables will be called \cpp{cos1}, \cpp{cos2} and \cpp{phi}.
	\end{itemize}
	\item \mathem{"d"} must be an integer. Note that for more than one angle, not every choice of \mathem{"d"} has been implemented.
	\item \mathem{"AD"} must be a boolean defining whether autodifferentiation code should be generated. This is necessary for flows using implicit time-steppers.
	\item \mathem{"ctype"} must be either \mathem{"double"} or \mathem{"float"}, which sets the internally used type for computations.
	\item \mathem{"Device"} must be either \mathem{"GPU"} if one wants to perform the integration on GPU or \mathem{"CPU"} if the integration should be done on the CPU. Note, that the GPU algorithm falls back to the CPU if no GPU is available during compilation.
\end{itemize}
%

\subsection{Parameter definitions}
\noindent
The set of parameters passed to a kernel has to be specified using an association with the following entries:
\begin{itemize}
	\item \mathem{"Name"} must be a string which gives the name of the parameter to be passed.
	\item \mathem{"Type"} must be one of the following:
		\begin{itemize}
			\item \mathem{Constant}: A constant of type \cpp{ctype}.
			\item \mathem{Variable}: A variable of type \cpp{ctype} which can be also \cpp{autodiff::real} if \mathem{"AD"} has been set to true.
			\item \mathem{FunctionTex1D} An interpolator object of the type
			\begin{lstlisting}[language=C++,style=myStyle,numbers=none]
DiFfRG::TexLinearInterpolator1D<double, LogarithmicCoordinates1D<float>>
\end{lstlisting}
			
			\item \mathem{FunctionTex3D} An interpolator object of the type
			\begin{lstlisting}[language=C++,style=myStyle,numbers=none]
DiFfRG::TexLinearInterpolator3D<double,CoordinatePackND<LogarithmicCoordinates1D<float>,LogarithmicCoordinates1D<float>,LinearCoordinates1D<float>>>
\end{lstlisting}
			
			\item \mathem{FunctionTex3DLogLinLin} An interpolator object of the type
			\begin{lstlisting}[language=C++,style=myStyle,numbers=none]
DiFfRG::TexLinearInterpolator3D<double,CoordinatePackND<LogarithmicCoordinates1D<float>,LinearCoordinates1D<float>,LinearCoordinates1D<float>>>
\end{lstlisting}
			\item \mathem{Function1D} An interpolator object of the type
			\begin{lstlisting}[language=C++,style=myStyle,numbers=none]
DiFfRG::LinearInterpolator1D<double, LogarithmicCoordinates1D<float>>
\end{lstlisting}
			\item \mathem{Function3D} An interpolator object of the type
			\begin{lstlisting}[language=C++,style=myStyle,numbers=none]
DiFfRG::LinearInterpolator3D<double,CoordinatePackND<LogarithmicCoordinates1D<float>,LogarithmicCoordinates1D<float>,LinearCoordinates1D<float>>>
\end{lstlisting}
			\item \mathem{Function3DLogLinLin} An interpolator object of the type
			\begin{lstlisting}[language=C++,style=myStyle,numbers=none]
DiFfRG::LinearInterpolator3D<double, CoordinatePackND<LogarithmicCoordinates1D<float>, LinearCoordinates1D<float>, LinearCoordinates1D<float>>>
\end{lstlisting}
			\item \mathem{FunctionTex1DBosonicFT} An interpolator object of the type
			\begin{lstlisting}[language=C++,style=myStyle,numbers=none]
DiFfRG::TexLinearInterpolator1DStack<double, BosonicCoordinates1DFiniteT<int, float>>
\end{lstlisting}
			\item \mathem{FunctionTex1DFermionicFT} An interpolator object of the type
			\begin{lstlisting}[language=C++,style=myStyle,numbers=none]
DiFfRG::TexLinearInterpolator1DStack<double, FermionicCoordinates1DFiniteT<int, float>>
\end{lstlisting}
		\end{itemize}
	\item \mathem{"AD"} must be a boolean. If the function is called in a context where automatic differentiation is requested, parameters flagged with \mathem{True} will take the \cpp{autodiff::real} type.
\end{itemize}

\section{Derivation of the four-Fermi system in \DiFfRG}
\label{app:fourFermi}
\noindent
In this section, we go through the Mathematica code for the four-fermi example in \Cref{ex:fourFermi} and explain the main points along the way.
The code presented here is slightly shortened to make it more readable, for the full code refer to \texttt{Examples/FourFermi/Four-Fermion.nb}.

As a first step, we import the \DiFfRG Mathematica package and set a few preliminaries. Importantly, we inform \texttt{FormTracer} of all variables and functions we would like it to recognise.
Finally, we define custom simplification routines for the flows, which perform the simplification of flow equations more efficiently. Using the commands \mathem{SetStandardSimplify} and \mathem{SetStandardQuickSimplify}, one can determine which simplification routines are used by DiFfRG while deriving flow equations.
{
	\footnotesize
	\begin{mmaCell}[index=1]{Code}
		Get["DiFfRG`"]
		SetDirectory[GetDirectory[]];
		
		DefineFormAutoDeclareFunctions[ZA, lambda, gA];
		AddExtraVars[k, p0f, m2A, muq, T,
		(*regulators*)RQ, RQdot, RA, RAdot,
		(*Angles*)cospq,
		(*wavefunction renormalisations*)ZA, ZQ,
		(*propagators*)GA, GAInv, GQ, MQ];
		
		PostTraceRules = { ... };
		PreTraceRules = { ... };
		
		CustomSimplify = ...
		CustomQuickSimplify = ...
		
		SetStandardSimplify[CustomSimplify];
		SetStandardQuickSimplify[CustomQuickSimplify];
		
	\end{mmaCell}
	\noindent\hspace{-7pt}
}
We define two sets of Diagrammatic rules in the above, one set to be used before tracing and one after.
Before the tracing process, all tensor structures are inserted, dressed with a single scalar factor. After the tracing, the dressings may be expanded to more complex terms. This splitting is useful as additional, superfluous terms will slow down FORM and inflate its memory usage. As an example, a bosonic propagator $G^{\phi\phi}$ with a diagonal group structure may be given by
\begin{equation}
	G^{\phi\phi}_{ab} = \delta_{ab} \lambda_{\phi\phi}\,,\qquad\textrm{with}\quad\lambda_{\phi\phi}=\frac{1}{\boldsymbol{p}^2 + p_0^2 + m^2 + R_k(\boldsymbol{p})^2}\,.
\end{equation}
In this case, the first expression is used in the rules before tracing, and after tracing the second expression is used for the parametrisation of the dressing.
The derivation of the actual flow equations is performed using \texttt{QMeS}, and we use the following truncation: (for details, see \cite{Pawlowski:2021tkk}):
{
	\footnotesize
	\begin{mmaCell}{Code}
		fRGEq = {"Prefactor" -> {1/2},
			<|"type" -> "Regulatordot", "indices" -> {i, j}|>,
			<|"type" -> "Propagator", "indices" -> {i, j}|>};
		fields = <|"bosonic" -> {A[p, {v, c}]},
		"fermionic" -> {{qb[p, {d, c, f}], q[p, {d, c, f}]}}|>;
		Truncation = {{q, qb}, {A, A},{qb, q, A},{qb, qb, q, q}};
		SetupfRG = <|"MasterEquation" -> fRGEq,
		"FieldSpace" -> fields,
		"Truncation" -> Truncation|>;
		DiagramStyle = {A -> Orange, q -> Black};
		
	\end{mmaCell}
	\noindent\hspace{-7pt}
}
Next, code generation is addressed. We define a kernel for each of the ten four-Fermi couplings and specify what the integration routine should do - integrate out one angle in four dimensions and use a 3+1 dimensional quadrature on the GPU. 
{
	\footnotesize
	\begin{mmaCell}{Code}
		kernelLambda1=<|"Path"->"qbqqbq","Name"->"lambda1","Type"->"Quadratureq0","Angles"->1,
		"d"->4,"AD"->False,"ctype"->"double","Device"->"GPU"|>;
		
		...
		
		kernels={kernelLambda1,kernelLambda2,kernelLambda3,kernelLambda4,kernelLambda5,kernelLambda6,kernelLambda7,kernelLambda8,kernelLambda9,kernelLambda10};
		
	\end{mmaCell}
	\noindent\hspace{-7pt}
}
We also define the full list of parameters to be passed to the integration routines - everything that appears in the diagrammatic rules:
{
	\footnotesize
	\begin{mmaCell}{Code}		
		kernelParameterList = {
			<|"Name" -> "p0f", "Type" -> "Constant", "AD" -> False|>,
			<|"Name" -> "T", "Type" -> "Constant", "AD" -> False|>,
			<|"Name" -> "muq", "Type" -> "Constant", "AD" -> False|>,
			...
			<|"Name" -> "lambda10", "Type" -> "Variable", "AD" -> False|>
		};
		
	\end{mmaCell}
	\noindent\hspace{-7pt}
}
Finally, we generate the overall code structure and add some functions to be optimised as explained in \Cref{sec:design3}.
{
	\footnotesize
	\begin{mmaCell}{Code}		
		MakeFlowClassFiniteT["FourFermi",kernels]
		AddCodeOptimizeFunctions[RB[__],RF[__],RBdot[__],RFdot[__]]
		
	\end{mmaCell}
	\noindent\hspace{-7pt}
}
The command \mathem{MakeFlowClassFiniteT["FourFermi",kernels]} creates a directory \texttt{flows/} in the current directory with a \texttt{CMakeLists.txt} and several header and source files. These can be directly used in the simulation by calling \cmake{add_flows(<target> flows)} in the \texttt{CMakeLists.txt} of the Four-Fermi project.

After these preliminaries, the actual derivation of the four-fermi vertex is performed. We use the QMeS facilities as set up above to derive the flow of a general four-point function and reduce the number of diagrams by exploiting a symmetry of the diagrams:
{
	\footnotesize
	\begin{mmaCell}{Code}
		(*Diagrams*)
		DerivativeListqbqqbq = {qb[-p,{d1,c1,f1}], q[p,{d2,c2,f2}], 
			qb[-p,{d3,c3,f3}], q[p,{d4,c4,f4}]};
		Diagramsqbqqbqsidx = DeriveFunctionalEquation[
		SetupfRG,DerivativeListqbqqbq,"OutputLevel"->"SuperindexDiagrams"];
		symmetries = {{{1, 3}, {2, 4}, Plus}};
		Diagramsqbqqbqsidx = ReduceIdenticalFlowDiagrams[
		Diagramsqbqqbqsidx, DerivativeListqbqqbq, symmetries];
		Diagramsqbqqbq = SuperindexToFullDiagrams[
		Diagramsqbqqbqsidx, SetupfRG, DerivativeListqbqqbq];
		Diagramsqbqqbq = splitDiagrams[Diagramsqbqqbq, {}];
		PlotSuperindexDiagram[Diagramsqbqqbqsidx, SetupfRG, "EdgeStyle" -> DiagramStyle]
		
	\end{mmaCell}
	\noindent\hspace{-7pt}
}
These diagrams need to be traced out. To do so, we utilise some additional FORM code, provided by the \DiFfRG library:
{
	\footnotesize
	\begin{mmaCell}{Code}
		preTraceRule = {{PreambleFormRule, "Vector q,p;"}};
		postTraceRule = MakeP0FormRule[q, {p}, {p0f - I muq}];
		p0Projection = (#//.vec[p,0]:>p0f-I muq)&;
		
	\end{mmaCell}
	\noindent\hspace{-7pt}
}
The above defines two lists which we will pass to FORMTracer - the \mathem{preTraceRule} makes sure that FORM is aware of all momenta used in the projection, whereas the \mathem{postTraceRule} as generated by the \mathem{MakeP0FormRule} library function is used to split the imaginary-time component of the flow and set it to $p_{0,f} - i \mu_q$ for all external fermion legs. We do not set $p_{0,f} = \pi T$, the lowest fermionic Matsubara frequency, but use $p_{0,f} = \pi T e^{-k/\pi T}$. This choice removes some artificial temperature dependence in the flow if not fully resolving the frequency spectrum, as discussed in \cite{Fu:2015naa}.

Finally, we iterate over all couplings and contract the diagrams with the corresponding projection operators. These projection operators are obtained using the TensorBases package \cite{TensorBases}. The full calculation of diagrams consists of two steps:
\mathem{TraceDiagrams} traces all given (projected) diagrams, as obtained from QMeS and saves the resulting expressions into \texttt{TraceBuffer/<name>/Diagram<number>.m}. These intermediate expressions are then summed by using the \mathem{SumDiagrams} function, which inserts the \mathem{PostTraceRules} and simplifies them before summing them to one expression.
In a final step, this expression is then exported to C++ code using the \mathem{MakeKernel} function.
{
	\footnotesize
	\begin{mmaCell}{Code}
		ProjectorLambdai = Table[0, 10];
		LambdaiLoops = Table[0, {i, 1, 10}];
		
		For[i=1, i<=10, i++,
			ProjectorLambdai[[i]] = (-TBGetProjector["FierzCompleteNf2Nc34D",i,
				{p,d2,c2,f2},{-p,d1,c1,f1},{p,d4,c4,f4},{-p,d3,c3,f3}]);
			ProjectionLambdai = (ProjectorLambdai[[i]] Diagramsqbqqbq//.PreTraceRules//
				p0Projection)//.p->0;
			TraceDiagrams[16, "lambda" <> ToString[i], ProjectionLambdai,
				preTraceRule,postTraceRule];
		
			LambdaiLoops[[i]] = SumDiagrams[16,"lambda"<>ToString[i],0,
				(#//.PostTraceRules//p0Projection//Finalize//CustomSimplify)&];
		
			MakeKernel[Symbol["kernelLambda"<>ToString[i]],kernelParameterList,
				Re[LambdaiLoops[[i]]],2*etaQ*Symbol["lambda"<>ToString[i]]];
		];
		Clear[i];
		
	\end{mmaCell}
	\noindent\hspace{-7pt}
}
With the above, the flow equations have been fully derived and the corresponding code generated - it can be now utilised in the numerical model by invoking the integrators, see also the code in \texttt{Examples/FourFermi/model.hh}:
\begin{lstlisting}[language=C++,style=myStyle]
// for convenience, tie the arguments used for the integration kernels
const auto arguments = std::tie(p0f, prm.T, prm.muq, gAqbq1, etaA, etaQ, MQ, 
                                lambdas[0], lambdas[1], lambdas[2], lambdas[3],
                                lambdas[4], lambdas[5], lambdas[6], lambdas[7],
                                lambdas[8], lambdas[9]);

// run all integration kernels and write the result into the residual
std::apply([&](const auto &...args) {
	residual[0] = flow_equations.lambda1_integrator.get<double>(k, args...);
	residual[1] = flow_equations.lambda2_integrator.get<double>(k, args...);
	residual[2] = flow_equations.lambda3_integrator.get<double>(k, args...);
	residual[3] = flow_equations.lambda4_integrator.get<double>(k, args...);
	residual[4] = flow_equations.lambda5_integrator.get<double>(k, args...);
	residual[5] = flow_equations.lambda6_integrator.get<double>(k, args...);
	residual[6] = flow_equations.lambda7_integrator.get<double>(k, args...);
	residual[7] = flow_equations.lambda8_integrator.get<double>(k, args...);
	residual[8] = flow_equations.lambda9_integrator.get<double>(k, args...);
	residual[9] = flow_equations.lambda10_integrator.get<double>(k, args...);
  }, arguments);
\end{lstlisting}

\endgroup

\bibliographystyle{elsarticle-num}
\bibliography{ref-lib}

\end{document}